\documentclass[aps,prd,twocolumn,showpacs,amsmath,amssymb]{revtex4-1}
\usepackage{amsmath}
\usepackage{graphicx}
\usepackage{subfigure}
\usepackage{epstopdf}
\usepackage{color}
\usepackage{multirow}
\usepackage{setspace}
\usepackage{overpic}
\usepackage{amssymb}
\usepackage{amsfonts}
\usepackage[bookmarksnumbered, pdfstartview=FitH,colorlinks,urlcolor=blue, citecolor=blue,linkcolor=blue] {hyperref}
\usepackage{lineno}
\usepackage{bm}
\usepackage{rotating}
\usepackage[utf8]{inputenc}

\usepackage{array}

\renewcommand{\thetable}{\arabic{table}}

\let\oldequation\equation
\let\oldendequation\endequation

\renewenvironment{equation}
{\linenomathNonumbers\oldequation}
{\oldendequation\endlinenomath}

\newcommand{\pilnu}{D \to \pi\ell^+\nu_{\ell}}
\newcommand{\pienu}{D^0\to \pi^-e^+\nu_e}
\newcommand{\pimunu}{D^0\to \pi^-\mu^+\nu_\mu}
\newcommand{\pizenu}{D^+\to \pi^0e^+\nu_e}
\newcommand{\pizmunu}{D^+\to \pi^0\mu^+\nu_\mu}

\newcommand{\ffpi}{f^{D\to\pi}_+(0)}

\newcommand{\BFpienu}{2.950\pm0.017_{\rm stat.}\pm 0.017_{\rm syst.}}
\newcommand{\BFpimunu}{2.817\pm0.037_{\rm stat.}\pm 0.019_{\rm syst.}}
\newcommand{\BFpizenu}{3.622\pm0.034_{\rm stat.}\pm 0.018_{\rm syst.}}
\newcommand{\BFpizmunu}{3.507\pm0.043_{\rm stat.}\pm 0.026_{\rm syst.}}

\newcommand{\bfpienu}{(2.950\pm0.017_{\rm stat.}\pm 0.017_{\rm syst.})\times10^{-3}}
\newcommand{\bfpimunu}{(2.817\pm0.037_{\rm stat.}\pm 0.019_{\rm syst.})\times10^{-3}}
\newcommand{\bfpizenu}{(3.622\pm0.034_{\rm stat.}\pm 0.018_{\rm syst.})\times10^{-3}}
\newcommand{\bfpizmunu}{(3.507\pm0.043_{\rm stat.}\pm 0.026_{\rm syst.})\times10^{-3}}

\newcommand{\lfudz}{0.947 \pm 0.014_{\rm stat.} \pm 0.005_{\rm syst.}}
\newcommand{\lfudp}{0.973 \pm 0.015_{\rm stat.} \pm 0.006_{\rm syst.}}

\newcommand{\pwre}{1.025 \pm 0.013_{\rm stat.} \pm 0.007_{\rm syst.}}
\newcommand{\pwrmu}{1.011 \pm 0.019_{\rm stat.} \pm 0.010_{\rm syst.}}

\newcommand{\fvcd}{0.1425\pm0.0005_{\rm stat.}\pm0.0003_{\rm syst.}}
\newcommand{\fpi}{0.6339\pm0.0024_{\rm stat.}\pm0.0014_{\rm syst.}}
\newcommand{\vcdpi}{0.2262\pm0.0008_{\rm stat.}\pm0.0005_{\rm syst.}\pm0.0018_{\rm LQCD.}}

\newcommand{\BESIIIorcid}[1]{\href{https://orcid.org/#1}{\hspace*{0.1em}\raisebox{-0.45ex}{\includegraphics[width=1em]{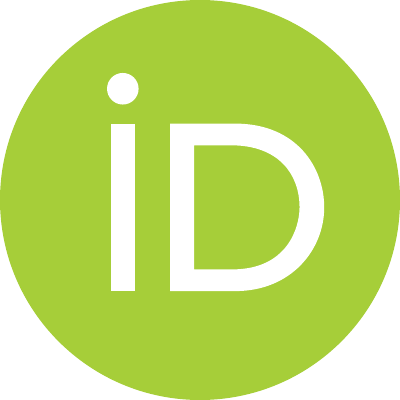}}}}

\begin{document}

%\linenumbers

\title{\boldmath Precision measurements of semleptonic decays $D^0 \to \pi^-\ell^+\nu_\ell$ and $D^+ \to \pi^0\ell^+\nu_\ell$ ($\ell =e,\mu$)}

%% Saved at => 2026-04-08
\author{M.~Ablikim$^{1}$\BESIIIorcid{0000-0002-3935-619X},
M.~N.~Achasov$^{4,c}$\BESIIIorcid{0000-0002-9400-8622},
P.~Adlarson$^{83}$\BESIIIorcid{0000-0001-6280-3851},
X.~C.~Ai$^{89}$\BESIIIorcid{0000-0003-3856-2415},
C.~S.~Akondi$^{31A,31B}$\BESIIIorcid{0000-0001-6303-5217},
R.~Aliberti$^{39}$\BESIIIorcid{0000-0003-3500-4012},
A.~Amoroso$^{82A,82C}$\BESIIIorcid{0000-0002-3095-8610},
Q.~An$^{78,65,\dagger}$,
Y.~H.~An$^{89}$\BESIIIorcid{0009-0008-3419-0849},
M.~S.~Anderson$^{39}$\BESIIIorcid{0009-0008-1550-2632}
% ...
\\
\vspace{0.2cm}
(BESIII Collaboration)\\
\vspace{0.2cm} {\it
$^{1}$ Institute of High Energy Physics, Beijing 100049, People's Republic of China\\
% ...
$^{4}$ Budker Institute of Nuclear Physics SB RAS (BINP), Novosibirsk 630090, Russia\\
% ...
$^{31}$ INFN Sezione di Ferrara, (A)INFN Sezione di Ferrara, I-44122, Ferrara, Italy; (B)University of Ferrara, I-44122, Ferrara, Italy\\
% ...
$^{39}$ Johannes Gutenberg University of Mainz, Johann-Joachim-Becher-Weg 45, D-55099 Mainz, Germany\\
% ...
$^{65}$ State Key Laboratory of Particle Detection and Electronics, Beijing 100049, Hefei 230026, People's Republic of China\\
% ...
$^{78}$ University of Science and Technology of China, Hefei 230026, People's Republic of China\\
% ...
$^{82}$ University of Turin and INFN, (A)University of Turin, I-10125, Turin, Italy; (B)University of Eastern Piedmont, I-15121, Alessandria, Italy; (C)INFN, I-10125, Turin, Italy\\
$^{83}$ Uppsala University, Box 516, SE-75120 Uppsala, Sweden\\
% ...
$^{89}$ Zhengzhou University, Zhengzhou 450001, People's Republic of China\\
\vspace{0.2cm}
$^{\dagger}$ Deceased\\
% ...
$^{c}$ Also at the Novosibirsk State University, Novosibirsk, 630090, Russia\\
% ...
}
}
%% ends here %%

\begin{abstract}
The branching fractions of $D^0\to \pi^-e^+\nu_e$, $D^0\to \pi^-\mu^+\nu_\mu$, $D^+\to \pi^0e^+\nu_e$, and $D^+\to \pi^0\mu^+\nu_\mu$ are measured to be $(2.950\pm0.017_{\rm stat.}\pm 0.017_{\rm syst.})\times10^{-3}$, $(2.817\pm0.037_{\rm stat.}\pm 0.019_{\rm syst.})\times10^{-3}$, $(3.622\pm0.034_{\rm stat.}\pm 0.018_{\rm syst.})\times10^{-3}$, and $(3.507\pm0.043_{\rm stat.}\pm 0.026_{\rm syst.})\times10^{-3}$ using $e^+e^-$ collision data with an integrated luminosity of 20.3 fb$^{-1}$ collected at the center-of-mass energy of 3.773 GeV with the BESIII detector. The partial decay rates of these four decays are measured with the best precision to date and their forward-backward asymmetries are determined for the first time. By performing a simultaneous fit to these results, the product of the hadronic transition form factor $f^{D\to\pi}_+(0)$ and the modulus of the $c\to d$ Cabibbo-Kobayashi-Maskawa matrix element $|V_{cd}|$ is given by $f^{D\to\pi}_+(0)|V_{cd}|=0.1425\pm0.0005_{\rm stat.}\pm0.0003_{\rm syst.}$. Taking the $|V_{cd}|$ provided by the standard model global fit and the $f^{D\to\pi}_+(0)$ calculated from the lattice quantum chromodynamics as input, we obtain $f^{D\to\pi}_+(0)=0.6339\pm0.0024_{\rm stat.}\pm0.0014_{\rm syst.}$ and $|V_{cd}|=0.2262\pm0.0008_{\rm stat.}\pm0.0005_{\rm syst.}\pm0.0018_{\rm LQCD.}$, respectively. The reported results have the best precision to date. We also search for the scalar current contribution in the $c\to d \ell^+\nu_{\ell}$ transition and determine Re$(C_S^\mu)=$ $0.022 \pm 0.023_{\rm stat.}\pm 0.003_{\rm syst.}$ and $|{\rm Im}(C_S^\mu)|=0.000 \pm $ $0.038_{\rm stat.} \pm 0.012_{\rm syst.}$. In addition, the lepton flavor universality is tested with the ratios of the decay rates between semimuonic and semielectronic decays in full and several $\ell^+\nu_\ell$ four-momentum transfer ranges.
\end{abstract}

\maketitle

\oddsidemargin  -0.2cm
\evensidemargin -0.2cm

\section{Motivation}

Experimental studies of the semileptonic decays of charmed mesons are important for understanding  the weak and strong
interactions in the charm sector.
Through analysis of the decay dynamics, one can extract the product of the modulus of the Cabibbo-Kobayashi-Maskawa (CKM) matrix
element $|V_{cs(d)}|$ and the hadronic transition form factor.
Taking $D\to \pi e^+\nu_e$ as an example, the hadronic transition form factors at the zero-momentum transfer $f^{D\to\pi}_+(0)$~\cite{Melikhov:2000yu,Verma:2011yw,Soni:2018adu,Ivanov:2019nqd,Faustov:2019mqr,Khodjamirian:2000ds,Wang:2002zba,Wu:2006rd,FermilabLattice:2004ncd,Na:2011mc,Lubicz:2017syv,FermilabLattice:2022gku} can be calculated via several theoretical approaches, e.g.,
constituent quark model (CQM)~\cite{Melikhov:2000yu}, covariant light-front quark model (LFQM)\cite{Verma:2011yw}, covariant confined quark model (CCQM) \cite{Soni:2018adu,Ivanov:2019nqd}, relativistic quark model (RQM) \cite{Faustov:2019mqr}, QCD light-cone sum rules
(LCSR)~\cite{Khodjamirian:2000ds,Wang:2002zba,Wu:2006rd}, and lattice quantum chromodynamics~(LQCD)~\cite{FermilabLattice:2004ncd,Na:2011mc,Lubicz:2017syv,FermilabLattice:2022gku}.
Using the $|V_{cd}|$ value provided by the CKMFitter group~\cite{pdg2024}, the hadronic transition form factors can be extracted, resulting in a stringent test of the theoretical predictions. Alternatively, assuming a $f^{D\to\pi}_+(0)$ value predicted by theory leads to $|V_{cd}|$, which is important for the test of CKM matrix unitarity.

In addition, lepton flavor universality (LFU), a fundamental principle predicted by the Standard Model (SM), asserts that the interactions between gauge bosons and the three generations of leptons should be equivalent. Consequently, any observed deviation from LFU implies new physics (NP) beyond the SM. There have been reported hints of LFU anomalies in the processes $b \to c \ell^+\bar{\nu}_{\ell}$, which deviate from the SM predictions by $(2\sim 3) \sigma$~\cite{Li:2018lxi,Bifani:2018zmi}. Various NP models, including two-Higgs-doublet models~\cite{Iguro:2022uzz,Blanke:2022pjy}, leptoquark scenarios~\cite{Sakaki:2013bfa,Becirevic:2018afm}, and model-independent investigations~\cite{Iguro:2024hyk}, suggest that these discrepancies may arise from scalar current contribution in weak interactions. 
Investigating the $c\to s(d)\ell^+\nu_\ell$ transition in the semileptonic decays of charmed mesons is crucial for understanding these discrepancies~\cite{Fajfer:2015ixa,Leng:2020fei,Jain:2025kqe}. Among the Cabibbo-suppressed semileptonic $D$ decays, the $D\to \pi\ell^+\nu_\ell$ processes are 
particularly promising for probing scalar current contributions in the $c\to d\ell^+\nu_\ell$ transition,
thanks to their relatively large branching fractions and high detection efficiencies. The relevant effective Lagrangian can be expressed as~\cite{Fajfer:2015ixa,Leng:2020fei,Jain:2025kqe}
\begin{equation}
	\mathcal{L}_{\rm eff}=-\frac{4 G_{F}}{\sqrt{2}} V_{c q}^{*} \sum_{\ell=e, \mu, \tau} \sum_{i = R, L} C_{i}^{\ell} \mathcal{O}_{i}^{\ell}+{\rm H.c. }
	\label{eqution:Leff}
\end{equation}
In the SM, the only permissible four-fermion operator is $\mathcal{O}_{\mathrm{SM}}^{\ell}= \left(\bar{q} \gamma_{\mu} P_{L} c\right)\left(\bar{\nu}_{\ell} \gamma^{\mu} P_{L} \ell\right)$ with  the coefficient $C_{\rm SM}^{\ell} = 1$. Conversely, the possible right(left)-handed scalar currents are represented by the NP operators $\mathcal{O}_{R(L)}^{\ell}=\left(\bar{q} P_{R(L)} c\right)\left(\bar{\nu}_{\ell} P_{R} \ell\right)$, where $C_{R(L)}^{\ell}$ are complex Wilson coefficients.
Here, $\gamma_{\mu}$ denotes the Dirac matrices, while $P_{R(L)}$ stand for the right-(left-) handed chirality projection operators.
Non-zero values of  $C_{R(L)}^{\ell}$ could lead to noticeable deviations from the SM in various observables, including both full and partial decay rates, as well as forward-backward asymmetries. Therefore, precise measurements of these observables are important to restrict the contributions of scalar currents in the $c\to d\ell^+\nu_\ell$ transition.

Previously, the branching fraction measurements and decay dynamics analyses of $D^0 \to \pi^-\ell^+\nu_\ell$ and $D^+ \to \pi^0\ell^+\nu_\ell$ came from BaBar~\cite{BaBar:2014xzf}, CLEO~\cite{CLEO:2005cuk,CLEO:2005rxg,CLEO:2007ntr,CLEO:2009svp}, Belle~\cite{Belle:2006idb} and BESIII~\cite{BESIII:2015tql,BESIII:2017ylw,BESIII:2018nzb} experiments.
With the exception of Belle, which investigated 
the $D^0\to \pi^-\mu^+\nu_\mu$ decay dynamics, other experiments only studied the positron channels.
The previous best measurements are from the BESIII analyses of $D^0 \to \pi^-e^+\nu_e$~\cite{BESIII:2015tql} and $D^+ \to \pi^0e^+\nu_e$~\cite{BESIII:2017ylw}, based on 2.93 fb$^{-1}$ of $e^+e^-$ collision data taken during 2010-2011 at the center-of-mass energy $\sqrt s=3.773$ GeV.
Moreover, no experimental investigations of the scalar current or the forward-backward asymmetry in $\pilnu$ decays have been reported.
Unless stated otherwise, charge-conjugate modes are always implied throughout this paper.
This paper reports the first experimental constraints on the scalar current in the $\pilnu$ transition, and the first measurements of the forward-backward asymmetries of $\pilnu$. These analyses are based on 20.3~fb$^{-1}$~\cite{Ablikim:2013ntc} of $e^+e^-$ collision data collected by the BESIII detector at $\sqrt s=3.773$ GeV during 2010-2011 and 2022-2024. Additionally, we present the precision measurements of the branching fractions, the hadronic transition form factor $f^{D\to\pi}_+(0)$, and the modulus of the CKM matrix element $|V_{cd}|$, as well as the LFU tests with $\pienu$, $\pimunu$, $\pizenu$, and $\pizmunu$.
All results reported in this paper improve upon the precision of previous measurement at BESIII by a factor of 2-3. Hence results in Refs.~\cite{BESIII:2015tql,BESIII:2017ylw,BESIII:2018nzb} are superceded.

\section{BESIII detector and Monte Carlo simulations}

The BESIII detector~\cite{Ablikim:2009aa} records symmetric $e^+e^-$ collisions provided by the BEPCII storage ring~\cite{Yu:IPAC2016-TUYA01} in the center-of-mass energy range from 1.84 to 4.95~GeV, with a peak luminosity of $1.1 \times 10^{33}\;\text{cm}^{-2}\text{s}^{-1}$ achieved at $\sqrt{s} = 3.773\;\text{GeV}$.
BESIII has collected large data samples in this energy region~\cite{Ablikim:2019hff,Li:2021iwf}. The cylindrical core of the BESIII detector covers 93\% of the full solid angle and consists of a helium-based multilayer drift chamber~(MDC), a plastic scintillator time-of-flight system~(TOF), and a CsI(Tl) electromagnetic calorimeter~(EMC), which are all enclosed in a superconducting solenoidal magnet providing a 1.0~T magnetic field. The solenoid is supported by an octagonal flux-return yoke with resistive plate counter muon identification modules interleaved with steel (MUC).
The charged-particle momentum resolution at $1~{\rm GeV}/c$ is $0.5\%$, and the ${\rm d}E/{\rm d}x$ resolution is $6\%$ for electrons from Bhabha scattering. The EMC measures photon energies with a resolution of $2.5\%$ ($5\%$) at $1$~GeV in the barrel (end cap) region. The time resolution in the TOF barrel region is 68~ps, while that in the end cap region was 110~ps. The end cap TOF system was upgraded in 2015 using multi-gap resistive plate chamber technology, providing a time resolution of 60~ps~\cite{etof}. Approximately 86\% of the data used here was collected after this upgrade.

Simulated samples produced with a {\sc geant4}-based~\cite{geant4} Monte Carlo (MC) package, which includes the geometric description~\cite{Huang:2022wuo} of the BESIII detector and the detector response, are used to determine detection efficiencies and to estimate backgrounds. The simulation models the beam
energy spread and initial state radiation (ISR) in the $e^+e^-$ annihilations with the generator {\sc kkmc}~\cite{ref:kkmc}. The ISR production of vector charmonium(-like) states and the continuum processes are incorporated in {\sc kkmc}~\cite{ref:kkmc}.
Simulation samples of the signal $D\to \pi\ell^+\nu_\ell$ are generated using the two-parameter series-expansion model, as detailed in Ref.~\cite{Becher:2005bg}, with parameter values estimated in Sec~\ref{sec:decayrates} for this analysis. The background is examined using an inclusive simulation sample with an integrated luminosity approximately 40 times that of the data, which includes the production of $D\bar{D}$ pairs, and emulates the quantum coherence for some neutral $D$ decay modes. Additionally, it accounts for the non-$D\bar{D}$ decays of the $\psi(3770)$, the initial state radiation from the production of $J/\psi$ and $\psi(3686)$ states, and continuum processes incorporated in {\sc kkmc}~\cite{ref:kkmc}. The known decay modes are modeled using {\sc evtgen}~\cite{ref:evtgen}, with branching fractions sourced from the Particle Data Group~(PDG)~\cite{pdg2024}.
The remaining unknown charmonium decays are modeled with {\sc lundcharm}~\cite{ref:lundcharm}. Final state radiation from charged final-state particles is incorporated using the {\sc photos} package~\cite{photos}.

\section{Double Tag Method}

At $\sqrt s=3.773$~GeV, the $D$ and $\bar D$ mesons are produced in pairs from the $e^+e^-\to \psi(3770)\to D\bar D$ process,
where $D$ denotes $D^0$ or $D^+$. This unique property allows us to measure the absolute branching fractions of $D$ meson decays using
the well established double-tag (DT) method~\cite{MARK-III:1985hbd}.
In this method, the single-tag (ST) candidate events are selected by reconstructing a $\bar D^0$ or $D^-$ in the hadronic final states
$\bar D^0 \to K^+\pi^-$, $K^+\pi^-\pi^0$, $K^+\pi^-\pi^-\pi^+$, $K^+\pi^-\pi^0\pi^0$, $K^+\pi^-\pi^-\pi^+\pi^0$ and
$D^- \to K^{+}\pi^{-}\pi^{-}$,
$K^0_{S}\pi^{-}$, $K^{+}\pi^{-}\pi^{-}\pi^{0}$, $K^0_{S}\pi^{-}\pi^{0}$, $K^0_{S}\pi^{+}\pi^{-}\pi^{-}$, $K^{+}K^{-}\pi^{-}$.
These inclusively selected candidates are referred to as ST $\bar D$ mesons.
In the presence of the ST $\bar D$ mesons, candidates for the signal decays
are selected to form DT events.
The branching fraction of each signal decay is determined by
\begin{equation}
\label{eq:bf}
{\mathcal B}_{\rm sig}=N_{\rm DT}/(N_{\rm ST}^{\rm tot}\cdot \bar{\varepsilon}_{\rm sig}),
\end{equation}
where $N_{\rm DT}$ is the DT yield in data, the total yield of ST $\bar D$ mesons is
\begin{equation}
N_{\rm ST}^{\rm tot} = \sum_{i=1}^{n=5,6}N_{\rm ST}^{i},
\end{equation}
where $N_{\rm ST}^{i}$ is the ST yield of the $i$th tag mode and
$\bar{\varepsilon}_{\rm sig}$ is the detection efficiency of the semileptonic decay in the presence of the ST $\bar D^0$ meson, which is weighted from
\begin{equation}
  \bar{\varepsilon}_{\rm sig} = \sum_{i=1}^{n=5,6}\frac{N_{\rm ST}^i \varepsilon_{\rm sig}^{i}}{N_{\rm ST}^{\rm tot}},
\end{equation}
where $\varepsilon_{\rm sig}^{i} = \varepsilon_{\rm DT}^{i}/\varepsilon_{\rm ST}^{i}$. The corresponding DT and ST efficiencies for the $i$th tag mode
are $\varepsilon_{\rm DT}^{i}$ and $\varepsilon_{\rm ST}^{i}$.

\section{Single-tag $\bar D$ candidates}

For all charged tracks used in this analysis except the lepton track, the polar angle with respect to the MDC axis ($\theta$) is required to satisfy $|\cos\theta|<0.93$.
Except for those used for $K^0_S$ reconstruction, the good charged tracks are required to come from the interaction region defined by $V_{xy} < 1$\,cm and $|V_{z}| < 10$\,cm, where $V_{xy}$ and $|V_{z}|$ are the distances of closest approach of the reconstructed track to the interaction point (IP) in the $xy$ plane and the $z$ direction (along the beam), respectively.
The charged tracks are identified by using the $dE/dx$ and TOF information, with which the combined confidence levels under the pion and kaon hypotheses ($\mathcal{L}_{\pi, K}$) are computed separately. Charged kaon and pion mesons are required to satisfy $\mathcal{L}_{K} > \mathcal{L}_{\pi}$ and $\mathcal{L}_{\pi} > \mathcal{L}_{K}$, respectively.

Candidate $K_S^0$ mesons are reconstructed from pairs of oppositely charged tracks.
The distance of closest approach of each charged track to the IP must be less than 20~cm along the MDC axis.
No limitation on the distance of closest approach in the transverse plane is required.
Additionally, no particle identification (PID) criteria are applied to the two charged tracks assigned as $\pi^+\pi^-$. 
The two charged tracks are constrained to originate from a common secondary vertex, which must be separated from the IP by a flight distance of at least twice the vertex resolution. To ensure high-quality vertex reconstruction, both the primary and secondary vertex fits are required to satisfy $\chi^2 < 100$.
The invariant mass of the $\pi^+\pi^-$ pair is required to be within $(0.487,0.511)$~GeV/$c^2$.

Neutral pion candidates are reconstructed via the $\pi^0\to\gamma\gamma$ decay. Photon candidates are chosen from the EMC showers. The EMC time deviation from the event start time is required to be within $[0,700]$~ns.  The energy deposited in the EMC  is required to be greater than 25 MeV in the barrel region ($|\cos\theta|<0.80$) and 50~MeV in the end cap region ($0.86<|\cos\theta|<0.92$).
The opening angle between the photon candidate and the nearest charged track is required to be greater than $10^{\circ}$. 
Photon pairs with an invariant mass within the range 
$(0.115,0.150)$~GeV$/c^2$ are considered as $\pi^0$ candidates. To improve momentum resolution, a mass-constrained~(1C) fit to the nominal $\pi^{0}$ mass~\cite{pdg2024} is imposed on the selected photon pair, and the fit is required to satisfy $\chi^2<50$.
The updated four-momentum of the $\pi^0$ candidate obtained from the kinematic fit is retained for further analysis.

The tag mode $\bar D^0\to K^+\pi^-$ suffers from the backgrounds of cosmic rays and Bhabha events.
These background events are suppressed using the same requirements as Ref.~\cite{BESIII:2014rtm}.

To identify the ST $\bar D$ mesons, we use two kinematic variables: the energy difference $\Delta E\equiv E_{\bar D}-E_{\rm beam}$ and the beam-constrained mass $M_{\rm BC}\equiv\sqrt{E_{\rm beam}^2/c^4-|\vec{p}_{\bar D}|^2/c^2}$, where $E_{\rm beam}$ is the beam energy, and $E_{\bar D}$ and $\vec{p}_{\bar D}$ are the total energy and momentum of the ST $\bar D$ meson in the $e^+e^-$ center-of-mass frame.
If there are multiple $\bar D$ candidates for a specific tag mode, the one giving the least $|\Delta E|$ is selected for further analysis.
To suppress combinatorial background in the $M_{\rm BC}$ distribution, the tag-mode dependent $\Delta E$ requirements are imposed on the ST candidates.
The detailed $\Delta E$ requirements and the efficiencies of detecting ST $\bar D$ mesons estimated with the inclusive MC sample are summarized in Table~\ref{table:styields}.

For each tag mode, the yield of ST $\bar D$ mesons is extracted by fitting the corresponding $M_{\rm BC}$ distribution.
In the fit, the $\bar D$ signal shape is described by the MC-simulated shape convolved with a double-Gaussian function to account for the resolution difference between data and MC simulation. 
To define this signal shape, the opening angle between the generated and reconstructed momenta of the ST daughter particles must be less than 15$^\circ$. Meanwhile, the background shape is parameterized as an ARGUS function~\cite{argus}, with its endpoint fixed at the $E_{\rm beam}$ value.

\begin{figure*}[htbp]
  \centering
  \includegraphics[width=\linewidth]{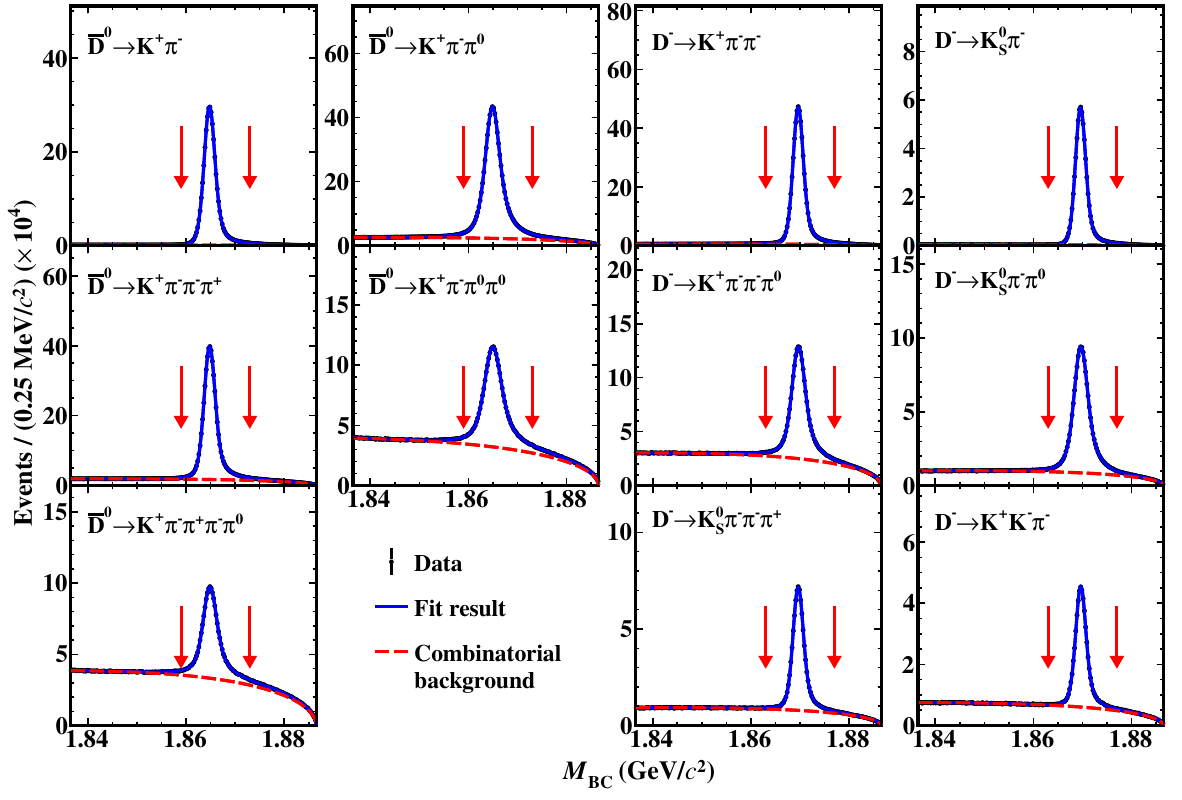}
  \caption{Fits to the $M_{\rm BC}$ distributions of the ST $\bar D$ candidates. The points with error bars are data, the blue curves are the best fits, and the red dashed curves are the fitted combinatorial background shapes. The pair of red arrows show the $M_{\rm BC}$ signal window.}
  \label{fig:datafitMassbc}
\end{figure*}

Figure~\ref{fig:datafitMassbc} shows the fit results for the $M_{\rm BC}$ distributions of the accepted ST candidates in data for individual tag modes. The candidates with $M_{\rm BC}$ in the range $(1.859,1.873)$ GeV/$c^2$ for $\bar D^0$ tags and $(1.863,1.877)$ GeV/$c^2$ for $D^-$ tags are kept for further analysis. Summing over all tag modes, the total yields of  ST $\bar D^0$ and $D^-$ mesons are determined to be $(19,057.6 \pm 5.4_{\rm stat})\times 10^3$ and $(10,646.9\pm3.8_{\rm stat})\times 10^3$, respectively.

\begin{table*}[htbp]
	\centering
	\caption {\label{tab:st}The $\Delta E$ requirements, the obtained ST $\bar D^0(D^-)$ yields ($N^{i}_{\rm ST}$) in data and the ST efficiencies ($\varepsilon^{i}_{\rm ST}$). The efficiencies do not include the branching fractions of the $K^0_S$ and $\pi^0$ decays. The uncertainties are statistical only.}
	\label{table:styields}
%	\resizebox{1.0\columnwidth}{!}{
	\begin{tabular}{lcr@{}lc}
		\hline\hline
		Tag mode 										        & $\Delta E$~(GeV)   &  \multicolumn{2}{c}{$N^{i}_{\rm ST}~(\times10^{3})$}     	&  $\varepsilon^{i}_{\rm ST}~(\%)$  \\\hline
		$\bar{D}^{0} \to K^{+} \pi^{-}$ 				        &  $(-0.027,0.027)$  & $3725.8$ &$ \pm 2.0$ 			& $65.09 \pm 0.01$ \\
		$\bar{D}^{0} \to K^{+} \pi^{-} \pi^{0}$ 		        &  $(-0.062,0.049)$  & $7419.9$ &$ \pm 3.2$ 			& $35.53 \pm 0.01$ \\
		$\bar{D}^{0} \to K^{+} \pi^{+} \pi^{-} \pi^{-}$         &  $(-0.026,0.024)$  & $4987.2$ &$ \pm 2.5$ 			& $40.70 \pm 0.01$ \\
		$\bar{D}^{0} \to K^{+} \pi^{-} \pi^{0} \pi^{0}$         &  $(-0.068,0.053)$  & $1771.8$ &$ \pm 2.2$             & $15.11 \pm 0.01$ \\
        $\bar{D}^{0} \to K^{+} \pi^{+} \pi^{-} \pi^{-} \pi^{0}$ &  $(-0.057,0.051)$  & $1152.8$ &$ \pm 1.8$             & $16.19 \pm 0.01$ \\
		\hline
		$D^{-} \to K^{+} \pi^{-} \pi^{-}$ 				        &  $(-0.025,0.024)$  & $5552.7$ &$ \pm 2.5$ 			& $51.01 \pm 0.01$ \\
		$D^{-} \to K_{S}^{0} \pi^{-}$ 					        &  $(-0.025,0.026)$  & $656.4$ &$  \pm 0.8$ 			& $51.41 \pm 0.01$ \\
		$D^{-} \to K^{+} \pi^{-} \pi^{-} \pi^{0}$ 		        &  $(-0.057,0.046)$  & $1723.7$ &$ \pm 1.8$ 			& $24.40 \pm 0.01$ \\
		$D^{-} \to K_{S}^{0} \pi^{-} \pi^{0}$ 			        &  $(-0.062,0.049)$  & $1442.3$ &$ \pm 1.5$ 			& $26.42 \pm 0.01$ \\
		$D^{-} \to K_{S}^{0} \pi^{-} \pi^{-} \pi^{+}$ 	        &  $(-0.028,0.027)$  & $790.6$ &$  \pm 1.1$ 			& $29.61 \pm 0.01$ \\
		$D^{-} \to K^{+} K^{-} \pi^{-}$ 				        &  $(-0.024,0.023)$  & $481.2$ &$  \pm 0.9$ 			& $40.87 \pm 0.01$ \\
		\hline\hline
	\end{tabular}
%	}
\end{table*}

\section{Double-tag events}

The candidates for the $\pienu$, $\pimunu$, $\pizenu$, and $\pizmunu$ decays are selected from the remaining tracks in events containing ST $\bar D$ candidates.
The $\pi^-$ and $\pi^{0}$ candidates are selected with the same criteria as those used in the tag selection. To ensure better data-MC consistencies and minimize systematic uncertainties, positron and muon candidates  are required to satisfy $|\cos\theta|<0.80$. 
Lepton identification is performed by combining measurements from the TOF, $dE/dx$, and the EMC to calculate joint likelihoods ($\mathcal{L}_e$, $\mathcal{L}_\mu$, $\mathcal{L}_K$, and $\mathcal{L}_\pi$) for the electron, muon, kaon, and pion hypotheses, respectively. 
The positron candidate is required to satisfy $\mathcal{L}_e/(\mathcal{L}_e+\mathcal{L}_\pi+\mathcal{L}_K)> 0.8$, and $\mathcal{L}_e >0.001$. 
Additionally, its deposited energy in the EMC have to be greater than 0.7 times its momentum reconstructed by the MDC.
The muon candidate is required to satisfy $\mathcal{L}_\mu > \mathcal{L}_e$, and $\mathcal{L}_\mu >0.001$, and
have an energy deposit in the EMC
of less than 0.3 GeV.
For muons with momentum greater than 0.5 GeV/$c$, a two-dimensional requirement on the hit depth ($\mathcal{D}_{\mu}$) in the MUC~\cite{BESIII:2009fln}, based on $|\cos\theta_\mu|$ and momentum ($p_\mu$), is imposed as listed in Table~\ref{table:mudepth}. 
These requirements are determined based on the MUC layout, taking into account the different absorber thicknesses and layer structures in the barrel and end-cap regions.

\begin{table}[htbp]
	\caption{Identification criteria for muon candidates.}
	\label{table:mudepth}
	\centering
		\begin{tabular}{lcl}
			\hline\hline
			$|\cos\theta_{\mu}|$                & $p_{\mu}$(GeV$/c$) & $\mathcal D_{\mu}$(cm)             \\ \hline
			\multirow{5}{*}{$(0.0,0.2)$} & $(0.50,0.61)$   & $>3.0$                \\
			& $(0.61,0.75)$   & $>100.0\times p_{\mu}-58.0$ \\
			& $(0.75,0.88)$   & $>17.0$               \\
			& $(0.88,1.04)$   & $>100.0\times p_{\mu}-71.0$ \\
			& $(1.04,1.20)$   & $>33.0$               \\ \hline
			\multirow{5}{*}{$(0.2,0.4)$} & $(0.50,0.64)$   & $>3.0$                \\
			& $(0.64,0.78)$   & $>100.0\times p_{\mu}-61.0$ \\
			& $(0.78,0.91)$   & $>17.0$               \\
			& $(0.91,1.07)$   & $>100.0\times p_{\mu}-74.0$ \\
			& $(1.07,1.20)$   & $>33.0$               \\ \hline
			\multirow{5}{*}{$(0.4,0.6$)} & $(0.50,0.67)$   & $>3.0$                \\
			& $(0.67,0.81)$   & $>100.0\times p_{\mu}-64.0$ \\
			& $(0.81,0.94)$   & $>17.0$               \\
			& $(0.94,1.10)$   & $>100.0\times p_{\mu}-77.0$ \\
			& $(1.10,1.20)$   & $>33.0$               \\ \hline
			\multirow{5}{*}{$(0.6,0.8)$} & $(0.50,0.67)$   & $>3.0$                \\
			& $(0.67,0.81)$   & $>100.0\times p_{\mu}-64.0$ \\
			& $(0.81,0.94)$   & $>17.0$               \\
			& $(0.94,1.10)$   & $>100.0\times p_{\mu}-77.0$ \\
			& $(1.10,1.20)$   & $>33.0$               \\ \hline
			\hline
		\end{tabular}
\end{table}

To suppress backgrounds from the hadronic $D$ decays, it is required that there is no additional good charged track in the signal side ($N_{\rm extra}^{\rm char}$).
To reject the backgrounds from the hadronic decays involving $\pi^0$ mesons, the maximum energy of extra photons ($E_{\text{extra~}\gamma}^{\rm max}$) which have not been used in the event selection is required to be less than 0.25 GeV.
The invariant mass of $\pi^{+(0)}$ and $\ell^{+}$ ($M_{\pi\ell}$) is used to suppress the backgrounds associated with the misidentification between $\pi^{+}$ and $\ell^{+}$. Specifically, we require $M_{\pi^{-(0)}e^{+}} < 1.80$ GeV/$c^2$ for $D^{0(+)}\to\pi^{-(0)}e^+\nu_{e}$ and $M_{\pi^{-(0)}\mu^{+}} < 1.70$ GeV/$c^2$ for $D^{0(+)}\to\pi^{-(0)}\mu^+\nu_{\mu}$ decay.
To further reject the peaking backgrounds from $D^0\to K_{S}^0(\pi^{+}\pi^{-})\pi^0$ and $D^{+}\to\bar{K}^{0}\pi^+$ in the $\pimunu$ and $\pizmunu$ channels, a $K_S^0$ veto is imposed by requiring the $M_{\pi^{-}\mu^{+}}$ and $M_{D^{-}\mu{+}}^{\rm rec}$ ($D^{-}\mu^{+}$ recoil mass) to be outside the ranges (0.455, 0.497) GeV/$c^2$ and (0.437, 0.580) GeV/$c^2$, respectively.
These veto windows correspond to approximately three times the experimental resolution around the fitted  peaks in data.

The undetected neutrino is inferred from the kinematic variable $M_{\rm miss}^{2} \equiv E_{\rm miss}^{2}/c^{2} - |\vec{p}_{\rm miss}|^{2}/c^{2}$, which peaks at zero for signal events. The $E_{\rm miss}$ and $\vec{p}_{\rm miss}$, given by $E_{\rm miss}\equiv E_{\rm beam}-E_{\pi}-E_{e^+}$ and $\vec{p}_{\rm miss}\equiv\vec{p}_{D}-\vec{p}_{\pi}-\vec{p}_{\ell^+}$, are the missing energy and momentum of the DT event in the $e^+e^-$ center-of-mass frame, respectively. Here, the $E_{\pi(\ell^+)}$ and $\vec{p}_{\pi(\ell^+)}$ are the measured energy and momentum of the $\pi(\ell^+)$ candidates, respectively, and $\vec{p}_{D}\equiv-\hat{p}_{\bar D} \sqrt{E_{\rm beam}^2/c^2-m_{\bar D}^2 c^2 }$. The $\hat{p}_{\bar D}$ is the unit vector in the momentum direction of the ST $\bar D$ meson and $m_{\bar D}$ is the nominal $\bar D$ mass~\cite{pdg2024}.
To improve the $M_{\rm miss}^{2}$ resolution, the beam energy and the nominal $\bar D$ mass are used to calculate the momentum magnitude of the ST $\bar D$ mesons.

\section{Branching fractions}

\subsection{Signal efficiencies and branching fractions}
\label{sec:bffit}
\begin{figure}[htbp]
  \centering
  \includegraphics[width=\columnwidth]{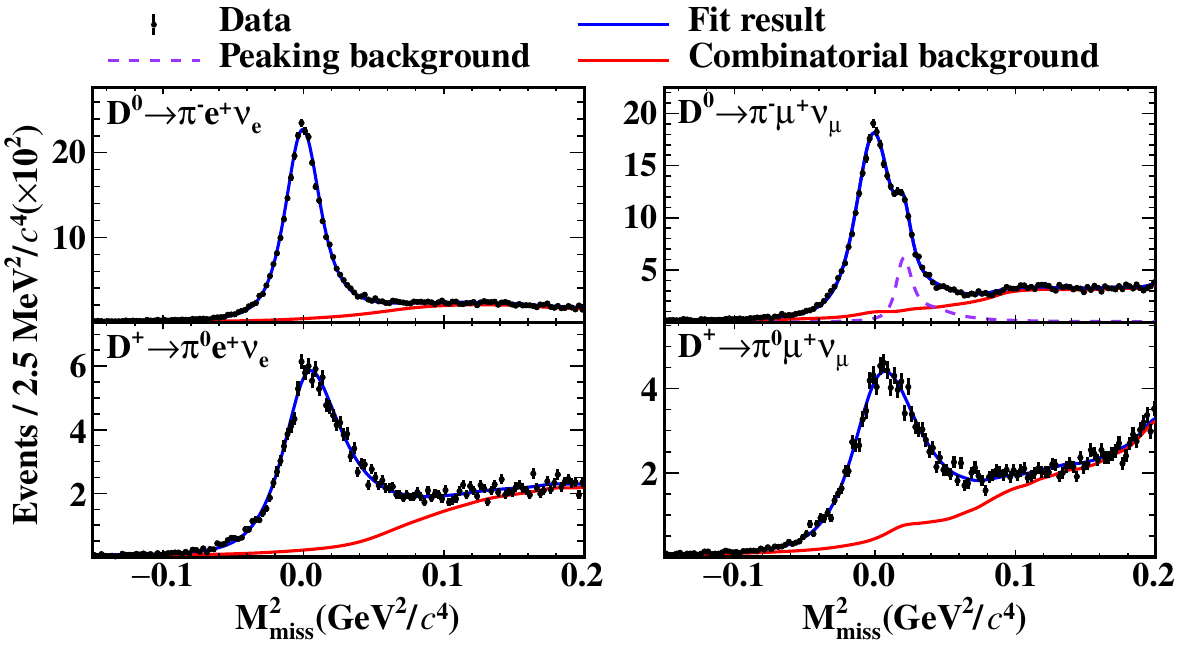}
  \caption{Fits to the $M_{\rm miss}^2$ distributions of the candidates for different semileptonic $D$ decays.
  	The points with error bars are data, the solid blue lines are the total fit, the dashed violet lines are the fitted peaking backgrounds, and the solid red lines are the fitted combinatorial background shapes.
  }
  \label{fig:fitmm2m}
\end{figure}

\begin{figure*}[htbp]
	\centering
	\includegraphics[width=\linewidth]{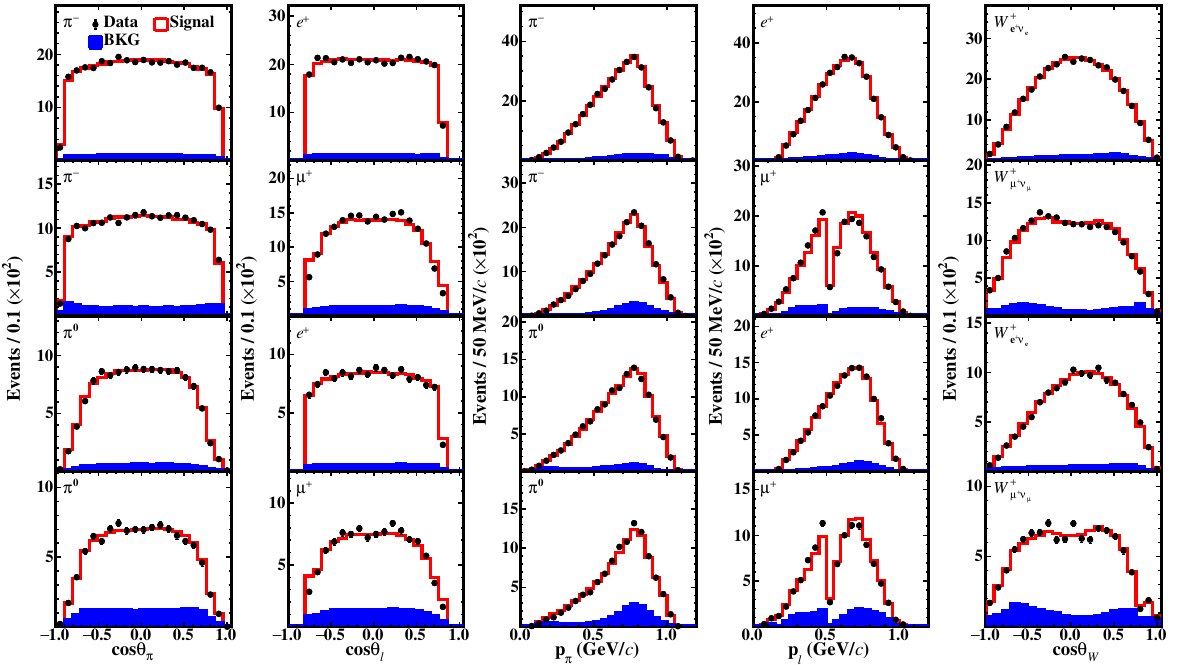}
	\caption{Comparisons of the distributions of $\cos\theta$ and momenta for pion and lepton candidates, as well as the $\cos\theta$ of $W$ boson of the candidates for $\pilnu$. The points with error bars are data, the filled blue histograms are the simulated backgrounds, and the red line histograms are the signal MC samples. These events have been required to satisfy $|M_{\rm miss}^{2}| < 0.05$ GeV$^{2}/c^{4}$ for $\pienu$, $\pizenu$ and $\pizmunu$, $-0.05<M_{\rm miss}^{2} < 0.01$ GeV$^{2}/c^{4}$ for $\pimunu$.
		\label{fig:pilnucompare}
	}
\end{figure*}

After imposing all selection criteria, the $M_{\rm miss}^{2}$ distributions of the signal candidates for the four semileptonic $D$ decays are obtained, as shown in Fig.~\ref{fig:fitmm2m}. For each signal decay, the signal yields are derived from the fits to the corresponding $M_{\rm miss}^{2}$ distributions.
In the fit, the signal and background components are described by the shapes derived from MC simulation with their yields left free. The signal shape is convolved with a Gaussian function with free parameters, which accounts for the difference of resolution and calibration between data and MC simulation. The peaking background $D^{0}\to\pi^{-}\pi^{+}\pi^{0}$ in the $M_{\rm miss}^{2}$ distribution of the $\pimunu$ process is  described by individual MC simulated shapes convolved with different Gaussian functions. Figure~\ref{fig:pilnucompare} shows good consistency
between data and inclusive MC sample
for the momentum and $\cos\theta$ distributions of pion and leptonic candidates for $\pienu$, $\pimunu$, $\pizenu$, and $\pizmunu$ , respectively.

With the signal yields in data~($N_{\rm DT}$), the weighted signal efficiencies~($\bar \varepsilon_{\rm sig}$), and the yields of ST $\bar D$ in data ($N^{\rm tot}_{\rm ST}$), we determine the branching fractions of $\pienu$, $\pimunu$, $\pizenu$ and $\pizmunu$ with Eq.~(\ref{eq:bf}). The obtained results are summarized in Table~\ref{table:bfsum}. The systematic uncertainties in the branching fraction measurements are discussed below.

\begin{table*}[htbp]
  \caption{The signal yields in data ($N_{\rm DT}$),
  	the weighted efficiencies of detecting the signal decays in the presence of the ST $\bar D$ ($\epsilon_{\rm sig}$).
  	and the obtained branching fractions of different signal decays ($\mathcal B_{\rm sig}$).
  	The signal efficiencies have been corrected by the data-MC differences of the pion tracking and PID,
  	the lepton tracking and PID, as well as the $\pi^{0}$ reconstruction. For $\mathcal B$, the first uncertainties are statistical and the second systematic.
  	For other quantities, the uncertainties are statistical only.}
  \label{table:bfsum}
  \centering
      \begin{tabular}{lccc}
          \hline\hline
          Decay                             & $\bar{\varepsilon}_{\rm sig}(\%)$  & $N_{\rm DT}$      & $\mathcal{B}_{\rm sig}(\times10^{-3})$ \\ \hline
          $D^{0}\to\pi^{-}e^{+}\nu_{e}$     & $61.97\pm0.02$                     & $34834\pm202$ & $\BFpienu$              \\
          $D^{0}\to\pi^{-}\mu^{+}\nu_{\mu}$ & $46.84\pm0.02$                     & $25140\pm330$ & $\BFpimunu$              \\
          $D^{+}\to\pi^{0}e^{+}\nu_{e}$     & $39.76\pm0.02$                     & $15330\pm145$ & $\BFpizenu$              \\
          $D^{+}\to\pi^{0}\mu^{+}\nu_{\mu}$ & $29.52\pm0.02$                     & $10890\pm135$ & $\BFpizmunu$              \\
          \hline\hline
      \end{tabular}
\end{table*}

\subsection{Systematic uncertainties in branching fractions}

Table~\ref{table:systot} summarizes the sources of the systematic uncertainties in the branching fraction measurements.
The uncertainties caused by ST $\bar{D}$ yields, tracking and PID of charged particles, $\pi^{0}$ reconstruction, $E_{{\rm extra} \gamma}^{\rm max}$ and $N_{\rm extra}^{\rm char}$ requirements, quoted branching fractions, and FSR effect are correlated between the corresponding decay channels in the simultaneous fit to the partial decay rates and the LFU calculation.

The total systematic uncertainties are 0.58\%, 0.66\%, 0.50\%, and 0.73\% for $\pienu$, $\pimunu$, $\pizenu$ and $\pizmunu$ decays, respectively, calculated by adding all the individual contributions in quadrature.

\paragraph{\boldmath \bf ST $\bar D$ yields}

The systematic uncertainty of the fits to the $M_{\rm BC}$ distributions is estimated by varying the signal and background shapes and repeating the fits for both data and inclusive MC sample. A variation of the signal shape is obtained by modifying the matching requirement between generated and reconstructed angles from 15$^\circ$ to 10$^\circ$ or 20$^\circ$.
The uncertainty related to the background shape is obtained by varying the endpoint by $\pm 0.2$ MeV.
The systematic uncertainty is assigned as the relative change of the yields of ST $\bar D$ mesons in data over the efficiencies of detecting ST $\bar D$ mesons.
An additional uncertainty due to the background fluctuation of the fitted yields of ST $\bar D$ is included.
Adding these three effects quadratically leads to 0.3\%, which is taken as the systematic uncertainty.

\paragraph{\boldmath \bf $\pi^-$ tracking and PID}

The $\pi^-$ tracking and PID efficiencies are studied by using the control sample of hadronic $D\bar D$ events ($\bar{D}^{0} \to K^{+} \pi^{-}$, $\bar{D}^{0} \to K^{+} \pi^{-} \pi^{0}$, and $\bar{D}^{0} \to K^{+} \pi^{+} \pi^{-} \pi^{-}$ vs. $D^{0}\to K^{-}\pi^{+}$, $D^{0}\to K^{-}\pi^{+}\pi^{+}\pi^{-}$, and $D^{0}\to K^{-}\pi^{+}\pi^{0}$, and $D^{-}\to K^{+}\pi^{-}\pi^{-}$ vs. $D^{+}\to K^{-}\pi^{+}\pi^{+}$).
The momentum-weighted data-MC correction factors for $\pi^-$ tracking and PID efficiencies are determined to be $0.9970 \pm 0.0020$ ($0.9905\pm 0.0020$) for $D^0\to \pi^-e^+\nu_e$ and $0.9969 \pm 0.0020$ ($0.9902 \pm 0.0020$) for $D^0\to \pi^-\mu^+\nu_{\mu}$. The signal efficiencies are corrected by these factors, and the remaining systematic uncertainties after these corrections are listed in Table~\ref{table:systot}.

\paragraph{\boldmath \bf $\pi^0$ reconstruction}
\label{sec:gamma}

The $\pi^0$ reconstruction efficiencies are examined by using DT events of $D^0\to K^-\pi^+\pi^0$. The polar angle distribution of the control sample is consistent with that in the signal decays, therefore its effect on the data-MC differences of $\pi^0$ reconstruction efficiency is negligible. The momentum-weighted data-MC differences of the $\pi^0$ reconstruction efficiency are 0.17\% for $\pizenu$ and 0.19\% for $\pizmunu$, which are taken as the systematic uncertainties.

\paragraph{\boldmath \bf $\ell^+$ tracking and PID}
\label{sec:electron}

The tracking and PID efficiencies of $e^+$ and $\mu^+$ are studied by using the control samples of $e^{+}e^{-}\rightarrow\gamma e^{+}e^{-}$ and $e^{+}e^{-}\rightarrow$$\gamma\mu^{+}\mu^{-}$, respectively.
The data-MC efficiency ratios are determined to be $0.9984\pm0.0020$ for $e^+$ tracking and $0.9875\pm0.0020$ for $e^+$ PID.
For $\mu^+$, the ratios are $1.0001\pm0.0020$ for tracking and $0.9376\pm0.0020$ ($0.9390 \pm 0.0020$) for PID.
The signal efficiencies are corrected using these factors, and their uncertainties are assigned as components of the systematic uncertainty, as listed in Table~\ref{table:systot}.

\paragraph{\boldmath \bf $E_{\rm extra~\gamma}^{\rm max}$ and $N_{\rm extra}^{\rm char}$ requirements}
\label{sec:em}

The systematic uncertainty in the $E_{\rm extra~\gamma}^{\rm max}$ and $N_{\rm extra}^{\rm char}$ requirements is estimated with hadronic DT $D\bar D$ samples ($\bar{D}^{0} \to K^{+} \pi^{-}$, $\bar{D}^{0} \to K^{+} \pi^{-} \pi^{0}$, and $\bar{D}^{0} \to K^{+} \pi^{+} \pi^{-} \pi^{-}$ vs. $D^{0}\to K^{-}\pi^{+}$, $D^{0}\to K^{-}\pi^{+}\pi^{+}\pi^{-}$, and $D^{0}\to K^{-}\pi^{+}\pi^{0}$, and $D^{-}\to K^{+}\pi^{-}\pi^{-}$, $D^-\to K_S^0\pi^-$, $D^-\to K^+\pi^-\pi^-\pi^0$, $D^-\to K_S^0\pi^-\pi^0$, $D^-\to K_S^0\pi^-\pi^-\pi^+$, and $D^-\to K^+K^-\pi^-$ vs. $D^{+}\to K^{-}\pi^{+}\pi^{+}$, $D^+\to K_S^0\pi^+$, $D^+\to K^-\pi^+\pi^+\pi^0$, $D^+\to K_S^0\pi^+\pi^0$, $D^+\to K_S^0\pi^+\pi^+\pi^-$, and $D^+\to K^-K^+\pi^+$).
The difference of the acceptance efficiencies between data and MC simulation is assigned as the systematic uncertainty.

\paragraph{\boldmath \bf Hadronic transition form factors}

The detection efficiencies are estimated by using signal MC events generated with the hadronic transition form factors measured in this work as inputs.
The corresponding systematic uncertainties are estimated by varying the form factor parameters by $\pm1\sigma$.

\paragraph{\boldmath \bf $M_{\pi \ell^+}$ requirement}

The systematic uncertainty associated with the $M_{\pi e^+}$ requirement is neglected, as the corresponding selection efficiency is almost 100\%.
The reliability of the MC modeling of the $M_{\pi\mu}$ requirement is evaluated using the positron channels as control samples. No significant variation in the measured branching fractions is observed, so this uncertainty is negligible.

\paragraph{\textbf{\textup{$K_{S}^{0}$ veto}}}
The systematic uncertainties due to the $K_S^0$ veto in $M_{\pi^{-}\mu^{+}}$ and $M_{D^{-}\mu^{+}}^{\rm rec}$ are evaluated by comparing the branching fractions measured with alternative $\pm 3\sigma$ mass windows which are determined by fitting the corresponding distributions in the inclusive MC samples. The resulting deviations from the nominal values are assigned as the systematic uncertainties.

\paragraph{\boldmath \bf $M_{\rm miss}^{2}$ fit}

The systematic uncertainty due to the $M_{\rm miss}^{2}$ fit is considered in two parts.
Since a Gaussian function has been convolved with the simulated signal shapes to account for the resolution difference between data and MC simulation already, the systematic uncertainty from the signal shape is ignored.
The systematic uncertainty due to the background shape is assigned by varying
the relative fractions of major backgrounds from $e^+e^-\to q\bar q$ according to the known cross section~\cite{BESqqbar} and the dominant background channels within $\pm 1\sigma$ of its input branching fractions in the inclusive MC sample.
The changes in the branching fractions are taken as the corresponding systematic uncertainties.

\paragraph{\boldmath \bf MC statistics}

The relative statistical uncertainties of the signal efficiencies are assigned as a source of systematic uncertainty due to MC statistics.

\paragraph{\boldmath \bf Quoted branching fractions}

For the $\pizenu$ and $\pizmunu$ decays,
the uncertainty of the quoted branching fraction of $\pi^{0}\to \gamma\gamma$ is 0.03\%~\cite{pdg2024}.

\paragraph{\boldmath \bf FSR effect}

The difference between the measured branching fractions obtained with and without FSR effect of the positron momentum is assigned as the conservative systematic uncertainty. The values are 0.26\% for $\pienu$ and 0.10\% for $\pizenu$.

\paragraph{\boldmath \bf Measurement method}

The nominal analysis method weighted the efficiencies with the ST yields of different tag modes. The associated systematic uncertainties are estimated with an alternative method, which are based on a simultaneous fit to different tag modes by sharing a common branching fraction. The changes in the measured branching fractions are taken as the systematic uncertainties.

\begin{table*}[htbp]
  \caption{Relative systematic uncertainties (in unit of \%) in the measurements of the branching fractions.}
  \label{table:systot}
  \centering
%  \resizebox{1.0\columnwidth}{!}{
      \begin{tabular}{lcccc}
          \hline\hline
          Source                                      & $D^{0}\to\pi^{-}e^{+}\nu_{e}$ & $D^{0}\to\pi^{-}\mu^{+}\nu_{\mu}$ & $D^{+}\to\pi^{0}e^{+}\nu_{e}$ & $D^{+}\to\pi^{0}\mu^{+}\nu_{\mu}$ \\ \hline
          $N_{\rm ST}^{\rm tot}$                      & 0.30                          & 0.30                              & 0.30                          & 0.30                               \\
          $\pi^-$ tracking                            & 0.20                          & 0.20                              & --                            & --                                 \\
          $\pi^-$ PID                                 & 0.20                          & 0.20                              & --                            & --                                 \\
          $\pi^{0}$ reconstruction                    & --                            & --                                & 0.17                          & 0.19                               \\
          $E_{{\rm extra}\gamma}^{\rm max} \&\& N_{\rm extra}^{\rm char}$          & 0.10                          & 0.10                                 & 0.10                          & 0.10                              \\
          $\ell^+$ tracking                           & 0.20                          & 0.20                              & 0.20                          & 0.20                               \\
          $\ell^+$ PID                                & 0.20                          & 0.20                              & 0.20                          & 0.20                               \\
          Quoted branching fraction                   & --                            & --                                & 0.03                          & 0.03                               \\
          MC model                                    & 0.03                          & 0.06                              & 0.11                          & 0.13                               \\
          FSR effect	                              & 0.26                          & --                                & 0.10                          & --                                 \\
          Measurement method						  & --  					      & 0.10							  & --  						  & 0.10							   \\ \hline
          $K_{S}^{0}$ veto                            & --                            & 0.06($M_{\pi^{-}\mu^{+}}$)        & --                            & 0.02($M_{D^{-}\mu^{+}}^{\rm rec}$) \\
          $M_{\rm miss}^2$ fit                        & 0.06                          & 0.41                              & 0.14                          & 0.54                               \\
          MC statistics                               & 0.02                          & 0.02                              & 0.03                          & 0.03                               \\ \hline
          Total                                       & 0.58                          & 0.66                              & 0.50                          & 0.73							   \\
          \hline\hline
      \end{tabular}
  %  }
\end{table*}

\section{Hadronic transition form factors}

\subsection{Theoretical formula}
\label{sec:FFthefor}
In theory, the partial   decay width of the semileptonic decay $D \to \pi \ell^+\nu_{\ell}$ can be expressed as~\cite{Faustov:2019mqr,Fajfer:2015ixa,Leng:2020fei,Jain:2025kqe}
\begin{equation}
	\small
	\begin{array}{l}
		\displaystyle
		\frac{{\rm d}\Gamma}{{\rm d}q^2} = X\mathcal{N}(q^2)\left(1-\frac{m_{\ell}^{2}}{q^{2}}\right)^{2} \\
		\times \left[\left(1+\frac{m_{\ell^{+}}^{2}}{2q^{2}}\right)\left|\mathcal{H}_{0}(q^2)\right|^2 + \frac{3m_{\ell^{+}}^{2}}{2 q^{2}}\left|\mathcal{H}_{t}(q^2)\right|^2\right].
	\end{array}
	\label{eq:pwr}
\end{equation}
Here, the term $\mathcal{N}(q^2) = \frac{G_{F}^{2}|V_{cd}|^{2}|\mathbf{q}|q^2}{96\pi^{3}m_{D}^{2}}$ represents the overall normalization factor, which incorporates the Fermi coupling constant $G_F$, the modulus of the CKM matrix element $V_{cd}$, and the mass of the $D$ meson, $m_D$~\cite{pdg2024}. The vector $\mathbf{q}$ denotes the four-momentum of the $\ell^+\nu_{\ell}$ system in the rest frame of the $D$ meson, while $m_\ell^+$ is the mass of the lepton.
In Eq.~\ref{eq:pwr}, $X$ is a multiplicative factor due to isospin, which equals 1 for the decay $D^{0}\to\pi^{-}\ell^{+}\nu_{\ell}$ and $1/2$ for the decay $D^{+}\to\pi^{0}\ell^{+}\nu_{\ell}$.
The hadronic helicity amplitudes $\mathcal{H}_{0(t)}(q^2)$ are expressed as follows:
\begin{equation}
	\small
	\mathcal{H}_{0}(q^2)=\frac{2m_D|\mathbf{q}|}{\sqrt{q^2}}f^{D\to\pi}_+(q^2),\\
	\mathcal{H}_{t}(q^2)=\frac{m_D^2-m_\pi^2}{\sqrt{q^2}}f^{D\to\pi}_0(q^2),
\end{equation}
where $f^{D\to\pi}_{+}(q^{2})$ and $f^{D\to\pi}_{0}(q^{2})$ are the vector and scalar form factors, respectively.

In the two-parameter series expansion~\cite{Becher:2005bg}, the hadronic transition form factors of semileptonic $D$ meson decays are among the most popular descriptions.
It is given by
\begin{equation}
	\begin{array}{l}
		f^{D\to\pi}_+(q^2)=\frac{1}{P(q^2)\Phi(q^2)}\frac{f^{D\to\pi}_+(0)P(0)\Phi(0)}{1+r_{1}(t_0)z(0,t_0)}\\
		\times\left(1+r_{1}(t_0)[z(q^2,t_0)]\right).
	\end{array}
\end{equation}
Here, $P(q^2)=1$, and $z(q^2,t_0)=\frac{\sqrt{t_+-q^2}-\sqrt{t_+-t_0}}{\sqrt{t_+-q^2}+\sqrt{t_+-t_0}}$. The symbol $\Phi$ represents
\begin{equation}
	\begin{array}{l}
		\displaystyle \Phi(q^2)=\sqrt{\frac{1}{24\pi\chi_{V}}}\left(\frac{t_+-q^2}{t_+-t_0}\right)^{1/4}\left(\sqrt{t_+-q^2}+\sqrt{t_+}\right)^{-5}\\
		\displaystyle \times\left(\sqrt{t_+-q^2}+\sqrt{t_+-t_0}\right)\left(\sqrt{t_+-q^2}+\sqrt{t_+-t_{-}}\right)^{3/2}\\
		\displaystyle \times\left(t_+-q^2\right)^{3/4},
	\end{array}
\end{equation}
where $t_{\pm}=m_{D}\pm m_\pi^2$,
$t_0=t_+(1-\sqrt{1-t_{-}/t_+})$,
$m_{D}$ and $m_\pi$ are the masses of $D$ and $\pi$,
and $\chi_{V}$ is obtained from dispersion
relations using perturbative QCD~\cite{chiV}.

\subsection{Partial decay rates in data}
\label{sec:decayrates}
To extract the hadronic transition form factors of semileptonic $D$ decays, the partial decay rates are measured
in ten $q^2$ intervals of ($m_\ell^2$,0.3), (0.3, 0.6), (0.6, 0.9), (0.9, 1.2), (1.2, 1.5), (1.5, 1.8), (1.8, 2.1), (2.1, 2.4), (2.4, 2.7), (2.7, $(m_D-m_\pi)^2$)GeV$^2/c^4$. The partial decay rate in the $i$-th $q^2$ interval is determined as
\begin{equation}
	\frac{{\rm d}\Gamma_{i}}{{\rm d}q_{i}^2}=\frac{\Delta\Gamma_{i}}{\Delta q^2_{i}},
\end{equation}
and the partial decay rate in the $i$-th $q^2$ interval is calculated by
\begin{equation}
	\Delta\Gamma_{i}=\frac{N_{\rm produced}^{i}}{\tau_{D}\cdot N_{\rm ST}^{\rm tot}}
\end{equation}
where $N_{\rm produced}^{i}$ is the number of events produced in the $i$-th $q^2$ interval,
$\tau_{D}$ is the $D$ lifetime~\cite{pdg2024} and $N_{\rm ST}^{\rm tot}$ is the number of the ST $\bar D$ mesons.

In the $i$-th $q^2$ interval, the number of events produced in data is calculated as
\begin{equation}
	N_{\rm produced}^{i}=\sum_j^{N_{\rm bins}}\left(\varepsilon^{-1}\right)_{ij}N_{\rm DT}^{j},
\end{equation}
where $N_{\rm DT}^{j}$ is the signal yield observed in the $j$-th $q^2$ interval and the efficiency matrix $\varepsilon_{ij}$, obtained by analyzing the signal MC events, given by
\begin{equation}
	\varepsilon_{ij}=\frac{N_{ij}^{\rm recontructed}}{N_j^{\rm generated}}\cdot \frac{1}{\varepsilon_{\rm ST}}.
	\label{equation:FFeff}
\end{equation}
Here, $N_{ij}^{\rm recontructed}$ is the number of events generated in the $j\text{-}$th $q^2$ interval and reconstructed in the $i$-th $q^2$ interval, $N_j^{\rm generated}$ is the total number of events generated in the $j\text{-}$th $q^2$ interval, and $\varepsilon_{\rm ST}$ is the detection efficiency of  ST $\bar D$ mesons.
The statistical uncertainty of $N_{\rm produced}^{i}$ is given by
\begin{equation}
	\left[\sigma\left(N_{\rm produced}^{i}\right)\right]^2=\sum_j^{N_{\rm bins}}\left(\varepsilon^{-1}\right)_{ij}^2\left[\sigma_{\rm stat}\left(N_{\rm DT}^{j}\right)\right]^2,
\end{equation}
where $\sigma_{\rm stat}(N_{\rm DT}^{j})$ is the statistical uncertainty of $N_{\rm DT}^{j}$.

Table~\ref{tab:pdweffmatrix} of~\hyperref[sec:appendix]{Appendix} gives the elements of the efficiency matrices weighted by the yields of ST $\bar D$ mesons in data.

\begin{figure}[htbp]
	\begin{center}
		\includegraphics[width=\columnwidth]{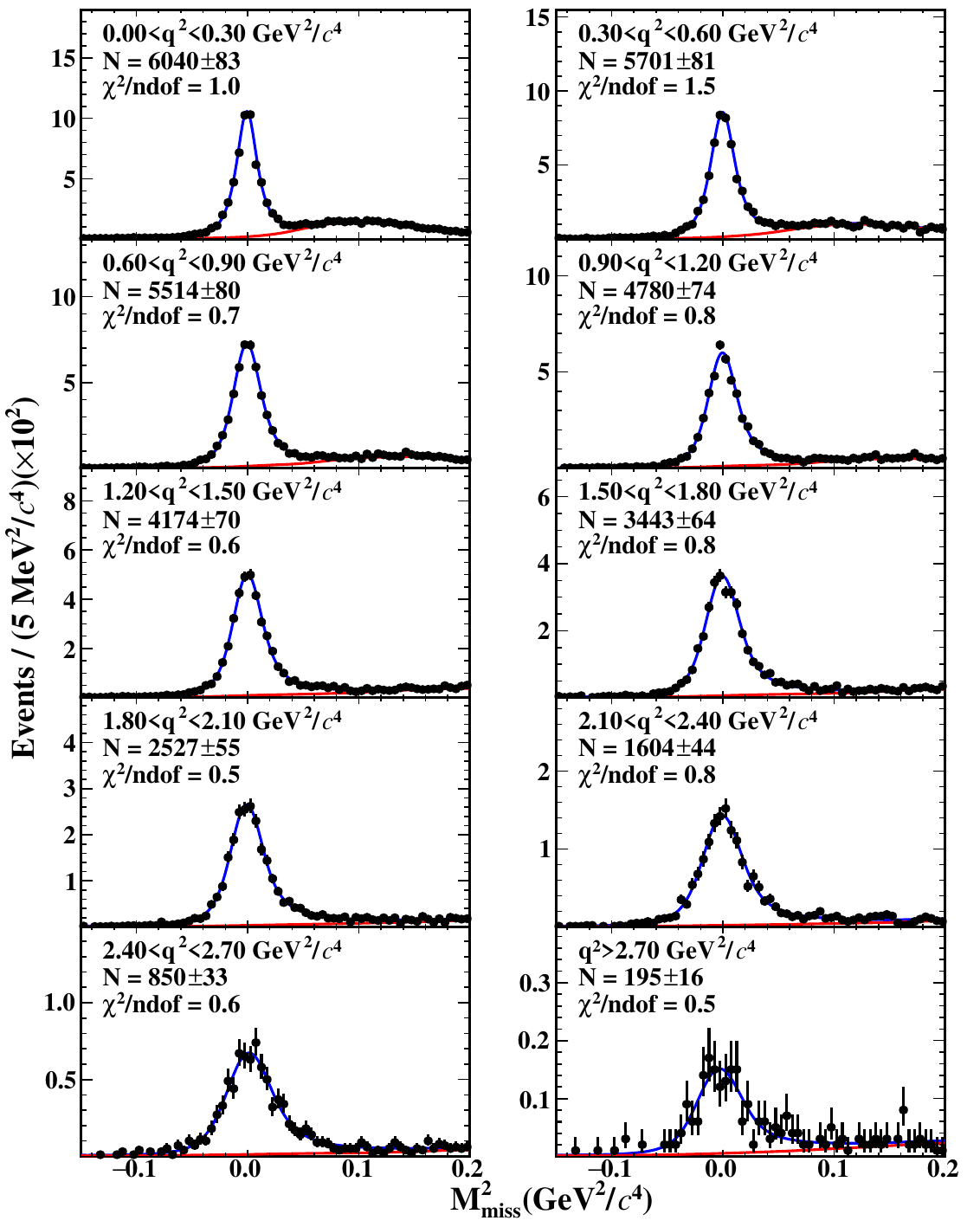}
		\caption{Fits to the $M_{\rm miss}^{2}$ distributions of the candidates for the semileptonic decay $\pienu$ with $q^2$ in various bins.
			The points with error bars are data, the blue curves are the fit results,
			and the red curves are the fitted combinatorial background shapes.
			\label{fig:pienumm2mq2}
		}
	\end{center}
\end{figure}

For each signal decay, the yield of signal events observed in each reconstructed $q^2$ interval is obtained
from a fit to the $M_{\rm miss}^{2}$ distribution, by using the same fit method mentioned in Sec.~\ref{sec:bffit}.
Figure~\ref{fig:pienumm2mq2} shows the results of the fits to the $M_{\rm miss}^{2}$ distributions in the reconstructed $q^2$ intervals for $\pienu$. Similar figures for other three signal channels are provided in Figs.~\ref{fig:pimunumm2mq2},~\ref{fig:pizenumm2mq2}, and~\ref{fig:pizmunumm2mq2} in~\hyperref[sec:appendix]{Appendix}.

Table~\ref{tab:decayrate} summarizes the $q^2$ ranges, the fitted numbers of observed DT events ($N_{\rm DT}$),
the numbers of produced events ($N_{\rm produced}$) calculated by the weighted efficiency matrix
and the decay rates of semileptonic $D$ decays ($\Delta\Gamma$) for $\pienu$, $\pimunu$, $\pizenu$, and $\pizmunu$ in the individual $q^2$ intervals, respectively. 

\begin{table*}[htbp]
	\caption{The measured partial decay rates of the semileptonic decays $\pienu$, $\pimunu$, $\pizenu$, and $\pizmunu$ in various $q^2$ bins, respectively, where
		$N_{\rm DT}$ is the observed yield, $N_{\rm produced}$ is the produced yield, and $\Delta\Gamma$ is the partial decay rate, and the uncertainties are statistical.}
		\label{tab:decayrate}
	\begin{tabular}{c|r@{}lr@{}lr@{}l|r@{}lr@{}lr@{}l}
		\hline\hline
		Decay &\multicolumn{6}{c|}{$\pienu$}                           & \multicolumn{6}{c}{$\pimunu$}         \\ \cline{2-13}
		$q^2~({\rm GeV}^{2}/c^4)$&\multicolumn{2}{c}{$N_{\rm DT}$}&\multicolumn{2}{c}{$N_{\rm produced}$}&\multicolumn{2}{c|}{$\Delta\Gamma~({\rm ns^{-1}})$} & \multicolumn{2}{c}{$N_{\rm DT}$}&\multicolumn{2}{c}{$N_{\rm produced}$}&\multicolumn{2}{c}{$\Delta\Gamma~({\rm ns^{-1}})$} \\ \hline
		$(m_{\ell^{+}}^{2},0.30)$       &   6040    &$\pm     83$   &     10490 &$\pm       149$    &  1.342    &$\pm  0.019$   &   4472    &$\pm     97$   &      9106 &$\pm       201$    &  1.165    &$\pm  0.026$   \\
		$(0.30,0.60)$                   &   5701    &$\pm     81$   &      9406 &$\pm       143$    &  1.203    &$\pm  0.018$   &   4740    &$\pm    109$   &      9485 &$\pm       228$    &  1.213    &$\pm  0.029$   \\
		$(0.60,0.90)$                   &   5514    &$\pm     80$   &      8767 &$\pm       135$    &  1.121    &$\pm  0.017$   &   4097    &$\pm     97$   &      8273 &$\pm       207$    &  1.058    &$\pm  0.026$   \\
		$(0.90,1.20)$                   &   4780    &$\pm     74$   &      7394 &$\pm       124$    &  0.946    &$\pm  0.016$   &   3442    &$\pm     82$   &      7271 &$\pm       183$    &  0.930    &$\pm  0.023$   \\
		$(1.20,1.50)$                   &   4174    &$\pm     70$   &      6419 &$\pm       117$    &  0.821    &$\pm  0.015$   &   2714    &$\pm     67$   &      6055 &$\pm       160$    &  0.774    &$\pm  0.020$   \\
		$(1.50,1.80)$                   &   3443    &$\pm     64$   &      5325 &$\pm       107$    &  0.681    &$\pm  0.014$   &   2180    &$\pm     58$   &      5128 &$\pm       146$    &  0.656    &$\pm  0.019$   \\
		$(1.80,2.10)$                   &   2527    &$\pm     55$   &      4002 &$\pm        94$    &  0.512    &$\pm  0.012$   &   1589    &$\pm     43$   &      3780 &$\pm       109$    &  0.483    &$\pm  0.014$   \\
		$(2.10,2.40)$                   &   1604    &$\pm     44$   &      2629 &$\pm        79$    &  0.336    &$\pm  0.010$   &   1025    &$\pm     67$   &      2488 &$\pm       175$    &  0.318    &$\pm  0.022$   \\
		$(2.40,2.70)$                   &    850    &$\pm     33$   &      1444 &$\pm        60$    &  0.185    &$\pm  0.008$   &    583    &$\pm     65$   &      1488 &$\pm       177$    &  0.190    &$\pm  0.023$   \\
		$(2.70,2.98)$                   &    195    &$\pm     16$   &       341 &$\pm        32$    &  0.044    &$\pm  0.004$   &    123    &$\pm     17$   &       339 &$\pm        51$    &  0.043    &$\pm  0.007$   \\
		\hline
		Decay &\multicolumn{6}{c|}{$\pizenu$}                           & \multicolumn{6}{c}{$\pizmunu$}        \\ \cline{2-13}
		$q^2~({\rm GeV}^{2}/c^4)$&\multicolumn{2}{c}{$N_{\rm DT}$}&\multicolumn{2}{c}{$N_{\rm produced}$}&\multicolumn{2}{c|}{$\Delta\Gamma~({\rm ns^{-1}})$} & \multicolumn{2}{c}{$N_{\rm DT}$}&\multicolumn{2}{c}{$N_{\rm produced}$}&\multicolumn{2}{c}{$\Delta\Gamma~({\rm ns^{-1}})$} \\ \hline
		$(m_{\ell^{+}}^{2},0.30)$       &   2795    &$\pm     59$   &      6849 &$\pm       147$    &  0.623    &$\pm  0.013$   &   2221    &$\pm     61$   &      6313 &$\pm       177$    &  0.574    &$\pm  0.016$   \\
		$(0.30,0.60)$                   &   2657    &$\pm     58$   &      6494 &$\pm       160$    &  0.590    &$\pm  0.015$   &   2081    &$\pm     63$   &      6159 &$\pm       207$    &  0.560    &$\pm  0.019$   \\
		$(0.60,0.90)$                   &   2396    &$\pm     56$   &      5861 &$\pm       162$    &  0.533    &$\pm  0.015$   &   1876    &$\pm     59$   &      6003 &$\pm       216$    &  0.546    &$\pm  0.020$   \\
		$(0.90,1.20)$                   &   2161    &$\pm     53$   &      5410 &$\pm       159$    &  0.492    &$\pm  0.014$   &   1509    &$\pm     49$   &      5241 &$\pm       200$    &  0.477    &$\pm  0.018$   \\
		$(1.20,1.50)$                   &   1739    &$\pm     47$   &      4407 &$\pm       146$    &  0.401    &$\pm  0.013$   &   1167    &$\pm     41$   &      4492 &$\pm       187$    &  0.408    &$\pm  0.017$   \\
		$(1.50,1.80)$                   &   1350    &$\pm     41$   &      3533 &$\pm       131$    &  0.321    &$\pm  0.012$   &    808    &$\pm    117$   &      3364 &$\pm       584$    &  0.306    &$\pm  0.053$   \\
		$(1.80,2.10)$                   &   1047    &$\pm     37$   &      2875 &$\pm       121$    &  0.261    &$\pm  0.011$   &    620    &$\pm     27$   &      2778 &$\pm       140$    &  0.253    &$\pm  0.013$   \\
		$(2.10,2.40)$                   &    690    &$\pm     32$   &      1928 &$\pm       107$    &  0.175    &$\pm  0.010$   &    382    &$\pm     24$   &      1735 &$\pm       130$    &  0.158    &$\pm  0.012$   \\
		$(2.40,2.70)$                   &    385    &$\pm     27$   &      1016 &$\pm        93$    &  0.092    &$\pm  0.008$   &    203    &$\pm     22$   &       936 &$\pm       126$    &  0.085    &$\pm  0.011$   \\
		$(2.70,3.01)$                   &    149    &$\pm     21$   &       295 &$\pm        68$    &  0.027    &$\pm  0.006$   &     46    &$\pm     17$   &       138 &$\pm       121$    &  0.013    &$\pm  0.011$   \\
		\hline\hline
	\end{tabular}
\end{table*}

\subsection{Construction of $\chi^2$ and statistical covariance matrices}
\label{sec:ffchifit}
To extract the hadronic transition form factor parameters and $|V_{cd}|$,
the least $\chi^2$ method is used to fit the partial decay rates of semileptonic $D$ decays.
Considering the correlations of the measured partial decay rates ($\Delta\Gamma_i^{\rm msr}$) among different $q^2$ bins, the $\chi^2$ is given by
\begin{equation}
	\label{eq:chi}
	\chi^2 = \sum_{i,j=1}^{N_{\rm bins}}\left(\Delta\Gamma_i^{\rm msr}-\Delta\Gamma_i^{\rm th}\right) C_{ij}^{-1}\left(\Delta\Gamma_j^{\rm msr}-\Delta\Gamma_j^{\rm th}\right),
\end{equation}
where $\Delta\Gamma_i^{\rm th}$ is the decay rate of the channel $i$ expected by theory, 
$C_{ij}$ is the element of the covariance matrix for the measured partial decay rates and divided into the statistical covariance matrix $C_{ij}^{\rm stat}$ and the systematic covariance matrix $C_{ij}^{\rm syst}$.
The elements of the statistical covariance matrix are defined as
\begin{equation}
	C_{ij}^{\rm stat} =\left (\frac{1}{\tau_{D}N_{\rm ST}}\right)^2\sum_{\alpha}\varepsilon_{i\alpha}^{-1}\varepsilon_{j\alpha}^{-1}\left(\sigma\left(N_{\rm DT}^{\alpha}\right)\right)^2.
	\label{equation:Covffstat}
\end{equation}
where $\sigma\left(N_{\rm DT}^{\alpha}\right)$ is the statistical uncertainty of the signal yield observed in the $\alpha$-th interval.
Table~\ref{tab:pdwstatmatrix} of~\hyperref[sec:appendix]{Appendix} gives the elements of the statistical correction density matrices for the $\pienu$, $\pimunu$, $\pizenu$ and $\pizmunu$ channels.

\subsection{Systematic uncertainties in hadronic transition form factor}
\label{sec:ffsys}
The systematic uncertainties in the measurement of hadronic transition form factor are discussed below.

\paragraph{\boldmath \bf ST $\bar D$ yields}

The systematic uncertainties associated with the total numbers of ST $\bar D^0(D^-)$ mesons are fully correlated across $q^2$ intervals.
The element of the related systematic covariance matrix is calculated by
\begin{equation}
	C_{ij}^{\mathrm{syst}}\left(N_{\rm ST}\right)=\Delta\Gamma_{i}\Delta\Gamma_{j}\left(\frac{\sigma\left(N_{\rm ST}\right)}{N_{\rm ST}}\right)^2,
\end{equation}
where $\sigma(N_{\rm ST})/N_{\rm ST}$ is the relative uncertainty on the number of $\bar D^0(D^-)$ tags.

\paragraph{\boldmath \bf  $D$ lifetime}

The systematic uncertainties associated with the $D$ lifetime are fully correlated across the $q^2$ bins. The element of the related systematic
covariance matrix is calculated by
\begin{equation}
	C_{ij}^{\rm syst}\left(\tau_{D}\right)=\sigma\left(\Delta\Gamma_i\right)\sigma\left(\Delta\Gamma_j\right),
\end{equation}
where $\sigma(\Delta\Gamma_i)=\sigma \tau_{D}\cdot\Delta\Gamma_i$ and $\sigma \tau_{D}$ is the uncertainty of the $D$ lifetime~\cite{pdg2024}.

\paragraph{\boldmath \bf  MC statistics}

The systematic uncertainties in efficiencies and correlations between the $q^2$ bins due to the limited MC size are calculated by
\begin{equation}
	C_{ij}^{\rm syst}=\left(\frac{1}{\tau_{D}N_{\rm ST}}\right)^2\sum_{\alpha\beta}N_{\rm DT}^{\alpha}N_{\rm DT}^{\beta}\mathrm{Cov}\left(\varepsilon_
	{i\alpha}^{-1},\varepsilon_{j\beta}^{-1}\right),
\end{equation}
where the covariances of the inverse efficiency matrix elements are given by~\cite{Lefebvre:1999yu}
\begin{equation}
	\mathrm{Cov}\left(\varepsilon_{i\alpha}^{-1},\varepsilon_{j\beta}^{-1}\right)=\sum\limits_{mn}\left(\varepsilon_{im}^{-1}\varepsilon_{j m}^{-1}\right)
	\left [\sigma\left (\epsilon_{mn}\right )\right]^2\left (\varepsilon_{\alpha n}^{-1}\varepsilon_{\beta n}^{-1}\right).
\end{equation}

\paragraph{\boldmath \bf Hadronic transition form factor}

The systematic uncertainties associated with the hadronic transition form factor used to generate signal MC events are estimated by varying the parameters of the two-parameters
series expansion model by $\pm1\sigma$ .
The partial decay rates are then calculated in different $q^2$ bins with the reweighted efficiency matrix, and their differences with the nominal values are taken as the systematic uncertainties. Then the element of the covariance matrix is
calculated with
\begin{equation}
	C_{ij}^{\rm syst}\left (\rm{FF}\right )=\delta\left(\Delta\Gamma_i\right)\delta\left(\Delta\Gamma_j\right),
\end{equation}
where $\delta(\Delta\Gamma_i)$ denotes the change of the partial decay rate in the $i$-th $q^2$ interval.

\paragraph{\textbf{\textup{Tracking, PID, and $\pi^0$ reconstruction}}}

The systematic uncertainties associated with the electron tracking and PID, pion tracking and PID, and $\pi^0$ reconstruction
are estimated by varying the corresponding correction factors of efficiencies within $\pm 1\sigma$. Using the new efficiency matrix, the element of the
corresponding systematic covariance matrix is calculated by
\begin{equation}
	C_{ij}^{\rm syst}\left (\rm{Tracking,~PID,~{\pi^0} rec.}\right)=\delta\left(\Delta\Gamma_i\right)\delta\left(\Delta\Gamma_j\right),
\end{equation}
where $\delta(\Delta\Gamma_i)$ denotes the change of the partial decay rate in the $i$-th $q^2$ interval.

\paragraph{\boldmath \bf $M_{\rm miss}^{2}$ fit}

The systematic covariance matrix arising from the uncertainty in the $M_{\rm miss}^{2}$ fit has elements
\begin{equation}
	C_{ij}^{\rm syst}\left(M_{\rm miss}^{2}~\rm{fit}\right)=\left(\frac{1}{\tau_{D}N_{\rm ST}}\right)^2\sum_{\alpha}\varepsilon_{i\alpha}^{-1}\varepsilon_
	{j\alpha}^{-1}\left(\sigma_{\alpha}^{\rm{fit}}\right)^2,
\end{equation}
where $\sigma_{\alpha}^{\rm{fit}}$ is the systematic uncertainty of the signal yield observed in the interval $\alpha$ obtained by varying the background
shape in the $M_{\rm miss}^{2}$ fit.

\paragraph{\boldmath \bf Remaining uncertainties}

The remaining systematic uncertainties are assumed to be  fully correlated across $q^2$ bins and the element of the corresponding systematic covariance
matrix is calculated by
\begin{equation}
	C_{ij}^{\rm syst}=\sigma\left(\Delta\Gamma_i\right)\sigma\left(\Delta\Gamma_j\right),
\end{equation}
where $\sigma(\Delta\Gamma_i)=\sigma_{\rm syst}\cdot\Delta\Gamma_i$ and $\sigma_{\rm syst}$ is the corresponding uncertainty reported in Table~\ref{tab:sysff} of~\hyperref[sec:appendix]{Appendix} for $\pienu$, $\pimunu$, $\pizenu$, and $\pizmunu$ decay channels. Table~\ref{tab:pdwsysmatrix} of~\hyperref[sec:appendix]{Appendix} gives the elements of the systematic covariance
density matrices for $\pienu$, $\pimunu$, $\pizenu$, and $\pizmunu$ channels.

\subsection{Hadronic transition form factor from separate fits}

The central values of the form factor parameters are obtained by fitting the data using a combined statistical and systematic covariance matrix.
The systematic error of the measured form factor parameters is calculated as the quadratic difference between the uncertainties from the combined
covariance matrix and those from fits using only the statistical covariance matrix.

\begin{figure*}[htbp]
	\begin{center}
		\includegraphics[width=0.95\textwidth]{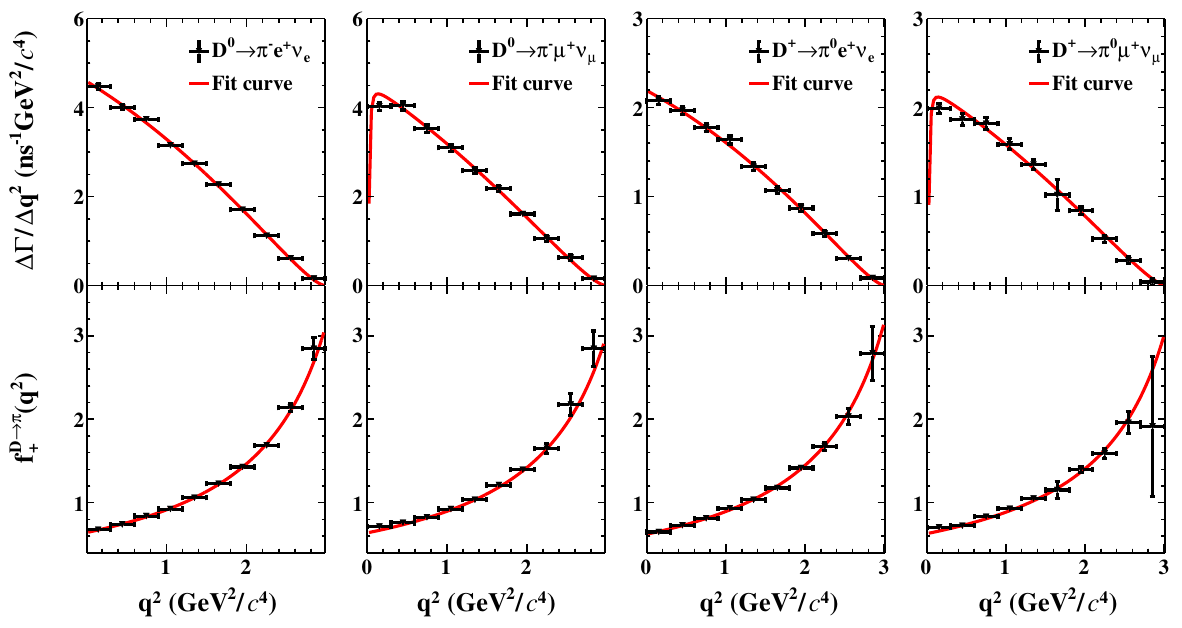}
		\caption{(Top) Separate fits to the partial decay rates of the four semileptonic decays $\pilnu$ and (Bottom) the projections on the hadronic transition form factor
			as a function of $q^2$. The dots with error bars are the measured partial decay rates and the solid curves are the best fits obtained with form factors parameterized by a two-parameter series expansion.
			\label{fig:ff_pilnu}
		}
	\end{center}
\end{figure*}

The top subfigures of Fig.~\ref{fig:ff_pilnu} show the results of the separate fits to the partial decay rates of $\pienu$, $\pimunu$, $\pizenu$, and
$\pizmunu$. The
bottom subfigures of Fig.~\ref{fig:ff_pilnu} show the projections of the fits on the hadronic transition form factor as a function of $q^2$, where the dots with
error bars show the measured values of the hadronic transition form factor, which are obtained with
\begin{equation}
	f^{D\to\pi}_+(q^2)=\sqrt{\frac{d\Gamma}{dq^2}\frac{|f^{D\to\pi}_+(q_i^2)|^2(q^2_{\rm max,{\it i}}-q^2_{\rm min,{\it i}})}{\int_{q^2_{\rm min,{\it
						i}}}^{q^2_{\rm max,{\it i}}}\frac{d\Gamma}{dq^2}dq^2}},
\end{equation}
where $q^2_{\rm{min,{\it i}}}$ and $q^2_{\rm max,{\it i}}$ are the low and high boundaries of the $i-$th $q^2$ bin. In the calculation, $f^{D\to\pi}_+(q^2)$ and
$f^{D\to\pi}_+(q_i^2)$ are calculated using the two-parameter series expansion model.

Table~\ref{tab:fit_par} gives the obtained results of the fit parameters from the individual fits to
the partial   decay rates of $\pienu$, $\pimunu$, $\pizenu$, and $\pizmunu$ decays.

\begin{table*}[htbp]
	\centering
	\caption{The parameters of hadronic transition form factor obtained by fitting the partial decay rates of the semileptonic decays $\pienu$, $\pimunu$, $\pizenu$, and $\pizmunu$.
		The first uncertainties are statistical and the second systematic. The ndf denotes the number of degrees of freedom.
		\label{tab:fit_par}}
		\footnotesize
		\begin{tabular}{lcccccc}%{C{2.4cm}C{2.2cm}C{3.3cm}C{2.8cm}C{3.3cm}C{1.5cm}C{1.3cm}}
			\hline\hline
			Case &Decay             & $f_{+}^{D\to\pi}(0)|V_{cd}|$    & $r^{+}_{1}$               & $f_{+}^{D\to\pi}(0)$        	 &  Coefficient   & $\chi^{2}/{\rm ndf}$ \\ \hline
			\multirow{4}{*}{Separate fit}  & $\pienu$       & $0.1441 \pm 0.0008 \pm 0.0004$  & $-2.02 \pm 0.05 \pm 0.01$ & $0.6408 \pm 0.0034 \pm 0.0018$ &  0.680         & 7.2/8          \\
			&$\pimunu$      & $0.1434 \pm 0.0012 \pm 0.0007$  & $-1.91 \pm 0.06 \pm 0.02$ & $0.6376 \pm 0.0055 \pm 0.0032$ &  0.806         & 3.1/8           \\
			&$\pizenu$      & $0.1400 \pm 0.0012 \pm 0.0005$  & $-2.05 \pm 0.06 \pm 0.02$ & $0.6228 \pm 0.0051 \pm 0.0024$ &  0.734         & 4.6/8           \\
			&$\pizmunu$     & $0.1415 \pm 0.0015 \pm 0.0009$  & $-1.90 \pm 0.08 \pm 0.03$ & $0.6291 \pm 0.0069 \pm 0.0039$ &  0.696         & 7.4/8           \\ \hline
			Simultaneous fit & $\pilnu$    & $0.1425 \pm 0.0005 \pm 0.0003$  & $-1.99 \pm 0.03 \pm 0.01$ & $0.6339 \pm 0.0024 \pm 0.0014$ &  0.695         & 44.1/38          \\ \hline
			\hline
		\end{tabular}
\end{table*}

\subsection{Hadronic transition form factor from simultaneous fit}

To take into account the correlation effects in the measurements of the hadronic transition form factor among these four semileptonic $D$ decays, we perform a
simultaneous fit to the partial   decay rates of the semileptonic decays $\pienu$, $\pimunu$, $\pizenu$, and $\pizmunu$ to extract the product of $\ffpi|V_{cd}|$. In the simultaneous fit, we still use the least $\chi^2$ method to obtain the hadronic transition form factor given in Eq.~(\ref{eq:chi}).
The $\Delta\Gamma_i$ for these four semileptonic decays are simultaneously inserted into one vector of length 40 and $C_{ij}$ becomes a $40\times40$ covariance
matrix for the $\Delta\Gamma_i$, which is redefined as $C_{ij}=C^{\rm stat}_{ij}+C^{\rm csyst}_{ij}+C^{\rm usyst}_{ij}$, $(i,j=1,2,3,...,39,
40)$, where $C^{\rm stat}_{ij}$ is the statistical covariance matrix defined as
\begin{equation}
	C^{\rm stat}_{ij}=
	\begin{pmatrix}
		A&0&0&0\\
		0&B&0&0\\
		0&0&C&0\\
		0&0&0&D
	\end{pmatrix},
\end{equation}
$A$, $B$, $C$, and $D$ here represent the statistical covariance matrices obtained for individual semileptonic decays.
The $C^{\rm csyst}_{ij}$ is the correlated systematic covariance matrix given by
\begin{equation}
	\label{eq:csys_matrix}
	C_{ij}^{\rm{csyst}}=\delta(\Delta\Gamma_i)\delta(\Delta\Gamma_j).
\end{equation}

For the uncorrelated systematic uncertainties, the systematic covariance matrix $C^{\rm usyst}_{ij}$ is defined as
\begin{equation}
	C^{\rm usyst}_{ij}=
	\begin{pmatrix}
		a&0&0&0\\
		0&b&0&0\\
		0&0&c&0\\
		0&0&0&d
	\end{pmatrix}
	,
\end{equation}
where $a$, $b$, $c$, and $d$ represent the uncorrelated systematic covariance matrices obtained for individual semileptonic decays. The
detailed covariance density matrices for the simultaneous fit are given in Tables~\ref{table:statcovellnu} and~\ref{table:systcovellnu} of~\hyperref[sec:appendix]{Appendix}.

\begin{figure}[htbp]
	\begin{center}
		\includegraphics[width=\columnwidth]{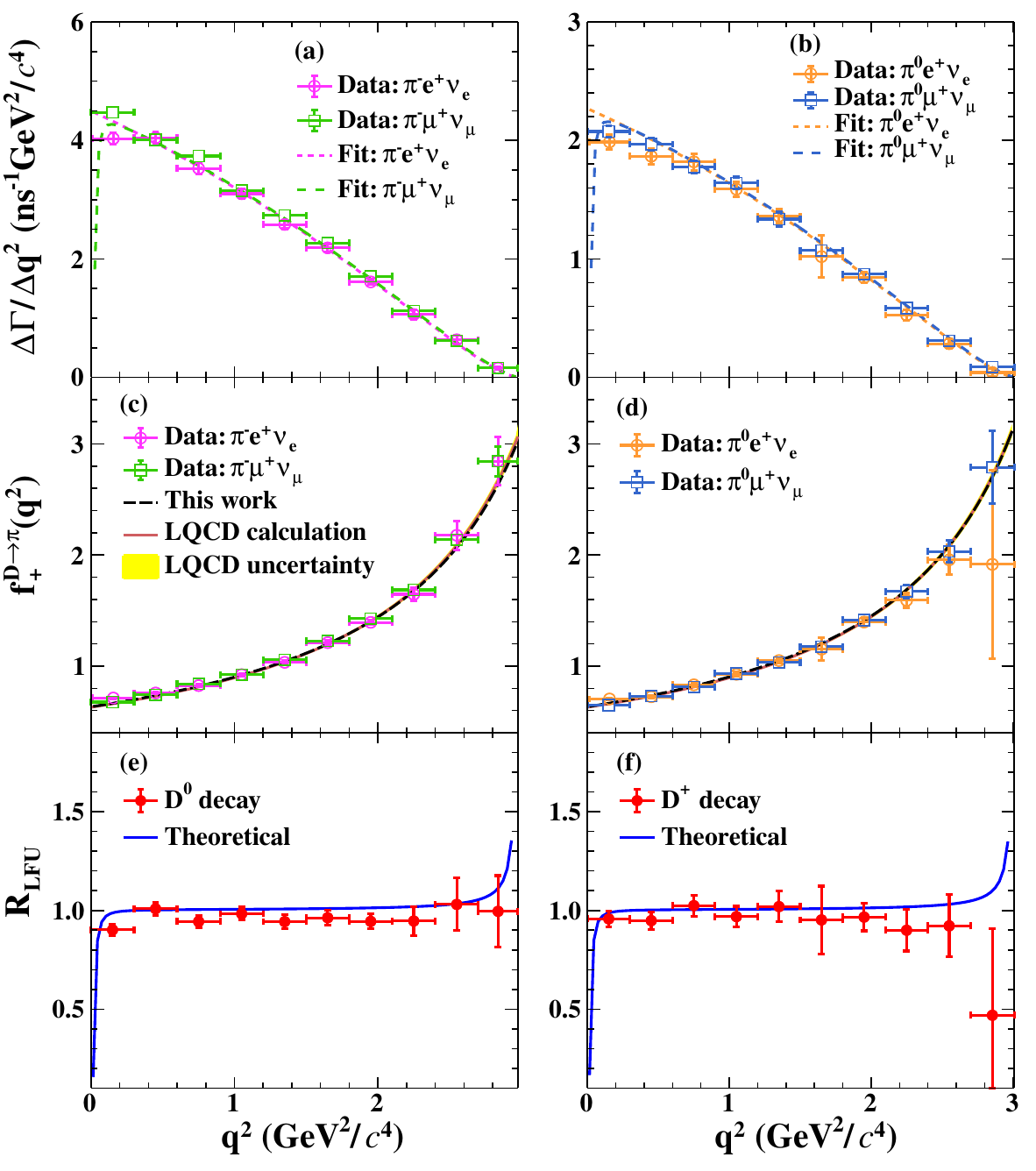}
		\caption{(a, b) Simultaneous fit to the partial   decay rates of the semileptonic decays $\pilnu$. (c, d) Projections to the hadronic transition form factor as function of $q^2$ of
			$\pilnu$. (e, f) The ratio of differential decay rates of $\pimunu$ over $\pienu$ and the ratio of differential decay rates of $\pizmunu$ over $\pizenu$ in each $q^{2}$ bin.
			 The dots with error bars are data, and the dash lines are the results with the parameters of the simultaneous fit.
			\label{ff_pilnu}}
	\end{center}
\end{figure}

A simultaneous fit is performed on the partial decay rates of the four semileptonic $D$ decays, with the modified $\Delta\Gamma_i$ and $C_{ij}$.
In the fit, the four semileptonic $D$ decays share the parameters of the hadronic transition form factor.
Figure~\ref{ff_pilnu} shows the fit results.
The fitted parameters are summarized in the last row of Table~\ref{tab:fit_par}.

\section{Forward and Backward asymmetries}

\subsection{Theoretical formula}

The angular distributions of the semileptonic decays $D\to\pi\ell^+\nu_{\ell}$ are of interest, in addition to the partial decay rate, due to its dependence on potential scalar current contribution in the $c\to d\ell^+\nu_\ell$ transition. As an angular observable, 
the forward-backward asymmetry is defined as~\cite{Faustov:2019mqr,Fajfer:2015ixa,Leng:2020fei,Jain:2025kqe}
\begin{equation}
	\mathcal{A}_{FB}(q^2)=\frac{d\Gamma^\ell(\cos\theta_W>0)-d\Gamma^\ell(\cos\theta_W<0)}{d\Gamma^\ell(\cos\theta_W>0)+d\Gamma^\ell(\cos\theta_W<0)},
\end{equation}
where $\theta_{W}$ is the angle between the direction of the lepton and the direction opposite to that of the $D$ meson in the $\ell^+\nu_{\ell}$ rest frame.  The theoretical expression for $\mathcal{A}^{\rm th}_{FB}(q^2)$ is given by~\cite{Faustov:2019mqr,Fajfer:2015ixa,Leng:2020fei,Jain:2025kqe}
\begin{equation}
	\frac{3\mathcal{N}(q^2)}{2}\frac{1}{d\Gamma/dq^2}\left(1-\frac{m_\ell^2}{q^2}\right)^2\frac{m_\ell^2}{q^2}{\rm Re}\left(\mathcal{H}_0(q^2)\mathcal{H}_t^*(q^2)\right),
\end{equation}
where the definitions of the variables align with those in Sec.~\ref{sec:FFthefor}. In the SM, this asymmetry depends on the lepton mass squared, $m_\ell^2$, resulting in a trivial distribution of $\mathcal{A}_{FB}^e(q^2) = 0$ for the positron channels over the full $q^2$ range. For the muon channels, the averaged asymmetry is predicted to be -0.040~\cite{Faustov:2019mqr}.

When scalar currents are considered, the hadronic helicity amplitude $\mathcal{H}_{t}(q^2)$ becomes
\begin{equation}
	\mathcal{H}_{t}(q^2)= \left(1+C_S^\ell\frac{q^2}{m_\ell(m_d-m_c)}\right)\frac{m_D^2-m_\pi^2}{\sqrt{q^2}}f^{D\to\pi}_0(q^2),
	\label{equation:heliamp}
\end{equation}
where $C_S^{\ell}=C_R^{\ell}+C_L^{\ell}$ represents the scalar combination of Wilson coefficients as defined in Eq.~\ref{eqution:Leff}. Therefore, any observed deviation of $\mathcal{A}_{FB}$ from the SM prediction may imply the presence of scalar current contribution in the decays $\pilnu$.
%In this paper, we will extract the $q^2$-binned asymmetries.

\subsection{$q^2$-binned forward-backward asymmetries}

\begin{figure*}[htbp]
	\centering
	\includegraphics[width=0.45\linewidth]{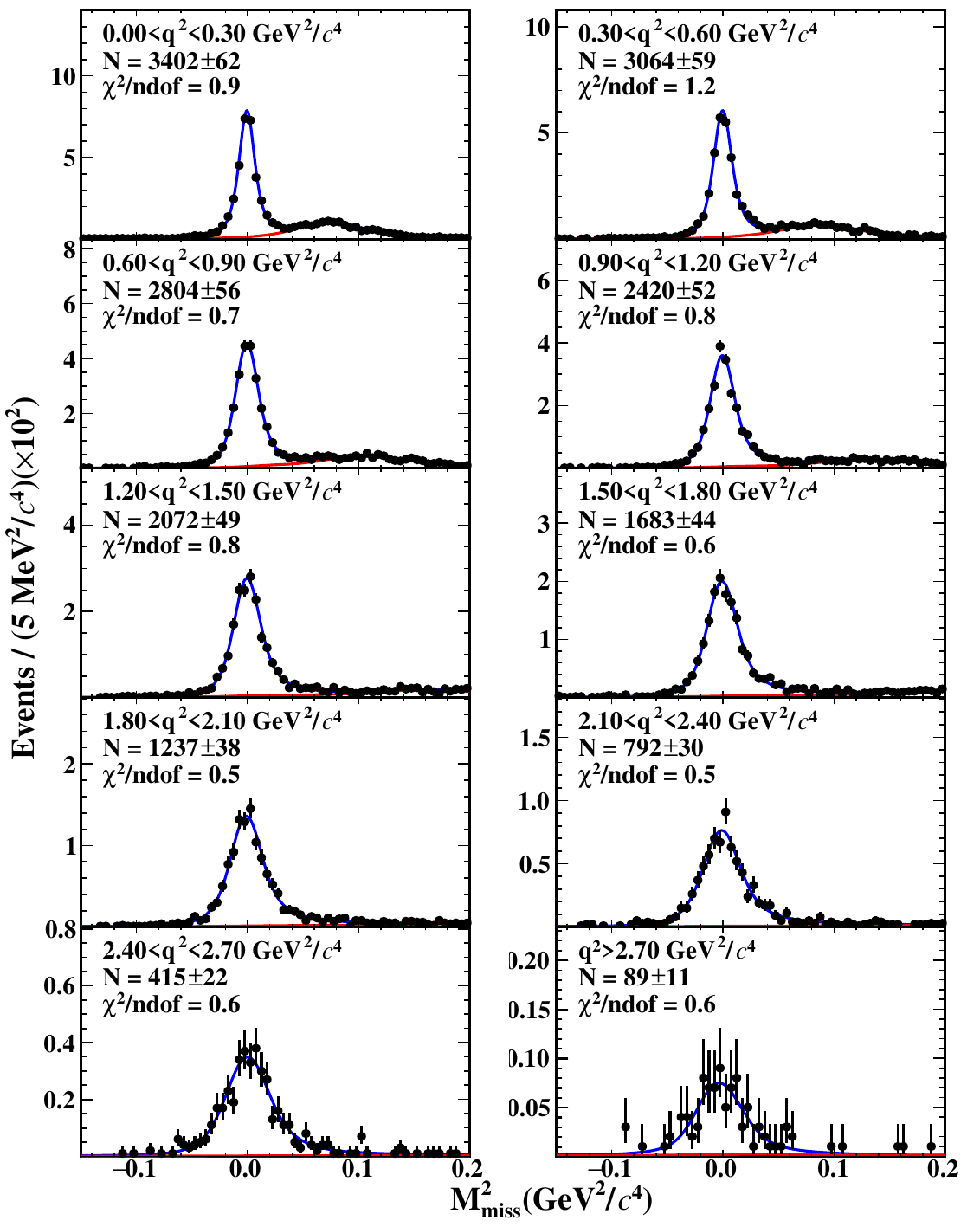}
	\includegraphics[width=0.45\linewidth]{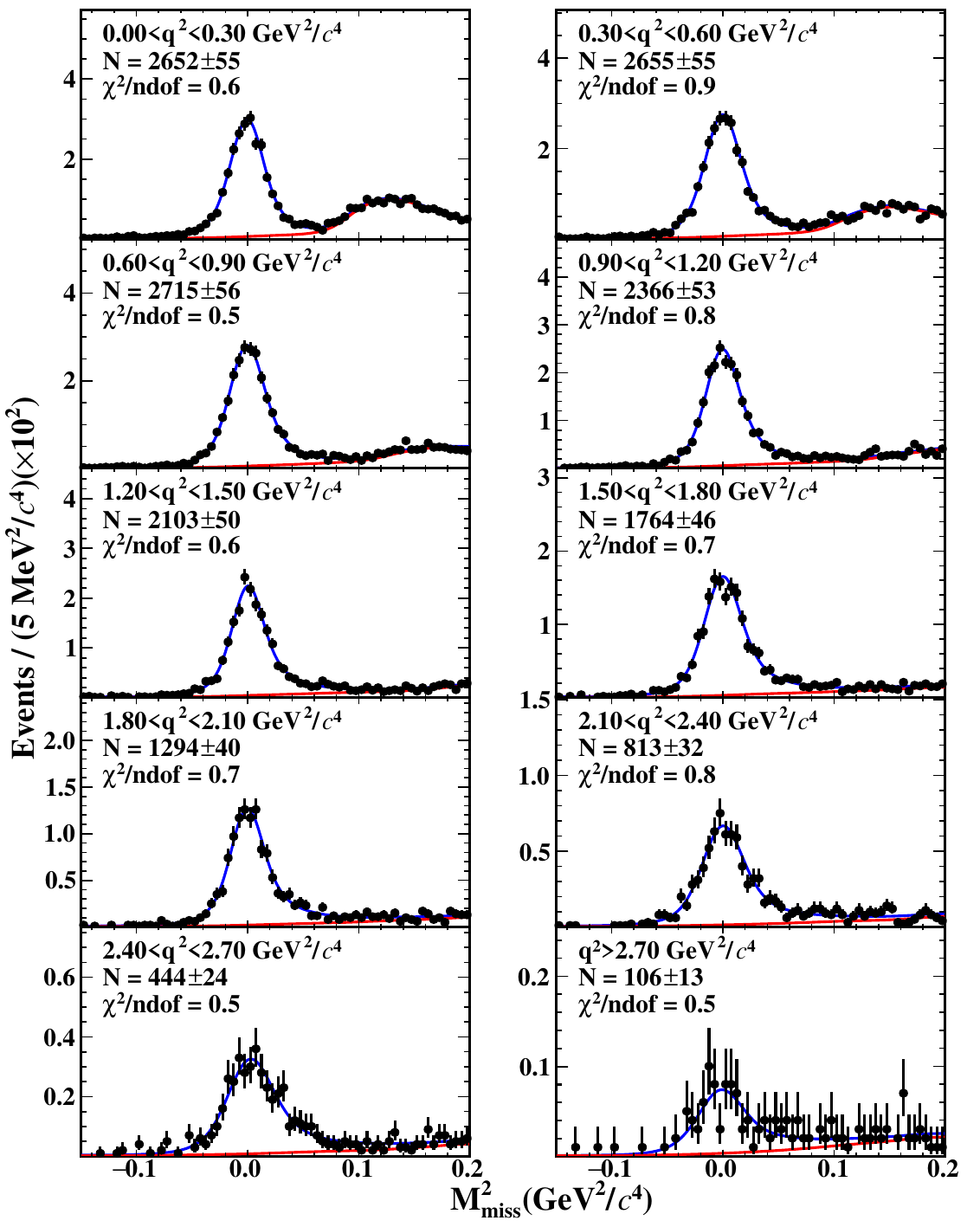}
	\caption{The $M_{\rm miss}^2$ distributions of the accepted forward (left) and backward (right) candidate events in different $q^{2}$ bins for $\pienu$. The points with error bars are data, the blue curves are the fit results and the red curves are the fitted combinatorial background shapes}
	\label{fig:forbackpienu}
\end{figure*}

The $q^{2}$-binned forward-backward asymmetry in the $i$-th $q^{2}$ interval is defined as
\begin{equation}
	\mathcal{A}_{FB,i}=\int_{q_{\text{min}(i)}^{2}}^{q_{\text{max}(i)}^{2}}\mathcal{A}_{FB}(q^{2})\frac{d\Gamma}{dq^{2}}dq^{2}\left/\int_{q_{\text{min}(i)}^{2}}^{q_{\text{max}(i)}^{2}}\frac{d\Gamma}{dq^{2}}dq^{2},\right.
\end{equation}
where the $q^2$ intervals are the same as that in Sec.~\ref{sec:decayrates}. They are measured with
\begin{equation}
	\footnotesize
	\mathcal{A}_{FB,i}^{\rm msr}=\frac{N_{\rm produced}(\cos\theta_{W}>0,q^2_i)-N_{\rm produced}(\cos\theta_{W}<0,q^2_i)}{N_{\rm produced}(\cos\theta_{W}>0,q^2_i)+N_{\rm produced}(\cos\theta_{W}<0,q^2_i)}.
\end{equation}
The number of produced events in the $i$-th $q^2$ interval is obtained with 
\begin{equation}
	N_{\rm produced}^{i,\alpha} = \sum_{(j,\beta)}(\varepsilon^{-1})_{(i,\alpha)(j,\beta)}N_{\rm DT}^{j,\beta},
\end{equation}
where the indices $\alpha$ and $\beta$ indicate forward and backward, respectively. Here, DT yields $N_{\rm DT}$ are obtained by fitting the $M_{\rm miss}^{2}$ distributions as shown in Fig.~\ref{fig:forbackpienu} for $\pienu$ with fitted yields summarized in Table~\ref{tab:forbackasy}. Results for other three signal decays are available in Figs.~\ref{fig:forbackpimunu},~\ref{fig:forbackpi0enu},~\ref{fig:forbackpi0munu} and Table~\ref{tab:forbackasy}.

The efficiency matrix $\varepsilon_{(i,\alpha)(j,\beta)}$ is determined in the same approach as Eq.~\ref{equation:FFeff} with detailed values shown in Tables~\ref{table:afbeffpienu},~\ref{table:afbeffpimunu},~\ref{table:afbeffpi0enu}, and~\ref{table:afbeffpi0munu} of~\hyperref[sec:appendix]{Appendix} for $\pienu$, $\pimunu$, $\pizenu$, and $\pizmunu$ channels, respectively. The statistical covariant matrices of $N_{\rm produced}$ are constructed with 
\begin{equation}
	C_{\alpha,\beta}^{\rm stat}=\sum_{\gamma}(\varepsilon^{-1})_{\alpha\gamma}(\varepsilon^{-1})_{\beta\gamma}\left(\sigma\left(N_{\mathrm{DT}}^{\gamma}\right)\right)^{2}
\end{equation}
as summarized in Table~\ref{tab:afbstatmatrix} in~\hyperref[sec:appendix]{Appendix}.

Except $\pi^-$ tracking and PID ($\pi^0$ reconstruction), $\ell^+$ tracking and PID, $M_{\rm miss}^2$ fit, and MC statistics, other systematic uncertainties of $N_{\rm produced}$ cancel in the calculation of $\mathcal{A}_{FB,i}$. All of these are estimated in the same approach as described in Sec.~\ref{sec:ffsys}. Table~\ref{tab:afbsysmatrix}  of~\hyperref[sec:appendix]{Appendix} gives the elements of the systematic covariance matrices for $\pienu$, $\pimunu$, $\pizenu$, and $\pizmunu$ channels.

\begin{figure}
	\centering
	\includegraphics[width=\columnwidth]{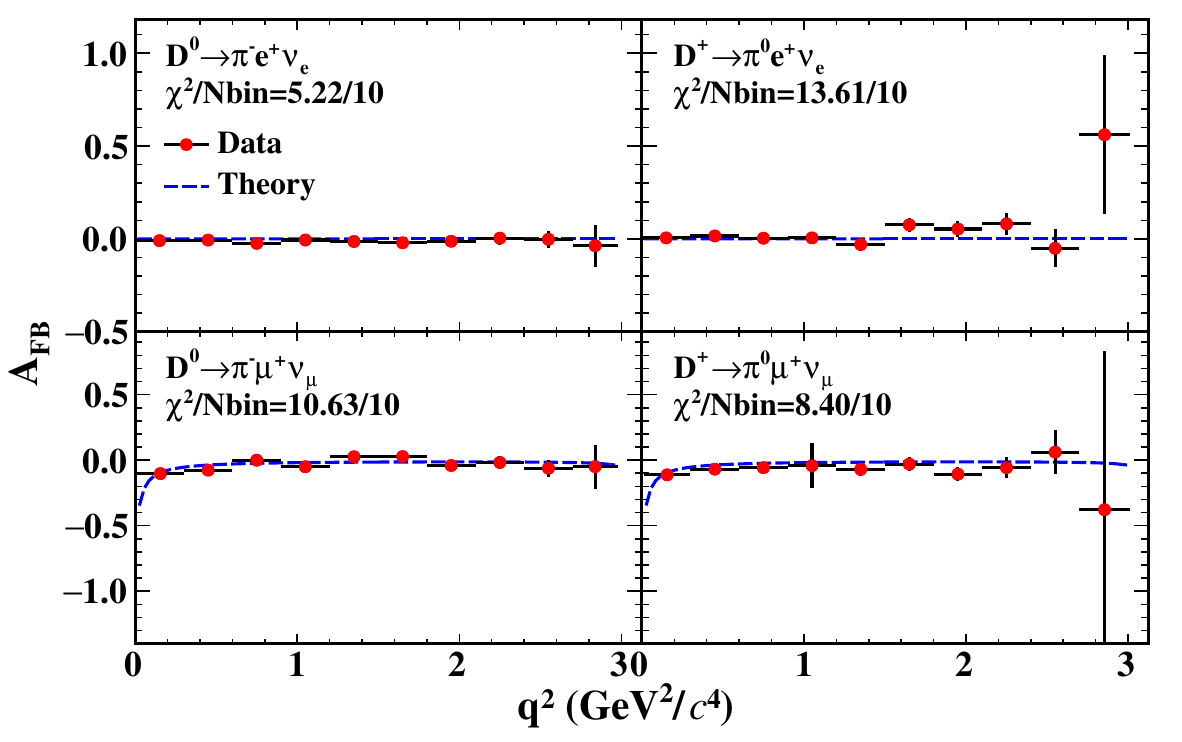}
	\caption{The measured $q^{2}$-binned forward-backward asymmetries. The red points with error bars are data. The dashed blue lines denote the theoretical curves.}
	\label{fig:afbsmfit}
\end{figure}

Table~\ref{tab:forbackasy} shows the determined $q^2$-binned forward-backward asymmetries of $\pienu$, $\pimunu$, $\pizenu$, and $\pizmunu$, respectively. Figure~\ref{fig:afbsmfit} shows the measured $q^2$-binned forward-backward asymmetries, which are consistent with the theoretical predictions.

\begin{table*}[htbp]
	\caption{The numbers of observed forward/backward events $N_{\rm DT}$, the produced forward/backward events $N_{\rm produced}$ in different $q^2$ intervals and the $q^2$-binned forward-backward asymmetries of $\pienu$, $\pimunu$, $\pizenu$, and $\pizmunu$ decays, respectively, where the first uncertainties are statistical and the second systematic.}
		\label{tab:forbackasy}
	\rotatebox{90}{
	\resizebox{0.95\textheight}{!}{
	\begin{tabular}{c|r@{}lr@{}lr@{}lr@{}lr@{}l|r@{}lr@{}lr@{}lr@{}lr@{}l}
		\hline\hline
		Decay & \multicolumn{10}{c|}{$\pienu$}                           & \multicolumn{10}{c}{$\pimunu$}         \\ \cline{2-21}
		$q^2$ (GeV$^2/c^4$)&\multicolumn{2}{c}{$N_{\rm DT}^{\rm Forward}$}&\multicolumn{2}{c}{$N_{\rm DT}^{\rm Backward}$}&\multicolumn{2}{c}{$N_{\rm produced}^{\rm Forward}$}&\multicolumn{2}{c}{$N_{\rm produced}^{\rm Backward}$}&\multicolumn{2}{c|}{$A_{FB}$} &\multicolumn{2}{c}{$N_{\rm DT}^{\rm Forward}$}&\multicolumn{2}{c}{$N_{\rm DT}^{\rm Backward}$}&\multicolumn{2}{c}{$N_{\rm produced}^{\rm Forward}$}&\multicolumn{2}{c}{$N_{\rm produced}^{\rm Backward}$}&\multicolumn{2}{c}{$A_{FB}$} \\ \hline
		$(m_{\ell^{+}}^{2},0.30)$       &  3402 &$\pm  62$  &  2652 &$\pm  55$  &   5217    &$\pm 100$  &   5317    &$\pm 115$  & -0.010    &$\pm  0.015$   $\pm$  0.002   &  1560 &$\pm  45$  &  2984 &$\pm  65$  &   4169    &$\pm 127$  &   5119    &$\pm 115$  & -0.102    &$\pm  0.019$   $\pm$  0.002    \\
		$(0.30,0.60)$                   &  3064 &$\pm  59$  &  2655 &$\pm  55$  &   4692    &$\pm  99$  &   4752    &$\pm 106$  & -0.006    &$\pm  0.016$   $\pm$  0.002   &  1869 &$\pm  51$  &  2936 &$\pm  68$  &   4412    &$\pm 129$  &   5141    &$\pm 124$  & -0.076    &$\pm  0.019$   $\pm$  0.002    \\
		$(0.60,0.90)$                   &  2804 &$\pm  56$  &  2715 &$\pm  56$  &   4285    &$\pm  95$  &   4495    &$\pm  99$  & -0.024    &$\pm  0.016$   $\pm$  0.002   &  1866 &$\pm  51$  &  2323 &$\pm  60$  &   4244    &$\pm 125$  &   4247    &$\pm 114$  & -0.000    &$\pm  0.020$   $\pm$  0.003    \\
		$(0.90,1.20)$                   &  2420 &$\pm  52$  &  2366 &$\pm  53$  &   3680    &$\pm  88$  &   3725    &$\pm  91$  & -0.006    &$\pm  0.017$   $\pm$  0.002   &  1599 &$\pm  47$  &  1919 &$\pm  54$  &   3523    &$\pm 112$  &   3896    &$\pm 114$  & -0.050    &$\pm  0.022$   $\pm$  0.003    \\
		$(1.20,1.50)$                   &  2072 &$\pm  49$  &  2103 &$\pm  50$  &   3165    &$\pm  83$  &   3256    &$\pm  85$  & -0.014    &$\pm  0.019$   $\pm$  0.003   &  1423 &$\pm  44$  &  1281 &$\pm  44$  &   3093    &$\pm 103$  &   2934    &$\pm 107$  &  0.026    &$\pm  0.025$   $\pm$  0.003    \\
		$(1.50,1.80)$                   &  1683 &$\pm  44$  &  1764 &$\pm  46$  &   2610    &$\pm  76$  &   2722    &$\pm  79$  & -0.021    &$\pm  0.021$   $\pm$  0.003   &  1190 &$\pm  37$  &   970 &$\pm  39$  &   2602    &$\pm  87$  &   2463    &$\pm 106$  &  0.027    &$\pm  0.028$   $\pm$  0.004    \\
		$(1.80,2.10)$                   &  1237 &$\pm  38$  &  1294 &$\pm  40$  &   1979    &$\pm  67$  &   2030    &$\pm  69$  & -0.013    &$\pm  0.025$   $\pm$  0.004   &   809 &$\pm  31$  &   757 &$\pm  36$  &   1788    &$\pm  74$  &   1940    &$\pm  99$  & -0.041    &$\pm  0.033$   $\pm$  0.004    \\
		$(2.10,2.40)$                   &   792 &$\pm  30$  &   813 &$\pm  32$  &   1320    &$\pm  57$  &   1310    &$\pm  60$  &  0.004    &$\pm  0.032$   $\pm$  0.005   &   532 &$\pm  31$  &   503 &$\pm  31$  &   1236    &$\pm  79$  &   1281    &$\pm  86$  & -0.018    &$\pm  0.047$   $\pm$  0.005    \\
		$(2.40,2.70)$                   &   415 &$\pm  22$  &   444 &$\pm  24$  &    728    &$\pm  43$  &    732    &$\pm  47$  & -0.003    &$\pm  0.046$   $\pm$  0.007   &   283 &$\pm  22$  &   283 &$\pm  24$  &    678    &$\pm  59$  &    768    &$\pm  72$  & -0.062    &$\pm  0.066$   $\pm$  0.007    \\
		$(2.70,2.98)$                   &    89 &$\pm  11$  &   106 &$\pm  13$  &    164    &$\pm  24$  &    177    &$\pm  27$  & -0.037    &$\pm  0.112$   $\pm$  0.016   &    58 &$\pm  12$  &    54 &$\pm   9$  &    146    &$\pm  35$  &    161    &$\pm  32$  & -0.049    &$\pm  0.168$   $\pm$  0.018    \\
		\hline
		Decay & \multicolumn{10}{c|}{$\pizenu$}                           & \multicolumn{10}{c}{$\pizmunu$}        \\ \cline{2-21}
		$q^2$ (GeV$^2/c^4$)&\multicolumn{2}{c}{$N_{\rm DT}^{\rm Forward}$}&\multicolumn{2}{c}{$N_{\rm DT}^{\rm Backward}$}&\multicolumn{2}{c}{$N_{\rm produced}^{\rm Forward}$}&\multicolumn{2}{c}{$N_{\rm produced}^{\rm Backward}$}&\multicolumn{2}{c|}{$A_{FB}$} &\multicolumn{2}{c}{$N_{\rm DT}^{\rm Forward}$}&\multicolumn{2}{c}{$N_{\rm DT}^{\rm Backward}$}&\multicolumn{2}{c}{$N_{\rm produced}^{\rm Forward}$}&\multicolumn{2}{c}{$N_{\rm produced}^{\rm Backward}$}&\multicolumn{2}{c}{$A_{FB}$} \\ \hline
		$(m_{\ell^{+}}^{2},0.30)$       &  1572 &$\pm  43$  &  1242 &$\pm  39$  &   3465    &$\pm  98$  &   3429    &$\pm 116$  &  0.005    &$\pm  0.023$   $\pm$  0.001   &   746 &$\pm  33$  &  1492 &$\pm  49$  &   2830    &$\pm 129$  &   3545    &$\pm 120$  & -0.112    &$\pm  0.029$   $\pm$  0.003    \\
		$(0.30,0.60)$                   &  1471 &$\pm  42$  &  1178 &$\pm  39$  &   3280    &$\pm 111$  &   3178    &$\pm 117$  &  0.016    &$\pm  0.026$   $\pm$  0.001   &   847 &$\pm  35$  &  1252 &$\pm  49$  &   2832    &$\pm 139$  &   3250    &$\pm 138$  & -0.069    &$\pm  0.033$   $\pm$  0.003    \\
		$(0.60,0.90)$                   &  1300 &$\pm  40$  &  1116 &$\pm  38$  &   2967    &$\pm 115$  &   2946    &$\pm 117$  &  0.004    &$\pm  0.028$   $\pm$  0.002   &   818 &$\pm  34$  &  1065 &$\pm  45$  &   2755    &$\pm 141$  &   3093    &$\pm 143$  & -0.058    &$\pm  0.035$   $\pm$  0.004    \\
		$(0.90,1.20)$                   &  1156 &$\pm  37$  &  1038 &$\pm  37$  &   2764    &$\pm 112$  &   2735    &$\pm 115$  &  0.005    &$\pm  0.030$   $\pm$  0.002   &   719 &$\pm 194$  &   794 &$\pm  36$  &   2431    &$\pm 822$  &   2643    &$\pm 138$  & -0.042    &$\pm  0.172$   $\pm$  0.004    \\
		$(1.20,1.50)$                   &   887 &$\pm  33$  &   863 &$\pm  34$  &   2148    &$\pm 103$  &   2282    &$\pm 107$  & -0.030    &$\pm  0.034$   $\pm$  0.002   &   580 &$\pm  27$  &   586 &$\pm  29$  &   1994    &$\pm 117$  &   2296    &$\pm 133$  & -0.070    &$\pm  0.051$   $\pm$  0.005    \\
		$(1.50,1.80)$                   &   727 &$\pm  29$  &   622 &$\pm  29$  &   1895    &$\pm  95$  &   1628    &$\pm  94$  &  0.076    &$\pm  0.039$   $\pm$  0.002   &   425 &$\pm  23$  &   380 &$\pm  23$  &   1526    &$\pm 102$  &   1621    &$\pm 118$  & -0.030    &$\pm  0.051$   $\pm$  0.005    \\
		$(1.80,2.10)$                   &   554 &$\pm  26$  &   491 &$\pm  26$  &   1506    &$\pm  87$  &   1354    &$\pm  90$  &  0.053    &$\pm  0.045$   $\pm$  0.002   &   302 &$\pm  19$  &   316 &$\pm  20$  &   1137    &$\pm  89$  &   1409    &$\pm 106$  & -0.107    &$\pm  0.055$   $\pm$  0.006    \\
		$(2.10,2.40)$                   &   371 &$\pm  22$  &   325 &$\pm  24$  &   1051    &$\pm  78$  &    894    &$\pm  83$  &  0.081    &$\pm  0.061$   $\pm$  0.003   &   188 &$\pm  16$  &   200 &$\pm  18$  &    734    &$\pm  76$  &    823    &$\pm  90$  & -0.057    &$\pm  0.077$   $\pm$  0.008    \\
		$(2.40,2.70)$                   &   183 &$\pm  16$  &   202 &$\pm  23$  &    481    &$\pm  59$  &    532    &$\pm  81$  & -0.051    &$\pm  0.102$   $\pm$  0.004   &    89 &$\pm  12$  &    84 &$\pm  16$  &    326    &$\pm  59$  &    288    &$\pm  79$  &  0.061    &$\pm  0.170$   $\pm$  0.011    \\
		$(2.70,3.01)$                   &    79 &$\pm  11$  &    62 &$\pm  20$  &    205    &$\pm  42$  &     58    &$\pm  70$  &  0.561    &$\pm  0.429$   $\pm$  0.012   &    15 &$\pm   7$  &    26 &$\pm  16$  &     18    &$\pm  32$  &     40    &$\pm  76$  & -0.378    &$\pm  1.212$   $\pm$  0.055    \\
		\hline\hline
	\end{tabular}
	}
	}
\end{table*}

\section{Constraint on scalar current in $c\to d\ell^+\nu_{\ell}$ transition}

According to Eq.~\ref{equation:heliamp}, the inclusion of potential scalar currents in the $c\to d\ell^{+}\nu_{\ell}$ transition leads to a modification of $\mathcal{H}_{t}(q^{2})$.  Therefore, a simultaneous fit to the $q^{2}$-binned partial decay rates and forward-backward asymmetries of $\pilnu$ is performed to impose stringent constraints on the parameter space of $C_S^{\ell}.$ This fit is performed in the least $\chi^{2}$ method, with the objective function constructed as
\begin{equation}
	\chi^{2}=\chi_{\Delta\Gamma}^{2}+\sum\chi_{\mathcal{A}_{\rm FB}}^{2}.
\end{equation}
Here, the decay rate component $\chi_{\Delta\Gamma}^{2}$ is identical to that described in Sec.~\ref{sec:ffchifit}. The forward-backward asymmetry component sums over all four decay channels, expressed as $\sum\chi^{2}_{\mathcal{A}_{{\rm FB},\pilnu}}$, with 
\begin{equation}
	\chi^2_{\mathcal{A}_{FB}} = \sum_{i,j}(\mathcal{A}_{FB,i}^{\rm msr}-\mathcal{A}_{FB,i}^{\rm th})(C_{FB}^{-1})_{ij}(\mathcal{A}_{FB,j}^{\rm msr}-\mathcal{A}_{FB,j}^{\rm th}).
\end{equation}

In this fit, along with the parameters $\ffpi|V_{cd}|$, we include both the real and imaginary parts of the complex coefficient $C_S^{\mu}$ as floating parameters for the muon channels. For the positron channels, the partial decay rate is sensitive only to the modulus $|C_S^e|$, and the forward-backward asymmetry does not respond to the scalar current contribution; thus, only the parameter $|C_S^e|$ is considered.

\begin{figure*}[htbp]
	\centering
	\includegraphics[width=\linewidth]{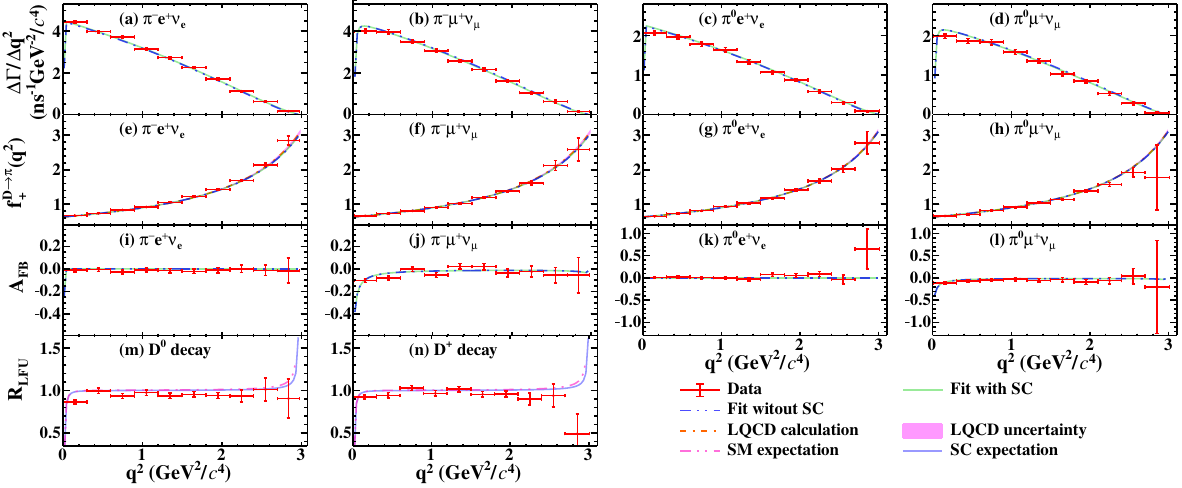}
	\caption{Simultaneous fit to the measured partial decay rates (a,b,c,d) and forward-backward asymmetries (i,j,k,l), the projections to the hadronic transition form factor as function (e,f,g,h) and the determined $R_{\rm LFU}$ (m,n).}
	\label{fig:fitafb}
\end{figure*}

\begin{figure}[htbp]
	\centering
	\includegraphics[width=\columnwidth]{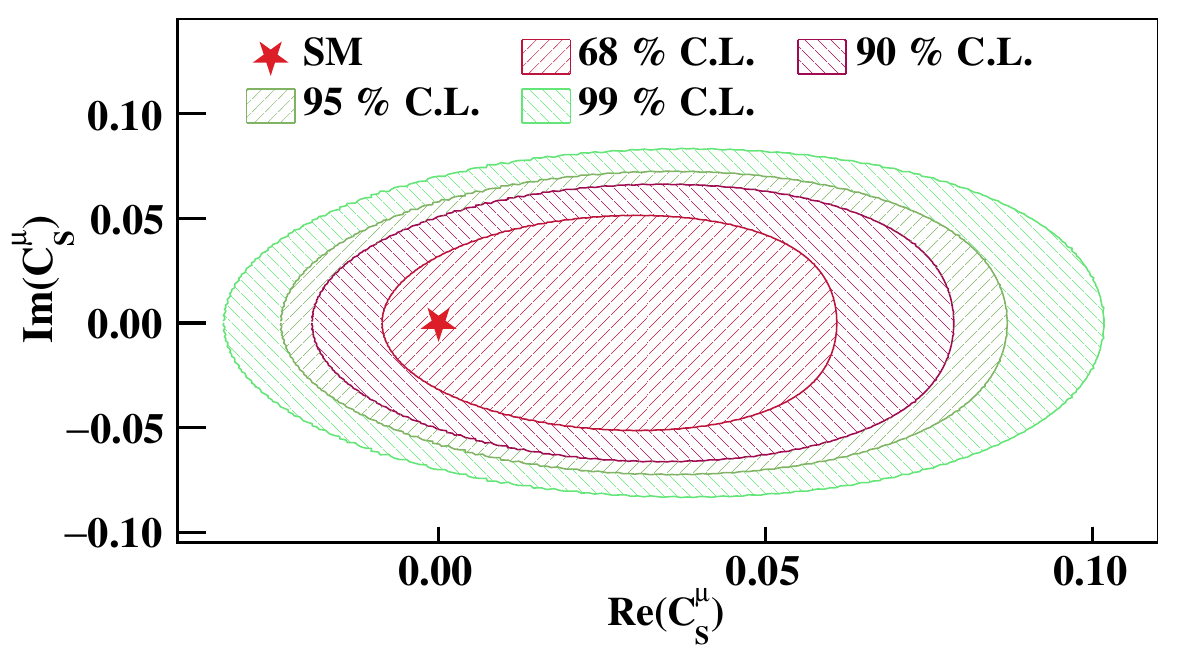}
	\caption{The confidence regions of the scalar combination of complex wilson coefficients $C_S^\mu$ with probabilities 68\%, 90\%, 95\%, and 99\%. The red dot indicates the SM value.}
	\label{fig:willson}
\end{figure}

Two separate fits are conducted under the hypotheses with and without scalar current (SC) contribution. The corresponding fit projections are shown in Fig.~\ref{fig:fitafb}, while the fitted parameters are summarized in Table~\ref{table:afbpar}. 
The scalar current in the $c \to d \ell^{+} \nu_{\ell}$ transition is estimated, and the $C_S^{\mu}$ with different confidence level are shown in Fig.~\ref{fig:willson}.
Additionally, since the LFU may not apply for the scalar current, which could depend on lepton mass, its coupling to the muon could be significantly stronger than to the positron. The significance of a non-zero $C_S^\mu$ is estimated by performing a fit with $C_S^\mu$ floating and $|C_S^e|$ fixed to zero. Compared to the null hypothesis without scalar current, the changes in $\Delta\chi^2$ and $\Delta{\rm ndf}$ correspond to a 1.6 $\sigma$ significance.
Based on the fit that includes scalar current contribution, no evidence beyond the SM is found. 

\begin{table}[htbp]
	\centering
	\caption{The obtained results from the fits with and without scalar current (SC). where the first uncertainties are statistical and the second systematic.}
	\label{table:afbpar}
	\begin{tabular}{lcc}
		\hline\hline
		Variable                  & With SC             & Without SC     \\
		\hline
		$f_{+}^{D\to\pi}|V_{cd}|$     & 0.1430(06)(04)      & 0.1425(05)(03) \\
		$r_{1}$                   & -1.97(03)(01)       & -1.99(03)(01)  \\
		$|C_S^e|$                  & 0.059(13)(09)       & --            \\
		${\rm Re}(C_S^\mu)$       & 0.022(23)(03)       & --            \\
		${\rm Im}(C_S^\mu)$       & 0.000(38)(12)     & --            \\
		$\chi^2/\rm ndf$              & $75.5/75$           & $81.0/78$      \\
		\hline
		\hline
	\end{tabular}
\end{table}

\section{Summary}

In summary, by analyzing 20.3 fb$^{-1}$ of $e^+e^-$ collision data collected at $\sqrt{s}=3.773$~GeV with the BESIII detector, precision measurements of
the semileptonic decays $\pilnu$ are reported. The branching fractions are measured to be
\begin{equation}
	\footnotesize
	\mathcal{B}(\pienu) = \bfpienu,
\end{equation}
\begin{equation}
	\footnotesize
	\mathcal{B}(\pimunu) = \bfpimunu,
\end{equation}
\begin{equation}
	\footnotesize
	\mathcal{B}(\pizenu) = \bfpizenu,
\end{equation}
and 
\begin{equation}
	\footnotesize
	\mathcal{B}(\pizmunu) = \bfpizmunu,
\end{equation}
respectively.
Subsequently, the ratios of the branching fractions between semielectronic and semimuonic $D$ decays are determined to be 
\begin{equation}
	\footnotesize
	\mathcal{R}_{\rm LFU}^{0} = \frac{\mathcal B(\pimunu)}{\mathcal B(\pienu)} = \lfudz
\end{equation}
and
\begin{equation}
	\footnotesize
	\mathcal{R}_{\rm LFU}^{+} = \frac{\mathcal B(\pizmunu)}{\mathcal B(\pizenu)} = \lfudp.
\end{equation}
Here, the systematic uncertainties in ST yields, $\pi^-/\ell^+$ tracking and PID, and $\pi^0$ reconstruction cancel.
They are consistent with the SM prediction, $0.985 \pm 0.002$~\cite{Riggio:2017zwh} within 2.5$\sigma$ and 0.7$\sigma$, respectively.
Compared to the previous best measurements~\cite{BESIII:2015tql,BESIII:2017ylw,BESIII:2018nzb}, the precision is improved upon by factors of 2.1-3.0 and 2.6 (2.8) for $\mathcal{R}_{\rm LFU}^{0(+)}$ results.
Combining with the $D^0$ and $D^+$ meson lifetimes $\tau_{D^0} = (410.3~\pm~1.0)$ fs and $\tau_{D^+} = (1033~\pm~5)$ fs~\cite{pdg2024},
we determine the ratios of branching fractions for $D^0\to \pi\ell^+\nu_\ell$ and $D^+\to \pi\ell^+\nu_\ell$ to be
\begin{equation}
	\footnotesize
	\mathcal{R}_{\rm IS}^{e} = \frac{\Gamma(\pienu)}{2\Gamma(\pizenu)} = \pwre	
\end{equation}
and
\begin{equation}
	\footnotesize
	\mathcal{R}_{\rm IS}^{\mu} = \frac{\Gamma(\pimunu)}{2\Gamma(\pizmunu)} = \pwrmu,
\end{equation}
where systematic uncertainties of ST yields, $\pi^-/\ell^+$ tracking and PID, and $\pi^0$ reconstruction are canceled, but those of $\tau_{D}$ are included. These ratios are consistent with unity based on isospin conservation assumption.

\begin{figure}[htbp]
	\centering
	\setlength{\abovecaptionskip}{-2pt}
	\setlength{\belowcaptionskip}{-3pt}
	\includegraphics[width=\columnwidth]{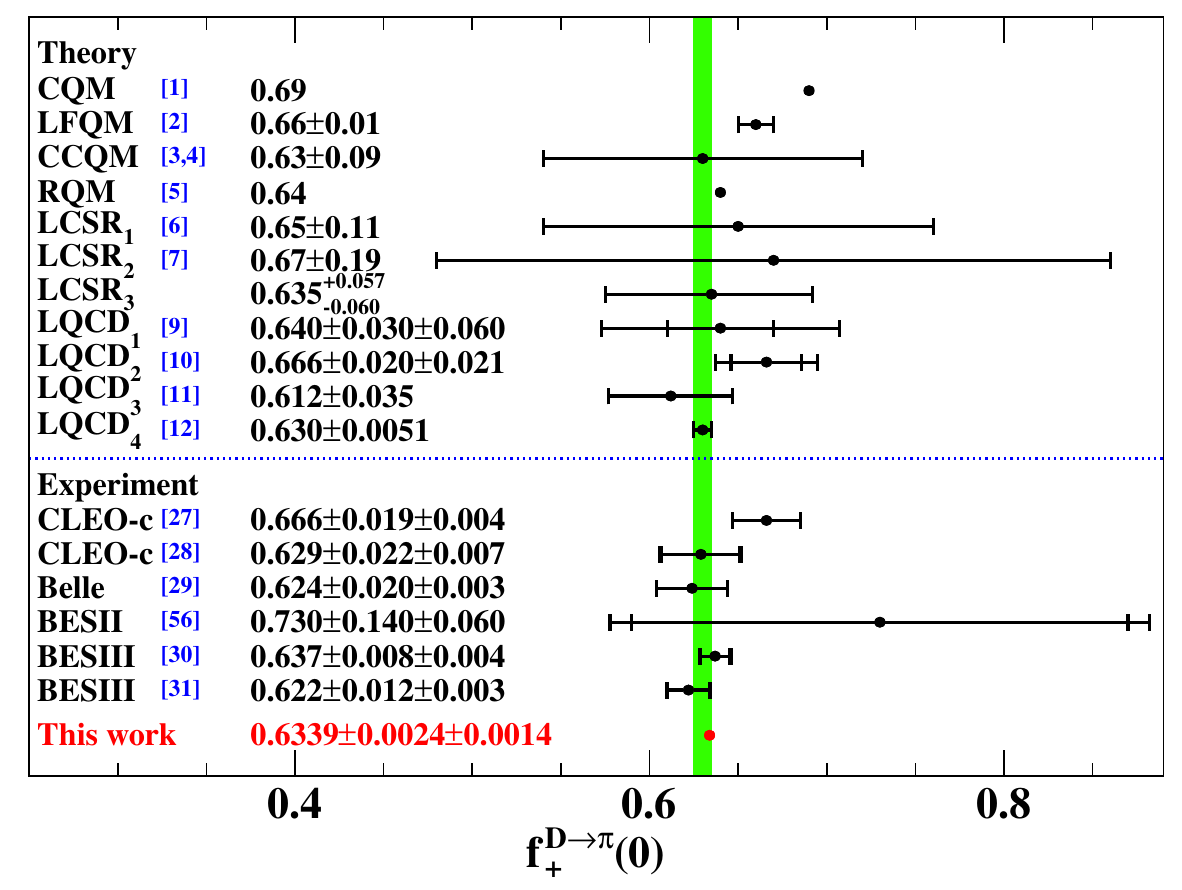}
	\caption{Comparison of the hadronic transition form factors $f^{D\to\pi}_+(0)$ measured by this work with other experiments and theoretical calculations.
		The first and second uncertainties are statistical and systematic, respectively.  The green band corresponds to the $\pm1\sigma$ limit of the LQCD$_4$~\cite{FermilabLattice:2022gku} calculation.
	}
	\label{compare_ff_pilnu}
\end{figure}

From the simultaneous fit to the partial decay rates of $\pienu$, $\pimunu$, $\pizenu$, and $\pizmunu$, the
product of the hadronic transition form factor $\ffpi$ and the modulus of the CKM matrix element $|V_{cd}|$ is determined to be
\begin{equation}
	\ffpi|V_{cd}| = \fvcd.
\end{equation}
Taking the $|V_{cd}|$ given by the PDG~\cite{pdg2024} as input, we obtain
\begin{equation}
	\ffpi = \fpi.
\end{equation}
The comparison with previous measurements and theoretical calculations is shown in Fig.~\ref{compare_ff_pilnu}.
Conversely, using the $\ffpi$ calculated in the LQCD~\cite{FermilabLattice:2022gku}, we obtain the CKM matrix element
\begin{equation}
	\small
	|V_{cd}| = \vcdpi,
\end{equation}
which facilitates crucial tests of the unitarity of the CKM matrix through direct measurements.
The hadronic transition form factor $\ffpi$ measured in this work is consistent with the previous measurements and
the theoretical calculations, but with precision improved by a factor of 3. 
This is important to test different theoretical calculations and to improve the precision of theoretical calculation.

Furthermore, the $q^{2}$-binned angular observable, forward-backward asymmetries $\mathcal{A}_{FB}$, in the semileptonic decays $\pilnu$ are measured for the first time. By conducting a simultaneous fit to the $q^{2}$-binned partial decay rates and forward-backward asymmetries, we search for scalar current contribution in the $c \to d \ell^{+} \nu_{\ell}$ transition. The first experimental constraints on the scalar combination of complex Wilson coefficients are determined to be Re$(C_S^\mu)=$ $0.022 \pm 0.023_{\rm stat.}\pm 0.003_{\rm syst.}$ and Im$(C_S^\mu)=\pm(0.000 \pm $ $0.038_{\rm stat.} \pm 0.012_{\rm syst.})$, which are important to restrict scalar current contribution in (semi)leptonic $D$ decays.

%% Saved at => 2026-04-08
\textbf{Acknowledgement}

The BESIII Collaboration thanks the staff of BEPCII (https://cstr.cn/31109.02.BEPC) and the IHEP computing center for their strong support. This work is supported in part by National Key R\&D Program of China under Contracts Nos. 2023YFA1606000, 2023YFA1606704, 2025YFA1613900; National Natural Science Foundation of China (NSFC) under Contracts Nos. 11635010, 11935015, 11935016, 11935018, 12025502, 12035009, 12035013, 12061131003, 12192260, 12192261, 12192262, 12192263, 12192264, 12192265, 12221005, 12225509, 12235017, 12342502, 12361141819, 12535005; the Chinese Academy of Sciences (CAS) Large-Scale Scientific Facility Program; the Strategic Priority Research Program of Chinese Academy of Sciences under Contract No. XDA0480600; CAS under Contract No. YSBR-101; 100 Talents Program of CAS; The Institute of Nuclear and Particle Physics (INPAC) and Shanghai Key Laboratory for Particle Physics and Cosmology; Agencia Nacional de Investigación y Desarrollo de Chile (ANID), Chile under Contract No. ANID CCTVal CIA250027; ERC under Contract No. 758462; German Research Foundation DFG under Contract No. FOR5327; Istituto Nazionale di Fisica Nucleare, Italy; Knut and Alice Wallenberg Foundation under Contracts Nos. 2021.0174, 2021.0299, 2023.0315; Ministry of Development of Turkey under Contract No. DPT2006K-120470; National Research Foundation of Korea under Contract No. RS-2026-25486791; National Science and Technology fund of Mongolia; Polish National Science Centre under Contract No. 2024/53/B/ST2/00975; STFC (United Kingdom); Swedish Research Council under Contract No. 2019.04595; U. S. Department of Energy under Contract No. DE-FG02-05ER41374

%% ends here %%

\clearpage
\onecolumngrid
\appendix
\section*{Appendix}
\label{sec:appendix}
\renewcommand{\thefigure}{A.\arabic{figure}} % 重新定义图的编号格式
\setcounter{figure}{0} % 重置图的计数器
\renewcommand{\thetable}{A.\arabic{table}} % 重新定义图的编号格式
\setcounter{table}{0} % 重置图的计数器

\begin{figure}[htbp]
	\centering
		\includegraphics[width=0.55\columnwidth]{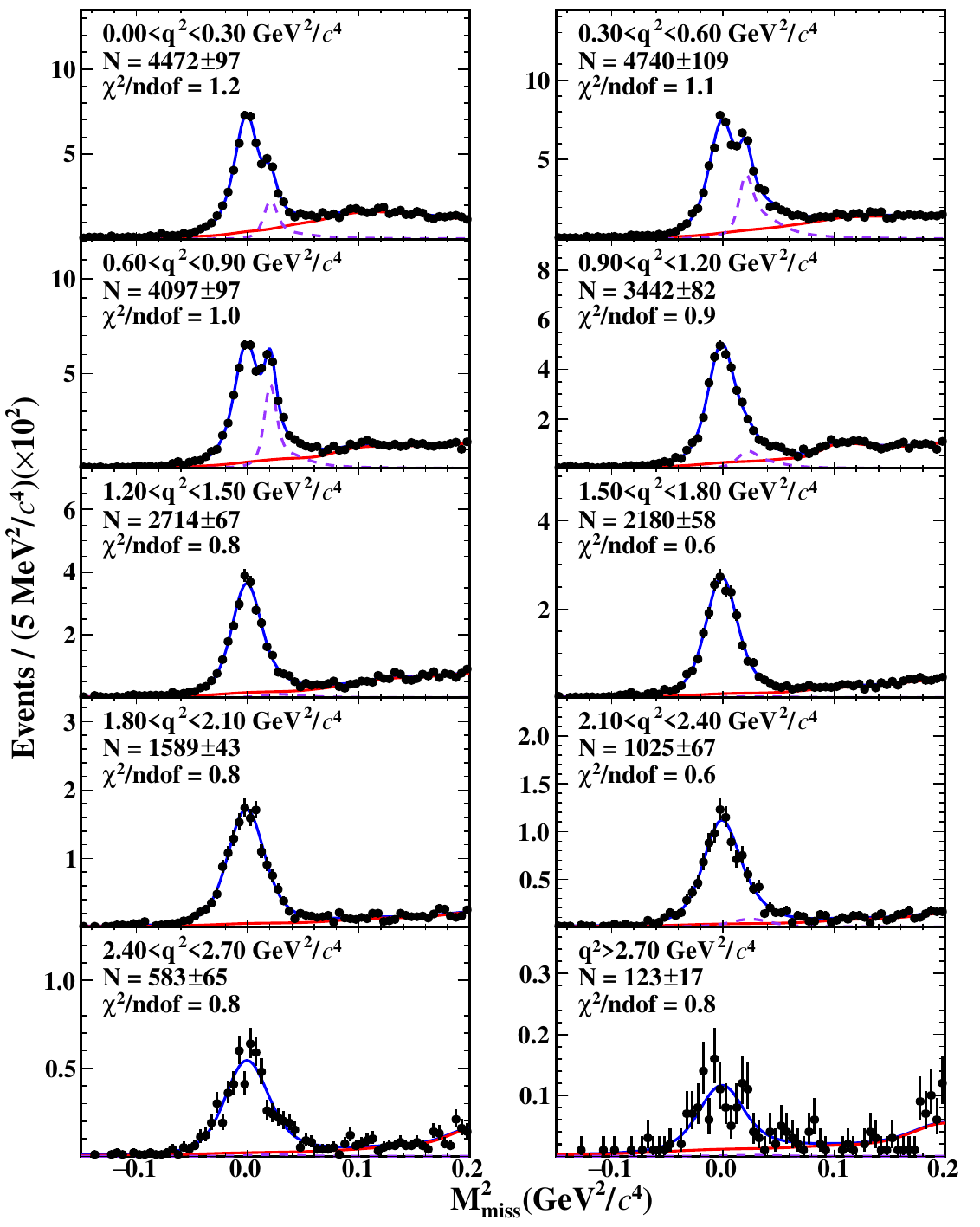}
		\caption{Fits to the $M_{\rm miss}^{2}$ distributions of the candidates for the semileptonic decay $\pimunu$ with $q^2$ in various bins.
			The points with error bars are data, the solid blue curves are the fit results,
			the dashed violet lines are the fitted peaking background shapes, and the solid red curves are the fitted combinatorial background shapes.
		}
		\phantomsection
		\label{fig:pimunumm2mq2}
\end{figure}

\begin{figure}[htbp]
	\centering
	\includegraphics[width=0.55\columnwidth]{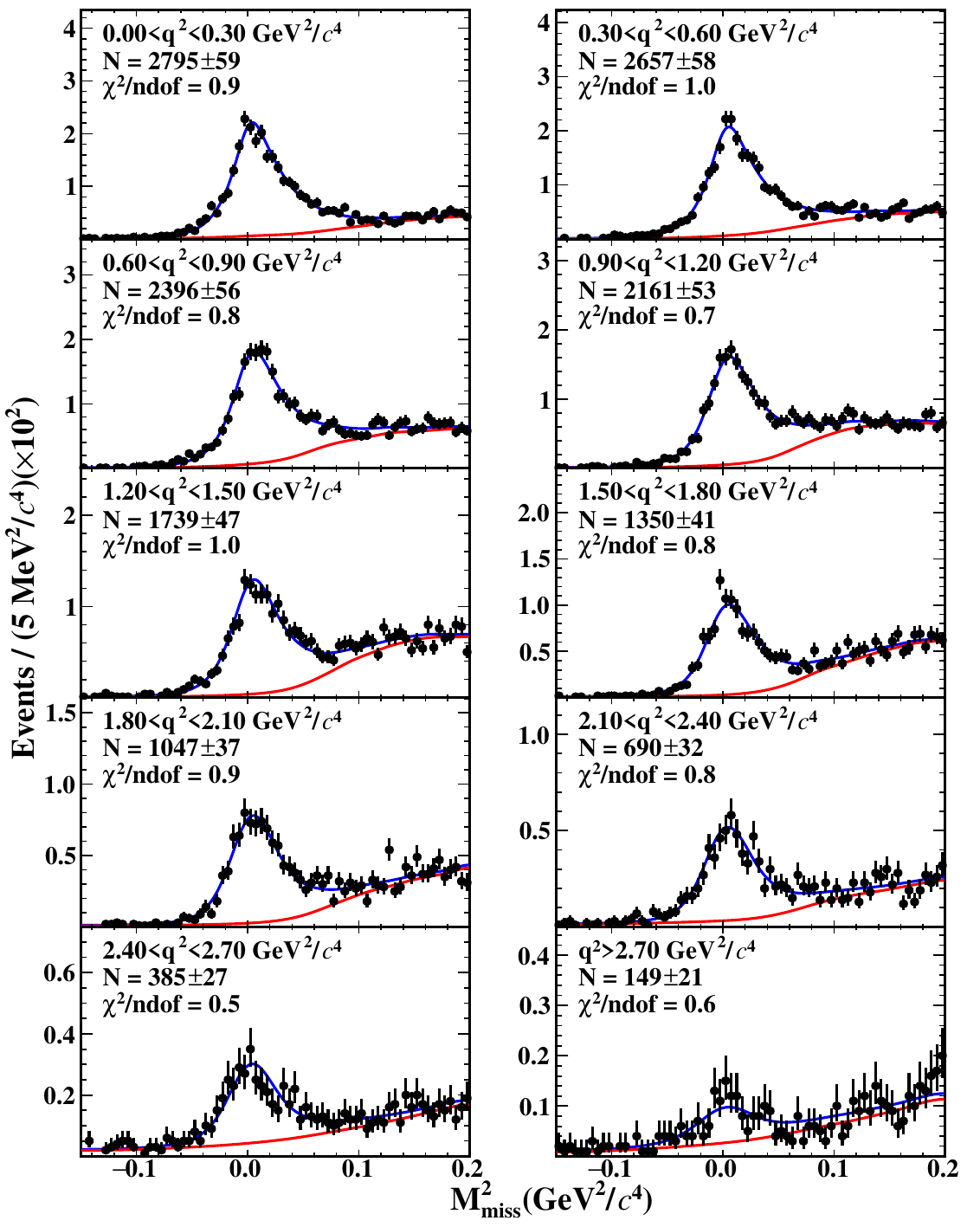}
	\caption{Fits to the $M_{\rm miss}^{2}$ distributions of the candidates for the semileptonic decay $\pizenu$ with $q^2$ in various bins.
		The points with error bars are data, the solid blue curves are the fit results,
		and the solid red curves are the fitted combinatorial background shapes.
	}
	\phantomsection
	\label{fig:pizenumm2mq2}
\end{figure}

\begin{figure}[htbp]
	\centering
	\includegraphics[width=0.55\columnwidth]{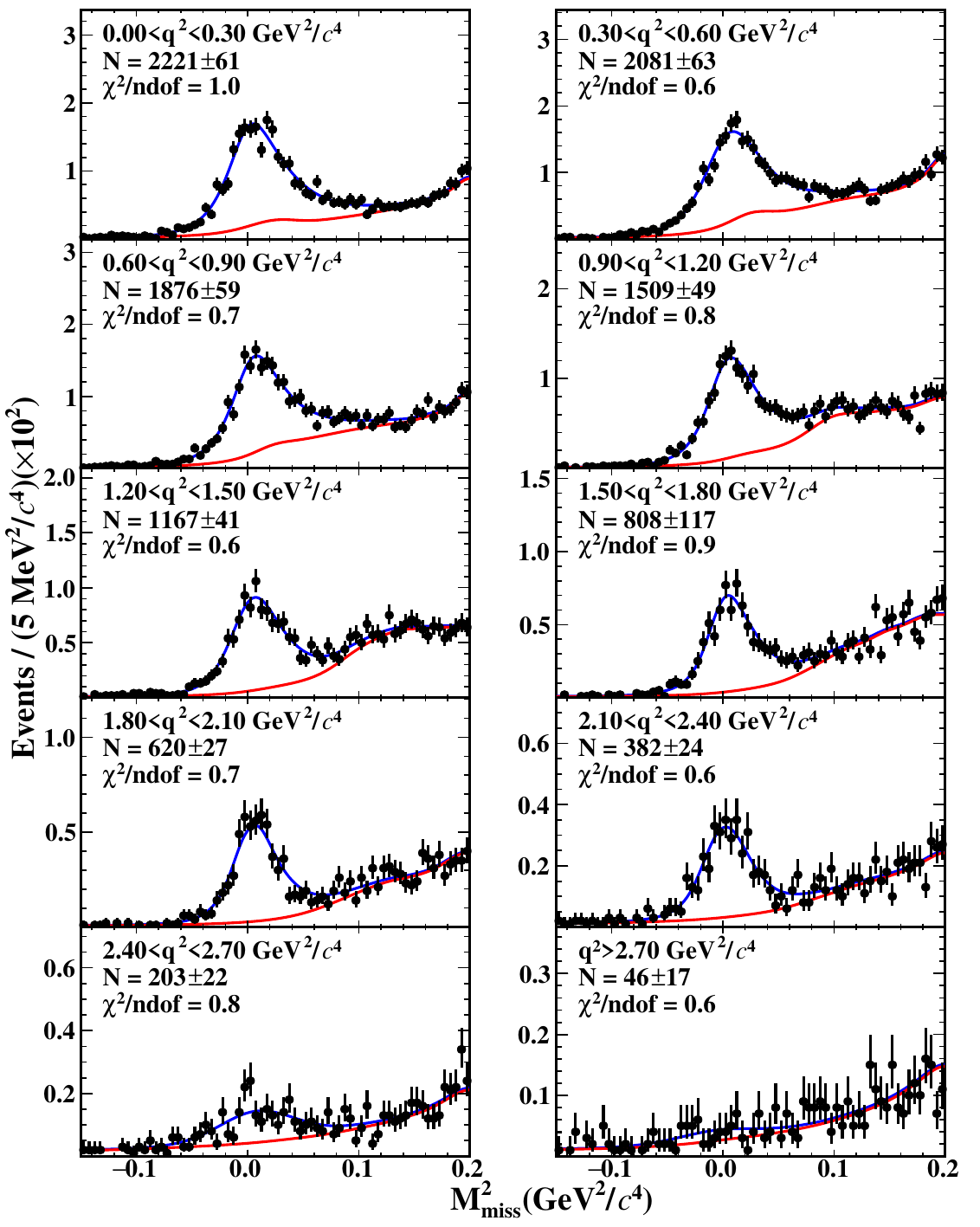}
	\caption{Fits to the $M_{\rm miss}^{2}$ distributions of the candidates for the semileptonic decay $\pizmunu$ with $q^2$ in various bins.
		The points with error bars are data, the solid blue curves are the fit results,
		and the solid red curves are the fitted combinatorial background shapes.
	}
	\phantomsection
	\label{fig:pizmunumm2mq2}
\end{figure}

Figures~\ref{fig:pimunumm2mq2},~\ref{fig:pizenumm2mq2}, and~\ref{fig:pizmunumm2mq2} show the results of the fits to the $M_{\rm miss}^2$ distributions in the reconstructed  $q^{2}$  intervals for $\pimunu$, $\pizenu$ and $\pizmunu$, respectively.

Table~\ref{tab:pdweffmatrix} reports the elements of the weighted efficiency matrices for $\pienu$, $\pimunu$, $\pizenu$ and $\pizmunu$, respectively.

Table~\ref{tab:pdwstatmatrix} gives the elements of the statistical covariance density matrices for $\pienu$, $\pimunu$, $\pizenu$ and $\pizmunu$ channels.

Table~\ref{tab:sysff} summarizes the systematic uncertainties  $\sigma_{\rm syst}$ of $\pienu$, $\pimunu$, $\pizenu$ and $\pizmunu$ in different $q^{2}$ intervals.

Table~\ref{tab:pdwsysmatrix} presents the elements of the systematic covariance density matrices for $\pienu$, $\pimunu$, $\pizenu$ and $\pizmunu$ channels.

Tables~\ref{table:statcovellnu} and~\ref{table:systcovellnu} provide the elements of the covariance density matrix  $\rho_{i j}(i=0,1,2, \ldots, 40)$ for the simultaneous fit.

\begin{figure}[htbp]
	\centering
	\includegraphics[width=0.45\linewidth]{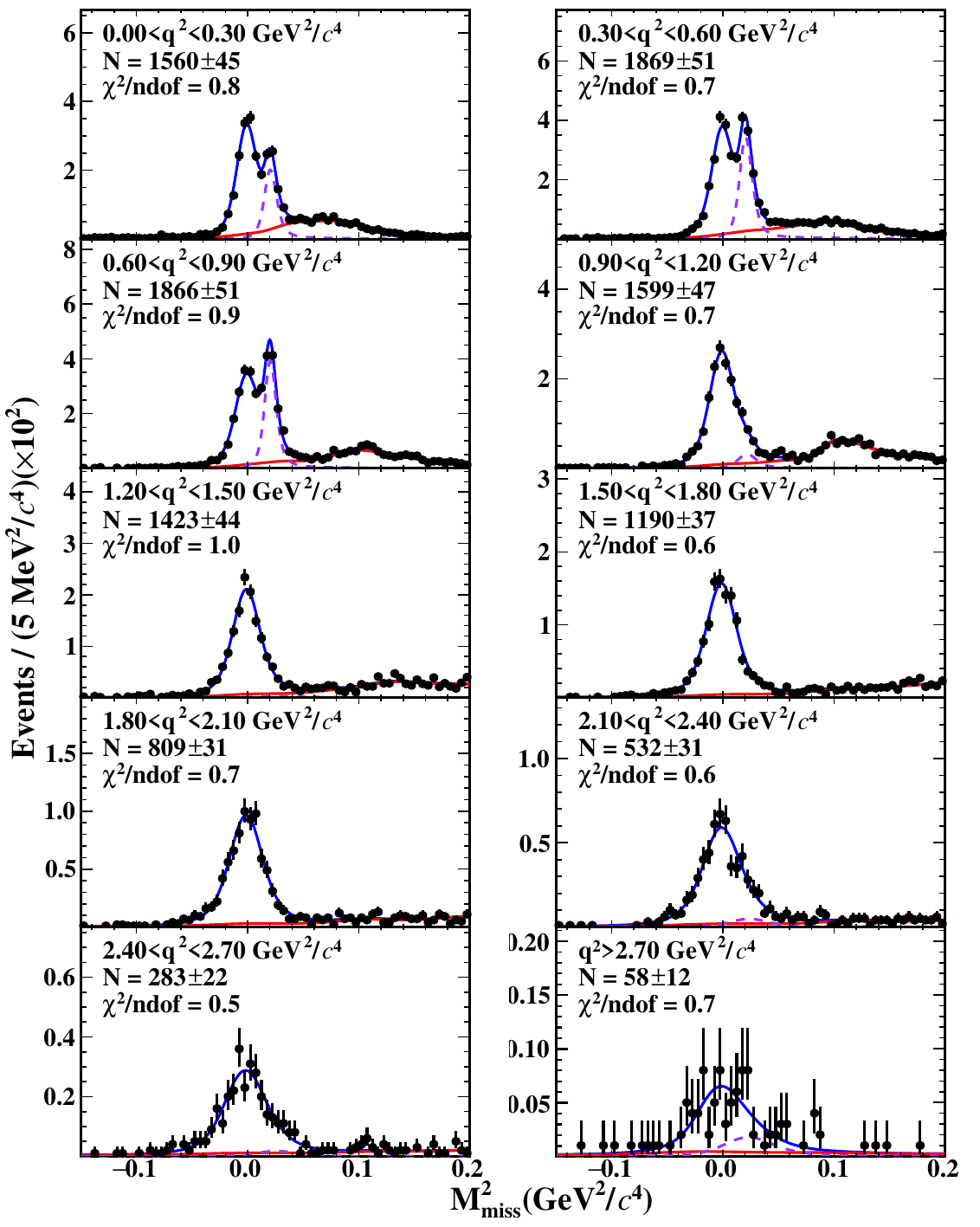}
	\includegraphics[width=0.45\linewidth]{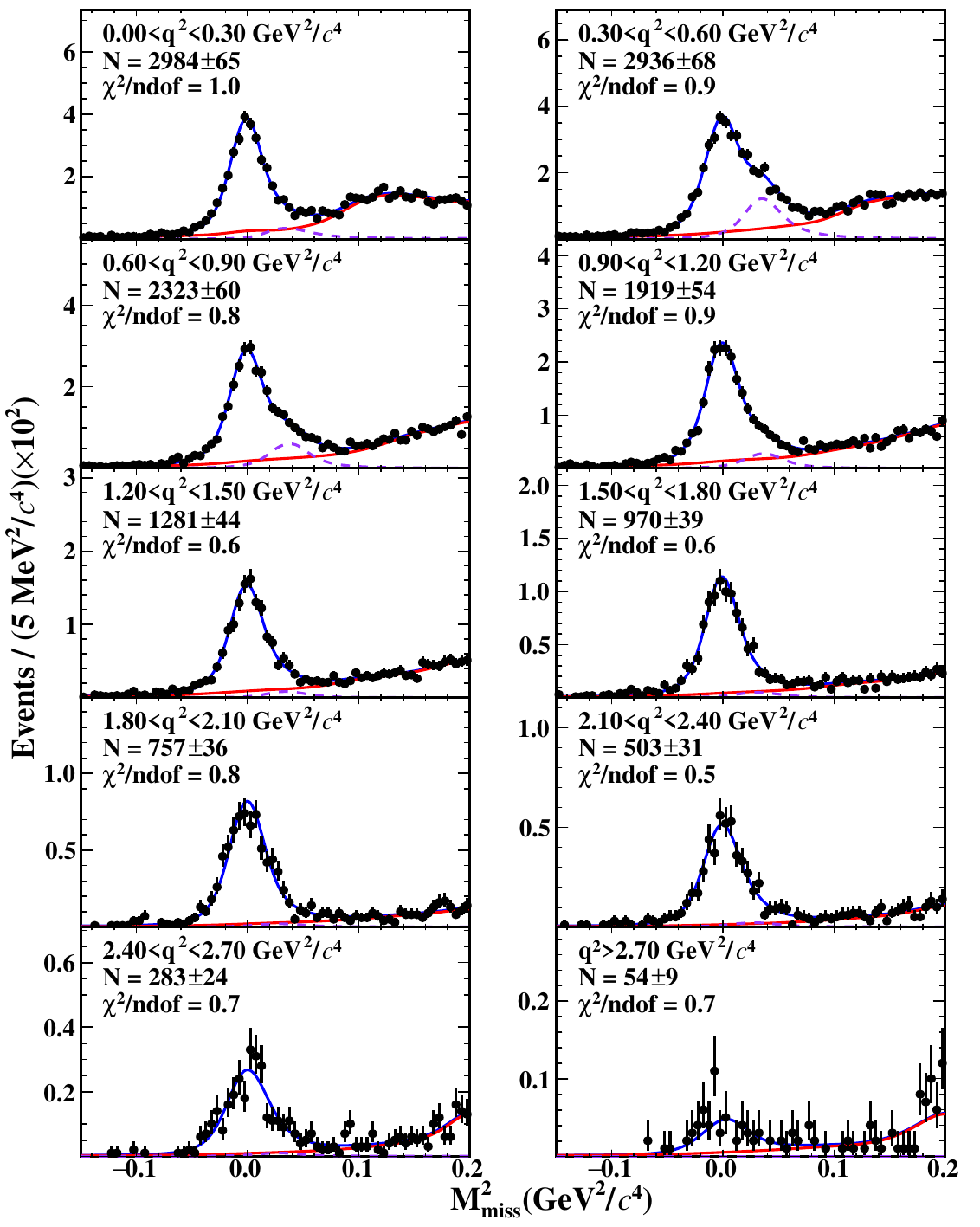}
	\caption{The $M_{\rm miss}^2$ distributions of the accepted forward (left) and backward (right) candidate events in different $q^{2}$ bins for $\pimunu$. The points with error bars are data, the solid blue curves are the fit results, the dashed violet lines are the fitted peaking background shapes, and the solid red curves are the fitted combinatorial background shapes.}
	\phantomsection
	\label{fig:forbackpimunu}
\end{figure}

\begin{figure}[htbp]
	\centering
	\includegraphics[width=0.45\linewidth]{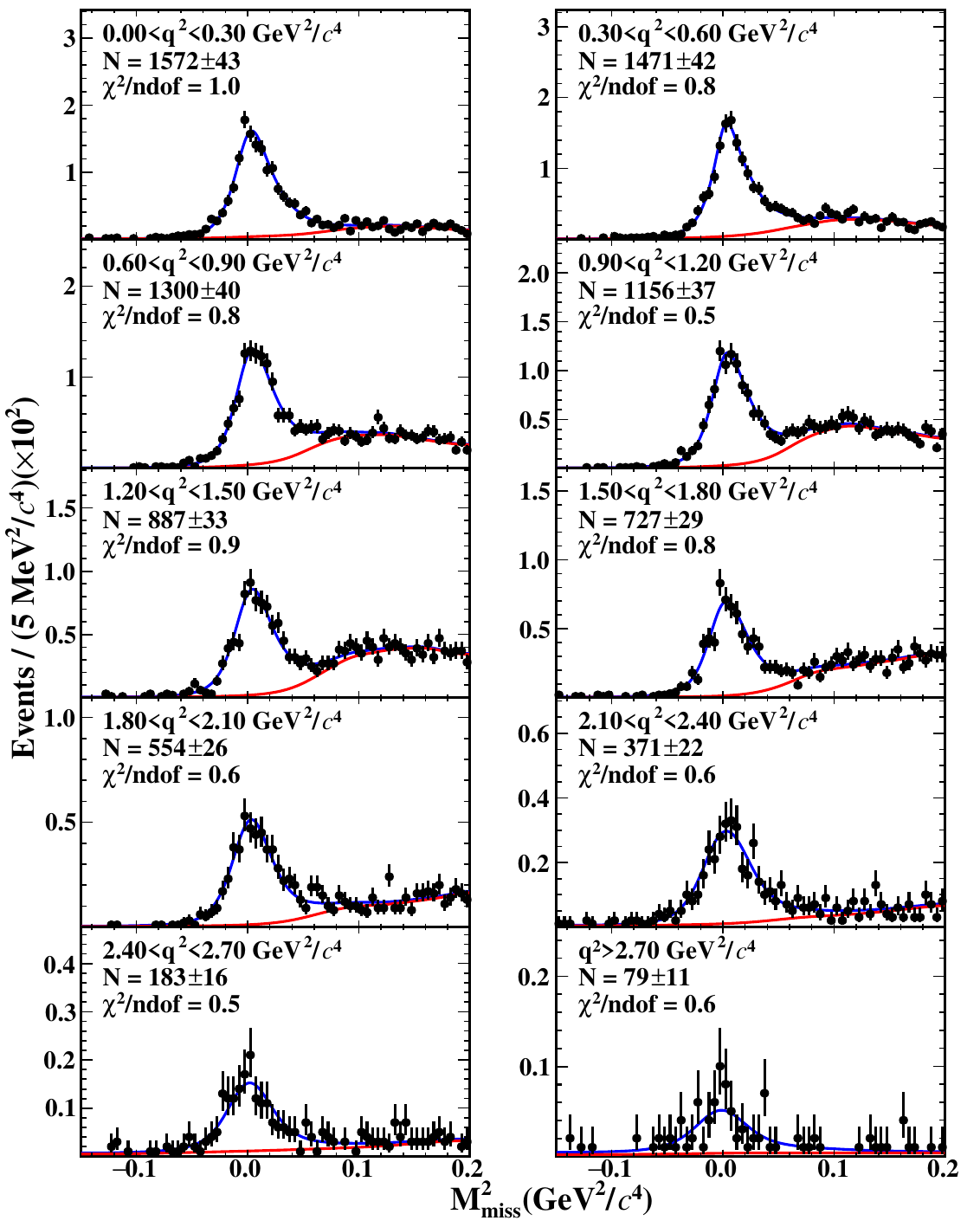}
	\includegraphics[width=0.45\linewidth]{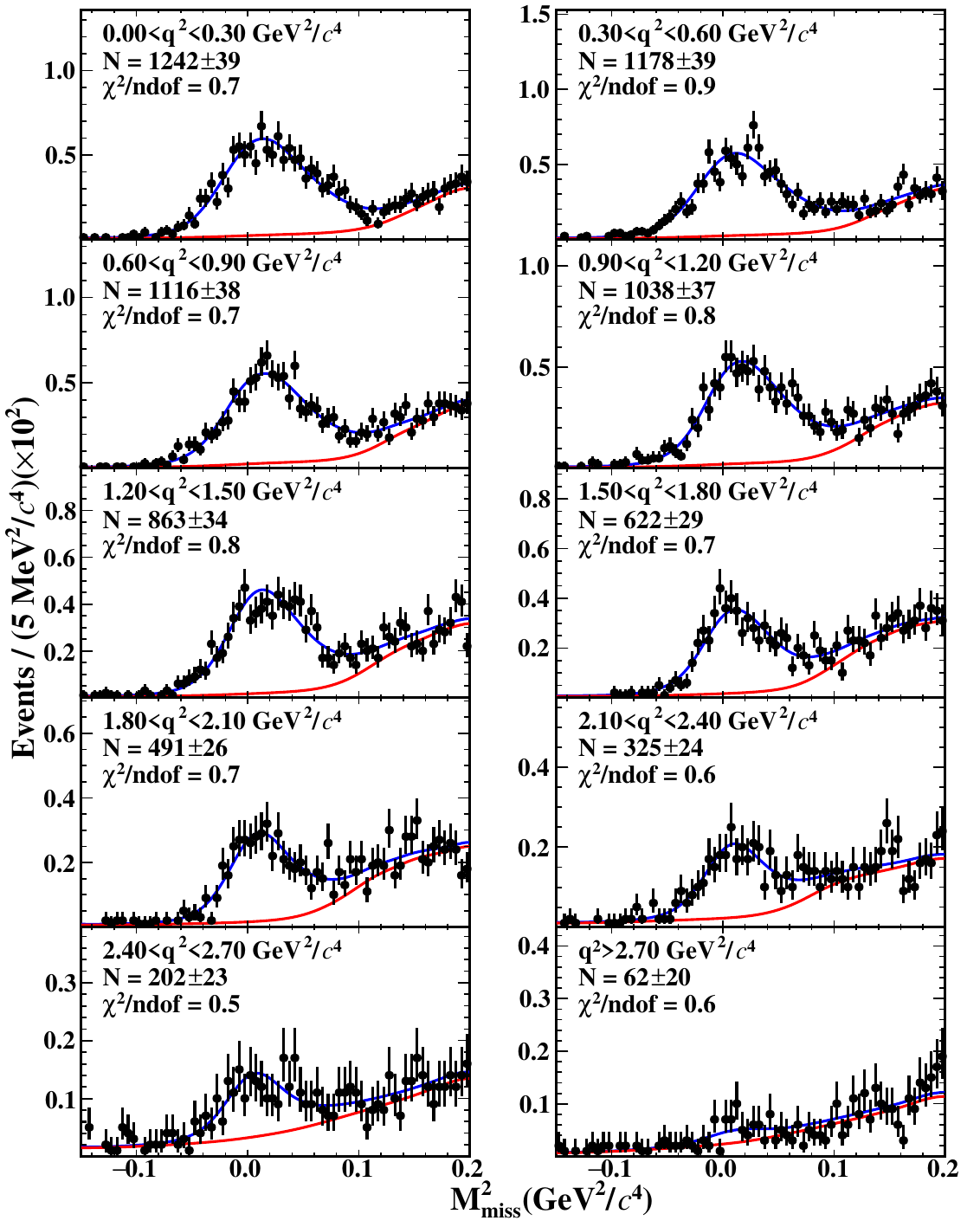}
	\caption{The $M_{\rm miss}^2$ distributions of the accepted forward (left) and backward (right) candidate events in different $q^{2}$ bins for $\pizenu$. The points with error bars are data, the solid blue curves are the fit results and the solid red curves are the fitted combinatorial background shapes.}
	\phantomsection
	\label{fig:forbackpi0enu}
\end{figure}

\begin{figure}[htbp]
	\centering
	\includegraphics[width=0.45\linewidth]{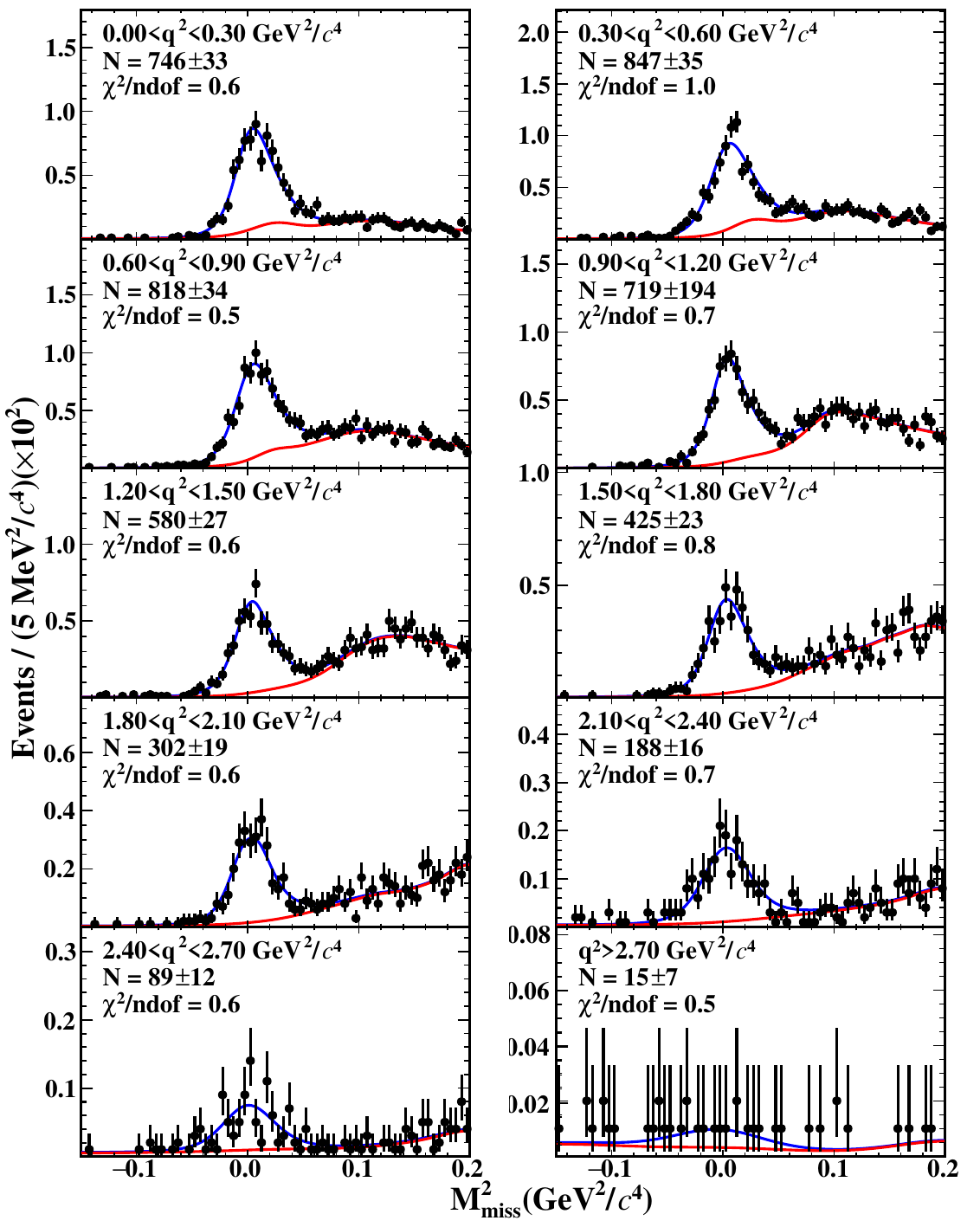}
	\includegraphics[width=0.45\linewidth]{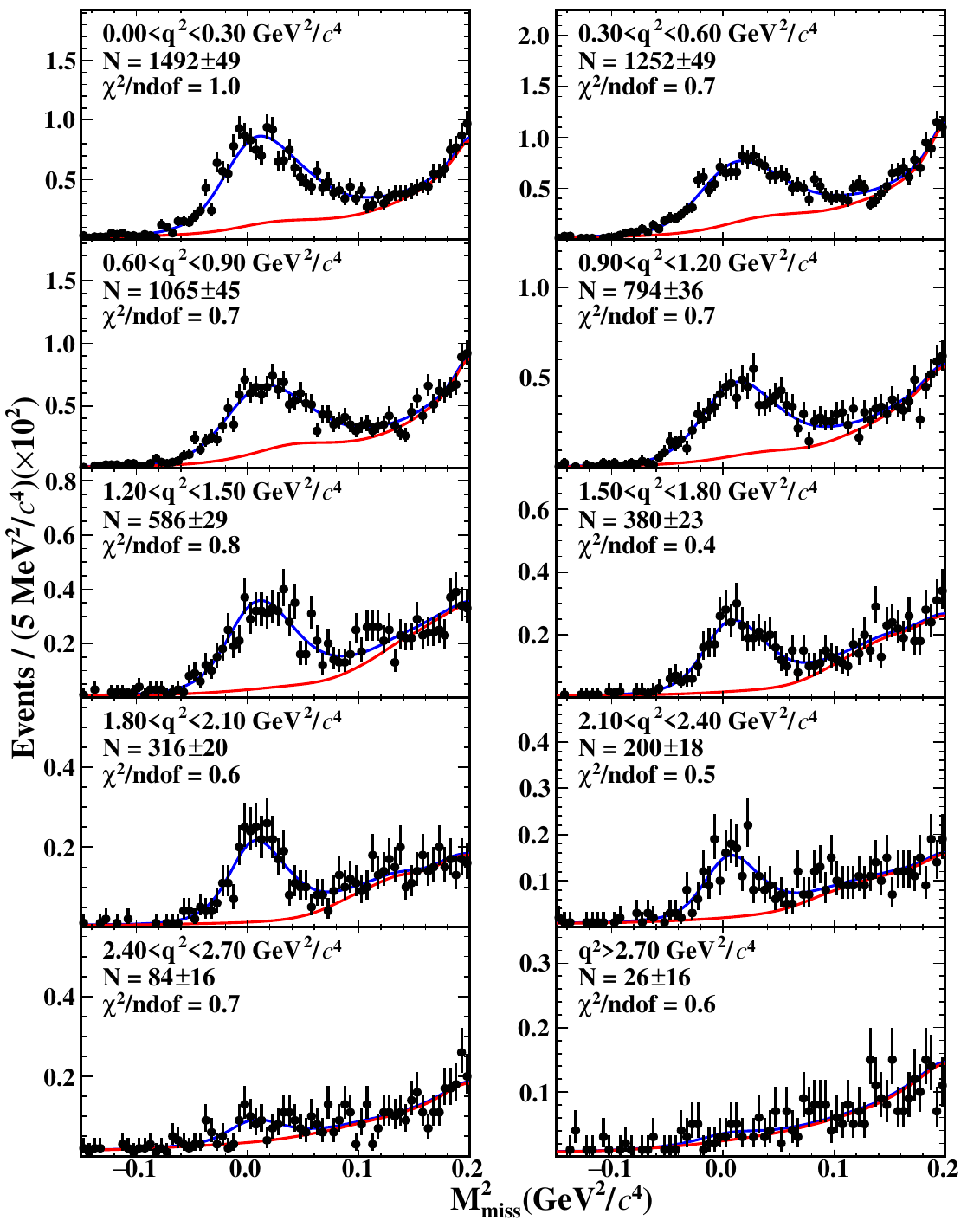}
	\caption{The $M_{\rm miss}^2$ distributions of the accepted forward (left) and backward (right) candidate events in different $q^{2}$ bins for $\pizmunu$. The points with error bars are data, the solid blue curves are the fit results and the solid red curves are the fitted combinatorial background shapes.}
	\phantomsection
	\label{fig:forbackpi0munu}
\end{figure}

Figures~\ref{fig:forbackpimunu},~\ref{fig:forbackpi0enu}, and~\ref{fig:forbackpi0munu} provide the results of the fits to the $M_{\rm miss}^2$ distributions of forward/backward events in the reconstructed $q^{2}$ intervals of $\pimunu$, $\pizenu$ and $\pizmunu$, respectively.

Tables~\ref{table:afbeffpienu},~\ref{table:afbeffpimunu},~\ref{table:afbeffpi0enu}, and~\ref{table:afbeffpi0munu} summarize the weighted efficiency matrices after considering correction of the forward-backward asymmetry in different $q^2$ intervals for $\pienu$, $\pimunu$, $\pizenu$ and $\pizmunu$, respectively.

Table~\ref{tab:afbstatmatrix} reports the statistical covariant matrices of $\mathcal{A}_{FB}(q^2_i)$ for $\pienu$, $\pimunu$, $\pizenu$ and $\pizmunu$.

Table~\ref{tab:afbsysmatrix} gives the systematic covariant matrices of $\mathcal{A}_{FB}(q^2_i)$ for $\pienu$, $\pimunu$, $\pizenu$ and $\pizmunu$.

\begin{table}[htbp]
	\caption{The weighted efficiency matrix for $\pienu$, $\pimunu$, $\pizenu$, and $\pizmunu$, respectively, where $\varepsilon_{ij}$ represents the efficiency (in unit of \%)  for events produced in the $j$-th $q^2$ interval and reconstructed in the $i$-th $q^2$ interval}
	\phantomsection
	\label{tab:pdweffmatrix}
	\resizebox{\linewidth}{!}{
\begin{tabular}{ccccccccccc|ccccccccccc}
		\hline\hline
		& \multicolumn{10}{c|}{$\pienu$}          &                        & \multicolumn{10}{c}{$\pimunu$}         \\ \cline{2-11} \cline{13-22} 
		$\varepsilon_{ij}$ & 1 & 2 & 3 & 4 & 5 & 6 & 7 & 8 & 9 & 10 & $\varepsilon_{ij}$ & 1 & 2 & 3 & 4 & 5 & 6 & 7 & 8 & 9 & 10 \\ \hline
		1  &56.00  &1.75   &0.02   &0.00   &0.00   &0.00   &0.00   &0.00   &0.00   &0.00   &  1    &48.15  &0.91   &0.01   &0.00   &0.00   &0.00   &0.00   &0.00   &0.00   &0.00   \\
		2  &1.54   &57.06  &1.95   &0.01   &0.00   &0.00   &0.00   &0.00   &0.00   &0.00   &  2    &1.13   &47.91  &1.12   &0.01   &0.00   &0.00   &0.00   &0.00   &0.00   &0.00   \\
		3  &0.09   &2.07   &58.92  &1.94   &0.02   &0.00   &0.00   &0.00   &0.00   &0.00   &  3    &0.06   &1.30   &46.98  &1.12   &0.01   &0.00   &0.00   &0.00   &0.00   &0.00   \\
		4  &0.03   &0.19   &2.34   &60.00  &1.80   &0.02   &0.00   &0.00   &0.00   &0.00   &  4    &0.03   &0.07   &1.38   &44.77  &1.05   &0.01   &0.00   &0.00   &0.00   &0.00   \\
		5  &0.02   &0.07   &0.25   &2.43   &60.34  &1.70   &0.02   &0.00   &0.00   &0.00   &  5    &0.01   &0.03   &0.09   &1.37   &42.16  &0.98   &0.01   &0.00   &0.00   &0.00   \\
		6  &0.01   &0.03   &0.10   &0.25   &2.38   &60.02  &1.57   &0.01   &0.00   &0.00   &  6    &0.00   &0.01   &0.03   &0.08   &1.26   &40.14  &0.94   &0.00   &0.00   &0.00   \\
\hline\hline
	\end{tabular}
	}
\end{table}

\begin{table}[htbp]
	\caption{The statistical covariance density matrix for the semileptonic decay $\pienu$, $\pimunu$, $\pizenu$, and $\pizmunu$, respectively. The $i$ and $j$ denote the produced and reconstructed bins, respectively.}
	\phantomsection
	\label{tab:pdwstatmatrix}
	\resizebox{\linewidth}{!}{
	\begin{tabular}{ccccccccccc|ccccccccccc}
		\hline\hline
		& \multicolumn{10}{c|}{$\pienu$}          &                        & \multicolumn{10}{c}{$\pimunu$}         \\ \cline{2-11} \cline{13-22} 
		$\rho_{ij}^{\rm stat}$ & 1 & 2 & 3 & 4 & 5 & 6 & 7 & 8 & 9 & 10 & $\rho_{ij}^{\rm stat}$ & 1 & 2 & 3 & 4 & 5 & 6 & 7 & 8 & 9 & 10 \\ \hline
		1   &1.000  &-0.105 &0.011  &-0.002 &0.000  &-0.000 &-0.000 &-0.000 &-0.000 &-0.000  &  1   &1.000  &-0.058 &0.001  &-0.001 &-0.000 &-0.000 &0.000  &-0.000 &-0.000 &-0.000 \\
		2   &       &1.000  &-0.146 &0.016  &-0.003 &-0.000 &-0.000 &-0.000 &-0.000 &-0.001  &  2   &       &1.000  &-0.069 &0.000  &-0.001 &-0.000 &-0.000 &-0.000 &-0.000 &-0.000 \\
		3   &       &       &1.000  &-0.164 &0.019  &-0.004 &-0.000 &-0.001 &-0.001 &-0.001  &  3   &       &       &1.000  &-0.072 &-0.001 &-0.002 &-0.001 &-0.000 &-0.000 &-0.000 \\
		4   &       &       &       &1.000  &-0.170 &0.019  &-0.004 &-0.001 &-0.001 &-0.001  &  4   &       &       &       &1.000  &-0.071 &-0.001 &-0.002 &-0.001 &-0.000 &-0.000 \\
		5   &       &       &       &       &1.000  &-0.166 &0.014  &-0.004 &-0.002 &-0.002  &  5   &       &       &       &       &1.000  &-0.069 &-0.001 &-0.001 &-0.001 &-0.000 \\
		6   &       &       &       &       &       &1.000  &-0.150 &0.009  &-0.005 &-0.003  &  6   &       &       &       &       &       &1.000  &-0.067 &-0.001 &-0.001 &-0.000 \\
\hline\hline
	\end{tabular}
	}
\end{table}

\begin{table}[htbp]
	\caption{The systematic uncertainties (in \%) of the measured decay rates of $\pienu$, $\pimunu$, $\pizenu$, and $\pizmunu$ in different $q^{2}$ bins, respectively.}
	\phantomsection
	\label{tab:sysff}
	\resizebox{\linewidth}{!}{
	\begin{tabular}{cccccccccccccccccccccc}
		\hline\hline
		& \multicolumn{10}{c}{$\pienu$}          &                        & \multicolumn{10}{c}{$\pizenu$}         \\ \cline{2-11} \cline{13-22} 
		$i$-th $q^2$ & 1 & 2 & 3 & 4 & 5 & 6 & 7 & 8 & 9 & 10 & $i$-th $q^2$ & 1 & 2 & 3 & 4 & 5 & 6 & 7 & 8 & 9 & 10 \\ \hline
		$N_{\rm ST}^{\rm tot}$                                   &0.30&0.30&0.30&0.30&0.30&0.30&0.30&0.30&0.30&0.30 &  $N_{\rm ST}^{\rm tot}$                                   &0.30&0.30&0.30&0.30&0.30&0.30&0.30&0.30&0.30&0.30\\
        $D^0$ lifetime                                           &0.24&0.24&0.24&0.24&0.24&0.24&0.24&0.24&0.24&0.24 &  $D^+$ lifetime                                           &0.48&0.48&0.48&0.48&0.48&0.48&0.48&0.48&0.48&0.48\\
        $\pi^{\pm}$ tracking                                     &0.06&0.06&0.06&0.07&0.07&0.08&0.10&0.12&0.17&0.36 &  $\pi^{0}$ reconstruction                                 &0.24&0.18&0.19&0.21&0.39&0.55&0.31&0.24&0.20&0.30\\
        $\pi^{\pm}$ PID                                          &0.20&0.20&0.20&0.20&0.20&0.20&0.20&0.20&0.20&0.20 &  $E_{\rm extra, \gamma}^{\rm max}$ requirement            &0.10&0.10&0.10&0.10&0.10&0.10&0.10&0.10&0.10&0.10\\
        $E_{\rm extra, \gamma}^{\rm max}$ requirement            &0.10&0.10&0.10&0.10&0.10&0.10&0.10&0.10&0.10&0.10 &  $e^\pm$ tracking                                         &0.20&0.20&0.20&0.20&0.20&0.20&0.20&0.20&0.20&0.38\\
        $e^\pm$ tracking                                         &0.07&0.07&0.08&0.08&0.09&0.10&0.11&0.14&0.20&0.43 &  $e^\pm$ PID                                              &0.20&0.20&0.20&0.20&0.20&0.20&0.20&0.20&0.20&0.20\\
\hline\hline
	\end{tabular}
	}
\end{table}

\begin{table}[htbp]
	\caption{The systematic covariance density matrix for the semileptonic decays $\pienu$, $\pimunu$, $\pizenu$, and $\pizmunu$, respectively.}
	\phantomsection
	\label{tab:pdwsysmatrix}
	\resizebox{\linewidth}{!}{
		\begin{tabular}{cccccccccccccccccccccc}
			\hline\hline
			& \multicolumn{10}{c}{$\pienu$}          &                        & \multicolumn{10}{c}{$\pimunu$}         \\ \cline{2-11} \cline{13-22} 
			$\rho_{ij}^{\rm sys}$ & 1 & 2 & 3 & 4 & 5 & 6 & 7 & 8 & 9 & 10 & $\rho_{ij}^{\rm sys}$ & 1 & 2 & 3 & 4 & 5 & 6 & 7 & 8 & 9 & 10 \\ \hline
			1   &1.000  &0.988  &0.990  &0.974  &0.969  &0.924  &0.831  &0.849  &0.656  &0.373  &  1   &1.000  &0.934  &0.864  &0.667  &0.498  &0.397  &0.628  &0.427  &-0.062 &-0.071 \\
			2   &       &1.000  &1.001  &0.985  &0.981  &0.934  &0.841  &0.859  &0.664  &0.377  &  2   &       &1.000  &0.854  &0.670  &0.499  &0.403  &0.627  &0.436  &-0.051 &-0.059 \\
			3   &       &       &1.000  &0.984  &0.979  &0.933  &0.840  &0.858  &0.663  &0.376  &  3   &       &       &1.000  &0.589  &0.490  &0.280  &0.640  &0.250  &-0.251 &-0.278 \\
			4   &       &       &       &1.000  &0.996  &0.949  &0.854  &0.872  &0.674  &0.383  &  4   &       &       &       &1.000  &0.844  &0.809  &0.811  &0.740  &0.484  &0.345  \\
			5   &       &       &       &       &1.000  &0.953  &0.857  &0.876  &0.677  &0.384  &  5   &       &       &       &       &1.000  &0.814  &0.803  &0.702  &0.535  &0.361  \\
			6   &       &       &       &       &       &1.000  &0.900  &0.919  &0.710  &0.403  &  6   &       &       &       &       &       &1.000  &0.690  &0.786  &0.714  &0.551  \\
\hline\hline
		\end{tabular}
	}
\end{table}

\begin{table}[htbp]
	\centering
	\caption{The statistical covariance density matrix for the simultaneous fit to four decay modes.}
	\phantomsection
	\label{table:statcovellnu}
	\rotatebox{90}{
		\resizebox{\textheight}{!}{
			\begin{tabular}{cc|cccccccccc|cccccccccc|cccccccccc|cccccccccc}
				\hline\hline
				& & \multicolumn{10}{c|}{$\pienu$} & \multicolumn{10}{c|}{$\pimunu$} & \multicolumn{10}{c|}{$\pizenu$} &  \multicolumn{10}{c}{$\pizmunu$} \\
				\hline
				&  $\rho_{ij}^{\rm stat}$&1&2&3&4&5&6&7&8&9&10&1&2&3&4&5&6&7&8&9&10&1&2&3&4&5&6&7&8&9&10&1&2&3&4&5&6&7&8&9&10\\
				\hline
				\multirow{10}{*}{\rotatebox{270}{$\pienu$}} &  1    &1.000  &-0.058 &0.001  &-0.001 &-0.000 &-0.000 &0.000  &-0.000 &-0.000 &-0.000 &0.000  &0.000  &0.000  &0.000  &0.000  &0.000  &0.000  &0.000  &0.000  &0.000  &0.000  &0.000  &0.000  &0.000  &0.000  &0.000  &0.000  &0.000  &0.000  &0.000  &0.000  &0.000  &0.000  &0.000  &0.000  &0.000  &0.000  &0.000  &0.000  &0.000  \\
				&  2   &       &1.000  &-0.069 &0.000  &-0.001 &-0.000 &-0.000 &-0.000 &-0.000 &-0.000 &0.000  &0.000  &0.000  &0.000  &0.000  &0.000  &0.000  &0.000  &0.000  &0.000  &0.000  &0.000  &0.000  &0.000  &0.000  &0.000  &0.000  &0.000  &0.000  &0.000  &0.000  &0.000  &0.000  &0.000  &0.000  &0.000  &0.000  &0.000  &0.000  &0.000  \\
				&  3   &       &       &1.000  &-0.072 &-0.001 &-0.002 &-0.001 &-0.000 &-0.000 &-0.000 &0.000  &0.000  &0.000  &0.000  &0.000  &0.000  &0.000  &0.000  &0.000  &0.000  &0.000  &0.000  &0.000  &0.000  &0.000  &0.000  &0.000  &0.000  &0.000  &0.000  &0.000  &0.000  &0.000  &0.000  &0.000  &0.000  &0.000  &0.000  &0.000  &0.000  \\
				&  4   &       &       &       &1.000  &-0.071 &-0.001 &-0.002 &-0.001 &-0.000 &-0.000 &0.000  &0.000  &0.000  &0.000  &0.000  &0.000  &0.000  &0.000  &0.000  &0.000  &0.000  &0.000  &0.000  &0.000  &0.000  &0.000  &0.000  &0.000  &0.000  &0.000  &0.000  &0.000  &0.000  &0.000  &0.000  &0.000  &0.000  &0.000  &0.000  &0.000  \\
\hline\hline
			\end{tabular}
		}
	}
\end{table}

\begin{table}[htbp]
	\centering
	\caption{The systematic covariance density matrix for the simultaneous fit to four decay modes.}
	\phantomsection
	\label{table:systcovellnu}
	\rotatebox{90}{
		\resizebox{\textheight}{!}{
			\begin{tabular}{cc|cccccccccc|cccccccccc|cccccccccc|cccccccccc}
				\hline\hline
				& & \multicolumn{10}{c|}{$\pienu$} & \multicolumn{10}{c|}{$\pimunu$} & \multicolumn{10}{c|}{$\pizenu$} &  \multicolumn{10}{c}{$\pizmunu$} \\
				\hline
				&  $\rho_{ij}^{\rm syst}$&1&2&3&4&5&6&7&8&9&10&1&2&3&4&5&6&7&8&9&10&1&2&3&4&5&6&7&8&9&10&1&2&3&4&5&6&7&8&9&10\\
				\hline
				\multirow{10}{*}{\rotatebox{270}{$\pienu$}} &  1    &1.000  &0.988  &0.990  &0.974  &0.969  &0.924  &0.831  &0.849  &0.656  &0.373  &0.188  &0.189  &0.167  &0.214  &0.195  &0.182  &0.196  &0.172  &0.092  &0.070  &0.160  &0.153  &0.108  &0.112  &0.035  &0.084  &0.120  &0.032  &0.042  &0.038  &0.000  &0.000  &0.000  &0.000  &0.000  &0.000  &0.000  &0.000  &0.000  &0.000  \\
				&  2   &       &1.000  &1.001  &0.985  &0.981  &0.934  &0.841  &0.859  &0.664  &0.377  &0.186  &0.187  &0.165  &0.211  &0.193  &0.180  &0.193  &0.170  &0.091  &0.069  &0.158  &0.151  &0.106  &0.111  &0.034  &0.083  &0.118  &0.032  &0.042  &0.038  &0.000  &0.000  &0.000  &0.000  &0.000  &0.000  &0.000  &0.000  &0.000  &0.000  \\
				&  3   &       &       &1.000  &0.984  &0.979  &0.933  &0.840  &0.858  &0.663  &0.376  &0.186  &0.187  &0.166  &0.211  &0.193  &0.180  &0.194  &0.170  &0.091  &0.069  &0.162  &0.155  &0.109  &0.113  &0.035  &0.085  &0.122  &0.033  &0.043  &0.039  &0.000  &0.000  &0.000  &0.000  &0.000  &0.000  &0.000  &0.000  &0.000  &0.000  \\
				&  4   &       &       &       &1.000  &0.996  &0.949  &0.854  &0.872  &0.674  &0.383  &0.186  &0.187  &0.166  &0.212  &0.193  &0.181  &0.194  &0.170  &0.091  &0.070  &0.159  &0.153  &0.107  &0.112  &0.034  &0.084  &0.120  &0.032  &0.042  &0.039  &0.000  &0.000  &0.000  &0.000  &0.000  &0.000  &0.000  &0.000  &0.000  &0.000  \\
\hline\hline
			\end{tabular}
		}
	}
\end{table}

\begin{table}[htbp]
	\caption{The efficiency matrix (in \%) of the forward-backward asymmetries in different $q^2$ intervals for $D^{0} \to \pi^{-} e^{+} \nu_{e}$.}
	\phantomsection
	\label{table:afbeffpienu}
	\resizebox{\linewidth}{!}{
		\centering
\begin{tabular}{cc|cccccccccc|cccccccccc}
			\hline\hline
			\multicolumn{2}{c|}{Reconstructed} & \multicolumn{10}{c|}{Forward} & \multicolumn{10}{c}{Backward} \\
			\multicolumn{2}{c|}{$q^2$ bin}&1&2&3&4&5&6&7&8&9&10&1&2&3&4&5&6&7&8&9&10\\
			\hline
			\multirow{10}{*}{\rotatebox{90}{Forward}} &1    &62.43  &2.30   &0.03   &0.00   &0.00   &0.00   &0.00   &0.00   &0.00   &0.00   &0.66   &0.01   &0.00   &0.00   &0.00   &0.00   &0.00   &0.00   &0.00   &0.00   \\
			&2 &1.97   &60.16  &2.34   &0.03   &0.00   &0.00   &0.00   &0.00   &0.00   &0.00   &0.16   &0.57   &0.02   &0.00   &0.00   &0.00   &0.00   &0.00   &0.00   &0.00   \\
			&3 &0.13   &2.44   &59.85  &2.16   &0.03   &0.00   &0.00   &0.00   &0.00   &0.00   &0.04   &0.20   &0.55   &0.02   &0.00   &0.00   &0.00   &0.00   &0.00   &0.00   \\
			&4 &0.06   &0.20   &2.64   &59.70  &1.86   &0.03   &0.00   &0.00   &0.00   &0.00   &0.01   &0.10   &0.16   &0.64   &0.03   &0.00   &0.00   &0.00   &0.00   &0.00   \\
			&5 &0.03   &0.07   &0.23   &2.63   &59.34  &1.69   &0.03   &0.00   &0.00   &0.00   &0.00   &0.05   &0.11   &0.14   &0.73   &0.04   &0.00   &0.00   &0.00   &0.00   \\
			&6 &0.01   &0.03   &0.07   &0.23   &2.50   &58.34  &1.54   &0.02   &0.00   &0.00   &0.00   &0.02   &0.06   &0.10   &0.13   &0.87   &0.06   &0.00   &0.00   &0.00   \\
\hline\hline
		\end{tabular}
	}
\end{table}

\begin{table}[htbp]
	\caption{The efficiency matrix (in \%) of the forward-backward asymmetries in different $q^2$ intervals for $D^{0} \to \pi^{-} \mu^{+} \nu_{\mu}$.}
	\phantomsection
	\label{table:afbeffpimunu}
	\resizebox{\linewidth}{!}{
		\centering
		\begin{tabular}{cc|cccccccccc|cccccccccc}
			\hline\hline
			\multicolumn{2}{c|}{Reconstructed} & \multicolumn{10}{c|}{Forward} & \multicolumn{10}{c}{Backward} \\
			\multicolumn{2}{c|}{$q^2$ bin}&1&2&3&4&5&6&7&8&9&10&1&2&3&4&5&6&7&8&9&10\\
			\hline
			\multirow{10}{*}{\rotatebox{90}{Forward}} &1    &35.76  &1.00   &0.01   &0.00   &0.00   &0.00   &0.00   &0.00   &0.00   &0.00   &0.48   &0.01   &0.00   &0.00   &0.00   &0.00   &0.00   &0.00   &0.00   &0.00   \\
			&2 &1.37   &39.46  &1.26   &0.02   &0.00   &0.00   &0.00   &0.00   &0.00   &0.00   &0.01   &0.31   &0.01   &0.00   &0.00   &0.00   &0.00   &0.00   &0.00   &0.00   \\
			&3 &0.10   &1.50   &40.87  &1.29   &0.02   &0.00   &0.00   &0.00   &0.00   &0.00   &0.00   &0.01   &0.32   &0.02   &0.00   &0.00   &0.00   &0.00   &0.00   &0.00   \\
			&4 &0.05   &0.12   &1.57   &41.74  &1.24   &0.02   &0.00   &0.00   &0.00   &0.00   &0.00   &0.00   &0.01   &0.36   &0.02   &0.00   &0.00   &0.00   &0.00   &0.00   \\
			&5 &0.03   &0.06   &0.15   &1.59   &42.39  &1.18   &0.02   &0.00   &0.00   &0.00   &0.00   &0.00   &0.00   &0.01   &0.47   &0.03   &0.00   &0.00   &0.00   &0.00   \\
			&6 &0.01   &0.02   &0.05   &0.14   &1.46   &42.31  &1.08   &0.01   &0.00   &0.00   &0.00   &0.00   &0.00   &0.00   &0.02   &0.61   &0.05   &0.00   &0.00   &0.00   \\
\hline\hline
		\end{tabular}
	}
\end{table}

\begin{table}[htbp]
	\caption{The efficiency matrix (in \%) of the forward-backward asymmetries in different $q^2$ intervals for $D^{+} \to \pi^{0} e^{+} \nu_{e}$.}
	\phantomsection
	\label{table:afbeffpi0enu}
	\resizebox{\linewidth}{!}{
		\centering
		\begin{tabular}{cc|cccccccccc|cccccccccc}
			\hline\hline
			\multicolumn{2}{c|}{Reconstructed} & \multicolumn{10}{c|}{Forward} & \multicolumn{10}{c}{Backward} \\
			\multicolumn{2}{c|}{$q^2$ bin}&1&2&3&4&5&6&7&8&9&10&1&2&3&4&5&6&7&8&9&10\\
			\hline
			\multirow{10}{*}{\rotatebox{90}{Forward}} &1    &43.57  &1.42   &0.01   &0.00   &0.00   &0.00   &0.00   &0.00   &0.00   &0.00   &0.44   &0.01   &0.00   &0.00   &0.00   &0.00   &0.00   &0.00   &0.00   &0.00   \\
			&2 &4.41   &38.42  &1.45   &0.01   &0.00   &0.00   &0.00   &0.00   &0.00   &0.00   &0.11   &0.33   &0.02   &0.00   &0.00   &0.00   &0.00   &0.00   &0.00   &0.00   \\
			&3 &0.13   &5.51   &35.79  &1.35   &0.01   &0.00   &0.00   &0.00   &0.00   &0.00   &0.03   &0.13   &0.31   &0.02   &0.00   &0.00   &0.00   &0.00   &0.00   &0.00   \\
			&4 &0.03   &0.17   &5.64   &34.07  &1.23   &0.01   &0.00   &0.00   &0.00   &0.00   &0.01   &0.07   &0.10   &0.28   &0.03   &0.00   &0.00   &0.00   &0.00   &0.00   \\
			&5 &0.01   &0.04   &0.18   &5.16   &32.75  &1.08   &0.03   &0.00   &0.00   &0.00   &0.00   &0.03   &0.07   &0.07   &0.32   &0.04   &0.00   &0.00   &0.00   &0.00   \\
			&6 &0.01   &0.02   &0.05   &0.18   &4.58   &31.37  &1.00   &0.03   &0.01   &0.01   &0.00   &0.02   &0.04   &0.06   &0.06   &0.35   &0.05   &0.00   &0.00   &0.00   \\
\hline\hline
		\end{tabular}
	}
\end{table}

\begin{table}[htbp]
	\caption{The efficiency matrix (in \%) of the forward-backward asymmetries in different $q^2$ intervals for $D^{+} \to \pi^{0} \mu^{+} \nu_{\mu}$.}
	\phantomsection
	\label{table:afbeffpi0munu}
	\resizebox{\linewidth}{!}{
		\centering
		\begin{tabular}{cc|cccccccccc|cccccccccc}
			\hline\hline
			\multicolumn{2}{c|}{Reconstructed} & \multicolumn{10}{c|}{Forward} & \multicolumn{10}{c}{Backward} \\
			\multicolumn{2}{c|}{$q^2$ bin}&1&2&3&4&5&6&7&8&9&10&1&2&3&4&5&6&7&8&9&10\\
			\hline
			\multirow{10}{*}{\rotatebox{90}{Forward}} &1    &25.29  &0.64   &0.00   &0.00   &0.00   &0.00   &0.00   &0.00   &0.00   &0.00   &0.33   &0.01   &0.00   &0.00   &0.00   &0.00   &0.00   &0.00   &0.00   &0.00   \\
			&2 &3.26   &25.60  &0.85   &0.01   &0.00   &0.00   &0.00   &0.00   &0.00   &0.00   &0.01   &0.18   &0.01   &0.00   &0.00   &0.00   &0.00   &0.00   &0.00   &0.00   \\
			&3 &0.10   &3.85   &24.66  &0.87   &0.01   &0.00   &0.00   &0.00   &0.00   &0.00   &0.00   &0.00   &0.15   &0.02   &0.00   &0.00   &0.00   &0.00   &0.00   &0.00   \\
			&4 &0.03   &0.11   &3.85   &24.13  &0.86   &0.01   &0.00   &0.00   &0.00   &0.01   &0.00   &0.00   &0.00   &0.17   &0.02   &0.00   &0.00   &0.00   &0.00   &0.00   \\
			&5 &0.01   &0.03   &0.11   &3.57   &23.61  &0.79   &0.01   &0.00   &0.00   &0.00   &0.00   &0.00   &0.00   &0.01   &0.21   &0.03   &0.00   &0.00   &0.00   &0.00   \\
			&6 &0.01   &0.01   &0.03   &0.11   &3.11   &22.68  &0.71   &0.02   &0.01   &0.00   &0.00   &0.00   &0.00   &0.00   &0.01   &0.24   &0.03   &0.00   &0.00   &0.00   \\
\hline\hline
		\end{tabular}
	}
\end{table}

\begin{table}[htbp]
	\caption{The statistical covariance density matrix of $\mathcal{A}_{FB}(q_{i}^2)$ in different $q^2$ intervals for $\pienu$, $\pimunu$, $\pizenu$, and $\pizmunu$, respectively.}
	\phantomsection
	\label{tab:afbstatmatrix}
	\resizebox{\linewidth}{!}{
	\begin{tabular}{ccccccccccc|ccccccccccc}
		\hline\hline
		& \multicolumn{10}{c|}{$\pienu$}          &                        & \multicolumn{10}{c}{$\pimunu$}         \\ \cline{2-11} \cline{13-22} 
		$\rho_{ij}^{\rm stat}$ & 1 & 2 & 3 & 4 & 5 & 6 & 7 & 8 & 9 & 10 & $\rho_{ij}^{\rm stat}$ & 1 & 2 & 3 & 4 & 5 & 6 & 7 & 8 & 9 & 10 \\ \hline
		1   &1.000  &-0.069 &0.002  &-0.001 &-0.001 &-0.000 &-0.000 &-0.000 &0.000  &-0.000 &  			1   &1.000  &-0.062 &0.001  &-0.001 &-0.001 &-0.000 &-0.000 &-0.000 &0.000  &0.000  \\
		2   &       &1.000  &-0.079 &0.001  &-0.001 &-0.000 &-0.000 &-0.000 &-0.000 &-0.000 &  			2   &       &1.000  &-0.068 &0.000  &-0.001 &-0.001 &-0.000 &-0.000 &-0.000 &-0.000 \\
		3   &       &       &1.000  &-0.081 &0.000  &-0.001 &-0.000 &0.000  &-0.000 &-0.000 &  			3   &       &       &1.000  &-0.070 &-0.001 &-0.001 &-0.000 &-0.000 &-0.000 &0.000  \\
		4   &       &       &       &1.000  &-0.076 &-0.001 &-0.001 &0.000  &0.000  &-0.000 &  			4   &       &       &       &1.000  &-0.068 &-0.001 &-0.001 &0.000  &-0.000 &-0.000 \\
		5   &       &       &       &       &1.000  &-0.072 &-0.001 &-0.001 &0.000  &0.000  &  			5   &       &       &       &       &1.000  &-0.064 &-0.002 &-0.001 &0.000  &-0.000 \\
		6   &       &       &       &       &       &1.000  &-0.068 &-0.001 &-0.000 &-0.000 &  			6   &       &       &       &       &       &1.000  &-0.060 &-0.001 &-0.000 &-0.000 \\
\hline\hline
	\end{tabular}
	}
\end{table}

\begin{table}[htbp]
	\caption{The systematic covariance density matrix of $\mathcal{A}_{FB}(q_{i}^2)$ in different $q^2$ intervals for $\pienu$, $\pimunu$, $\pizenu$, and $\pizmunu$, respectively.}
	\phantomsection
	\label{tab:afbsysmatrix}
	\resizebox{\linewidth}{!}{
	\begin{tabular}{ccccccccccc|ccccccccccc}
		\hline\hline
		& \multicolumn{10}{c|}{$\pienu$}          &                        & \multicolumn{10}{c}{$\pimunu$}         \\ \cline{2-11} \cline{13-22} 
		$\rho_{ij}^{\rm sys}$ & 1 & 2 & 3 & 4 & 5 & 6 & 7 & 8 & 9 & 10 & $\rho_{ij}^{\rm sys}$ & 1 & 2 & 3 & 4 & 5 & 6 & 7 & 8 & 9 & 10 \\ \hline
		1   &    1.0000 &   -0.0473 &    0.0015 &   -0.0004 &   -0.0003 &   -0.0001 &   -0.0000 &    0.0000 &    0.0000 &    0.0000 &  	1   &    1.0000 &   -0.0465 &    0.0038 &    0.0020 &    0.0009 &   -0.0053 &   -0.0052 &   -0.0026 &   -0.0008 &   -0.0001 \\
		2   &           &    1.0000 &   -0.0601 &    0.0016 &   -0.0003 &   -0.0001 &    0.0001 &    0.0000 &    0.0000 &    0.0000 &  	2   &           &    1.0000 &   -0.0519 &    0.0023 &    0.0002 &   -0.0038 &   -0.0036 &   -0.0017 &   -0.0006 &   -0.0001 \\
		3   &           &           &    1.0000 &   -0.0654 &    0.0007 &   -0.0006 &   -0.0000 &    0.0001 &    0.0000 &    0.0000 &  	3   &           &           &    1.0000 &   -0.0533 &    0.0010 &   -0.0043 &   -0.0040 &   -0.0018 &   -0.0006 &   -0.0001 \\
		4   &           &           &           &    1.0000 &   -0.0642 &    0.0005 &   -0.0006 &    0.0000 &   -0.0000 &    0.0000 &  	4   &           &           &           &    1.0000 &   -0.0531 &   -0.0029 &   -0.0038 &   -0.0015 &   -0.0005 &   -0.0000 \\
		5   &           &           &           &           &    1.0000 &   -0.0626 &   -0.0001 &   -0.0004 &    0.0000 &    0.0001 &  	5   &           &           &           &           &    1.0000 &   -0.0516 &   -0.0016 &   -0.0012 &   -0.0002 &   -0.0001 \\
		6   &           &           &           &           &           &    1.0000 &   -0.0594 &    0.0000 &   -0.0001 &   -0.0001 &  	6   &           &           &           &           &           &    1.0000 &   -0.0447 &    0.0029 &    0.0009 &    0.0003 \\
\hline\hline
	\end{tabular}
	}
\end{table}

\end{document}